\documentclass[10pt,final,doublecolumn]{IEEEtran}
\usepackage{amsthm}
\usepackage{amsmath}
\usepackage{latexsym}
\usepackage{graphicx}
\usepackage{bbding}
\usepackage{indentfirst}
\usepackage{algorithm,algorithmic}
\usepackage{setspace}
\usepackage{float}
\usepackage{epstopdf}
\usepackage{amssymb}
\usepackage{amsfonts}
\usepackage{enumerate}
\usepackage{multicol}
\usepackage{color}
\usepackage{slashbox}
\usepackage{bm}
\usepackage{amssymb}
\usepackage{stfloats}
\usepackage{epstopdf}
\usepackage{threeparttable}
\usepackage{epstopdf}
\usepackage{threeparttable}
\usepackage{subfigure}
\usepackage{makecell}
\usepackage{cite}
\usepackage{cancel} 
\usepackage{multirow}
\usepackage{array}     
\usepackage{hyperref}  
\usepackage{graphicx}  
\usepackage{float}  
\usepackage{subfigure}  

\usepackage{subcaption} 
\newcommand{\mv}[1]{\mbox{\boldmath{$ #1 $}}}
\newcommand{\bs}[1]{\boldsymbol{#1}}

\newcommand{\ib}[1]{\in\mathbb{#1}}
\newcommand{\ic}[1]{\in\mathcal{#1}}
\newcommand{\ca}[1]{\mathcal{#1}}

\IEEEoverridecommandlockouts
\allowdisplaybreaks[4]

\begin{document}
\title{A Tutorial on Six-Dimensional Movable Antenna for 6G Networks: Synergizing Positionable and Rotatable Antennas}

%Accessing Positionable and Rotatable Antennas: A Tutorial on Six-Dimensional Movable Antenna for Wireless Networks

%A Tutorial on Six-Dimensional Movable Antenna for Wireless Networks: Positionable and Rotatable Antennas

%Six-Dimensional Movable Antenna for Wireless Networks: A Tutorial
%A Tutorial on Six-Dimensional Movable Antenna (6DMA) for Wireless Networks: Positionable Antenna and Rotatable Antenna

\author{Xiaodan Shao, \IEEEmembership{Member,~IEEE}, Weidong Mei, \IEEEmembership{Member,~IEEE}, Changsheng You,
	\IEEEmembership{Member,~IEEE}, Qingqing Wu,
 \IEEEmembership{Member,~IEEE}, Beixiong Zheng,
	 \IEEEmembership{Member,~IEEE}, 
	 Cheng-Xiang Wang,
	 \IEEEmembership{Fellow, IEEE},
	 Junling Li,
	  \IEEEmembership{Member,~IEEE}, Rui Zhang, \IEEEmembership{Fellow, IEEE},
	  Robert Schober,
	\IEEEmembership{Fellow, IEEE},  Lipeng Zhu, \IEEEmembership{Member,~IEEE}, Weihua Zhuang, \IEEEmembership{Fellow, IEEE}, Xuemin (Sherman) Shen, \IEEEmembership{Fellow, IEEE}	
	
	\thanks{X. Shao, W. Zhuang, and X. Shen are with the Department of Electrical and Computer Engineering, University of Waterloo, Waterloo, ON N2L 3G1, Canada (e-mail: x6shao@uwaterloo.ca; wzhuang@uwaterloo.ca; sshen@uwaterloo.ca).}	\vspace{-20pt}
	
   \thanks{W. Mei is with the National Key Laboratory of Wireless
		Communications, University of Electronic Science and Technology of China (UESTC), Chengdu 611731, China (e-mail: wmei@uestc.edu.cn).}
	
	\thanks{C. You is with the Department of Electronic and Electrical Engineering, Southern University of Science and Technology (SUSTech), Shenzhen 518055,
		China (e-mail: youcs@sustech.edu.cn).}
	
	\thanks{Q. Wu is with the Department of Electronic Engineering, Shanghai Jiao
		Tong University, Shanghai 200240, China (e-mail: qingqingwu@sjtu.edu.cn).}

	\thanks{B. Zheng is with School of Microelectronics, South China University of
		Technology, Guangzhou 511442, China (e-mail: bxzheng@scut.edu.cn).}
	
	\thanks{C.-X. Wang and J. Li are with the National Mobile Communications Research Laboratory, School of Information Science and Engineering, Southeast University, Nanjing 211189, China (e-mail: chxwang@seu.edu.cn, junlingli@seu.edu.cn).}

	\thanks{R. Zhang is with School of Science and Engineering, Shenzhen Research Institute of Big Data, The Chinese University of Hong Kong, Shenzhen, Guangdong 518172, China. He is also with the Department of Electrical and Computer Engineering, National University of Singapore, Singapore 117583 (e-mail: elezhang@nus.edu.sg).}
	
			\thanks{R. Schober is with the Institute for Digital Communications, Friedrich-Alexander-University Erlangen-Nurnberg (FAU), 91054
		Erlangen, Germany (email: robert.schober@fau.de).}
	
	\thanks{L. Zhu is with the Department of Electrical and Computer
		Engineering, National University of Singapore, Singapore 117583 (e-mail:
		zhulp@nus.edu.sg).}

}

\maketitle

\IEEEpeerreviewmaketitle

\begin{abstract}
		Six-dimensional movable antenna (6DMA) is a new
	and revolutionary technique that fully exploits the wireless
	channel spatial variations at the transmitter/receiver by flexibly
	adjusting the three-dimensional (3D) positions and/or 3D rotations
	of antennas/antenna surfaces (sub-arrays), thereby improving  the performance of wireless
	networks cost-effectively without the need to deploy additional
	antennas. It is thus expected that
	the integration of new 6DMAs into future sixth-generation (6G) wireless networks will fundamentally enhance 
	antenna agility and adaptability, and introduce new degrees
	of freedom (DoFs) for system design. Despite its great potential,
	6DMA faces new challenges to be efficiently implemented in wireless
	networks, including corresponding architectures, antenna position and rotation optimization, channel estimation,
	and system design from both communication and sensing perspectives. In
	this paper, we provide a tutorial on 6DMA-enhanced wireless
	networks to address the above issues by unveiling associated new channel models, hardware implementations and
	practical position/rotation constraints, as well as various appealing applications in
	wireless networks. Moreover, we discuss two special cases of 6DMA, namely, rotatable 6DMA with fixed antenna position and positionable  6DMA with fixed antenna rotation, and highlight their respective design challenges and applications. 
	We further present prototypes developed for 6DMA-enhanced communication along with experimental results obtained with these prototypes. Finally, we outline promising directions for further investigation.
\end{abstract}
	
\section{Introduction}
\subsection{Motivation}
\begin{figure*}[!t]
	\centering
	\includegraphics[width=0.8\linewidth]{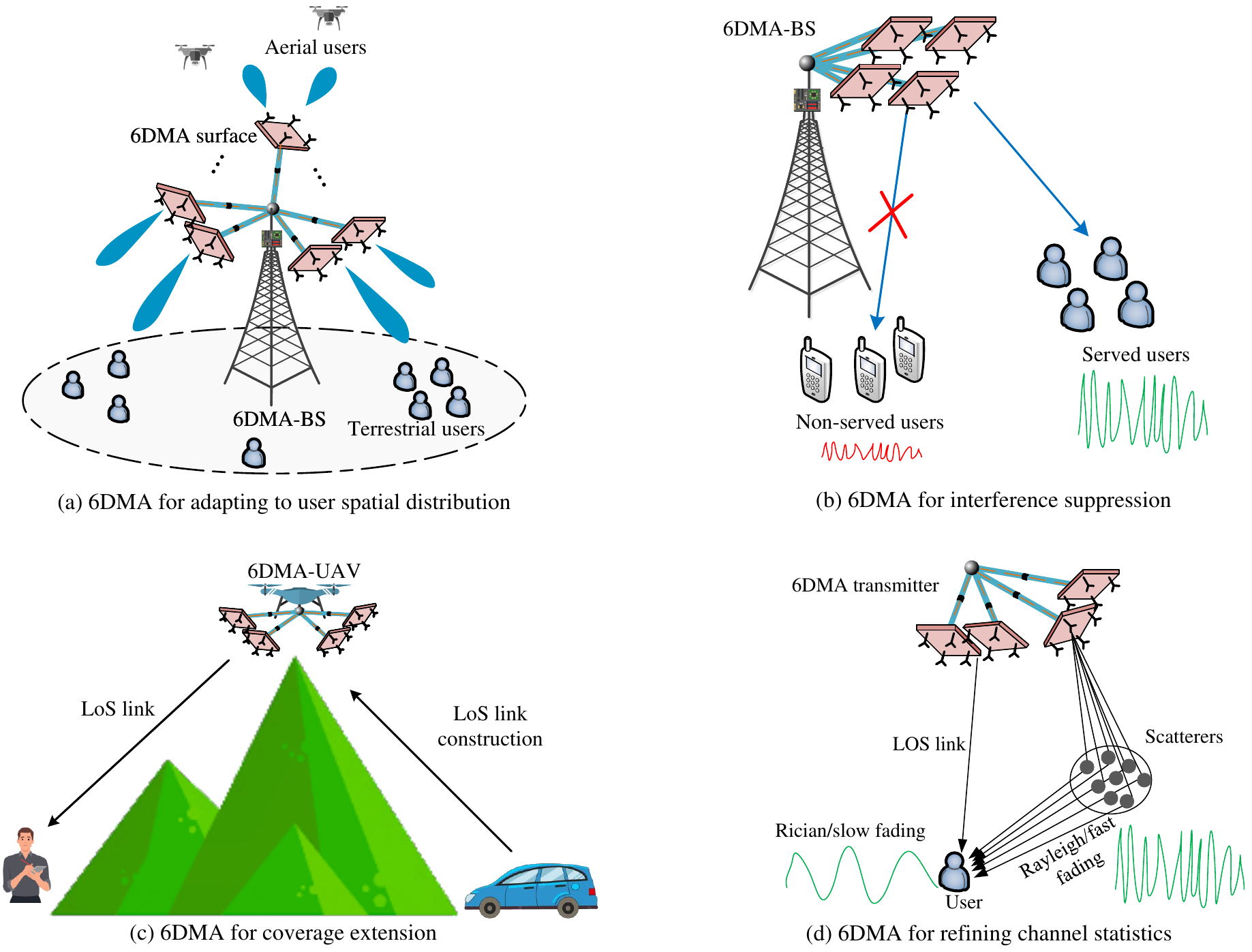}
	\caption{Main effects of 6DMA for wireless channel reconfiguration.}
	\label{function}
		\vspace{-0.59cm}
\end{figure*} 
While the fifth-generation (5G) wireless networks are in their final stages of standardization and commercial deployment, the global effort to develop the next/sixth-generation (6G) wireless networks and services has been in progress for several years \cite{wcx3}. Future 6G networks are envisioned to support a massive number of users/devices across diverse applications and services, demanding superior performance surpassing 5G's capabilities. For example, 6G will evolve from the Internet of Everything (IoE) to the Intelligence of Everything, connecting people, devices, and intelligence to drive transformative changes in our society \cite{Wang2023Realizing, wcx4, cyn}.
In terms of technical requirements, 6G will pursue global coverage (99\% coverage), ultra-high data rates (peak data rates at the Tbps level), ultra-low latency (0.1 millisecond), ultra-dense connections (up to $10^8$ devices/km$^2$), higher precision positioning (centimeter level) \cite{shao2024target,shao2022target}, and more intelligence compared to 5G \cite{Trichias20246G,Wang2023On}.
To achieve these ambitious goals, current trends in wireless technology are moving towards the allocation of substantially more radio spectrum and hardware resources. By shifting to higher frequency bands such as the millimeter wave (mmWave) and terahertz (THz) spectra, huge bandwidths and ultra-high data rates can be exploited to support emerging 6G applications and services \cite{Wang2023On}.
Additionally, with an even larger number of antennas than current massive multiple-input multiple-output (MIMO) systems, extremely large-scale MIMO can offer substantial spatial multiplexing gains and super-high resolution, leading to a dramatic increase in both channel capacity and spectrum efficiency \cite{exl}.

While wide bandwidth and large-scale arrays can offer significant performance gains, they come at the expense of prohibitive hardware costs and excessive power consumption. Specifically, hardware design complexity and implementation costs will escalate for higher frequencies and larger antenna arrays, thereby hindering progress towards future sustainable and energy-efficient wireless networks \cite{Larsson2014Massive}. Additionally, current MIMO technologies are mostly based on the fixed-position antenna (FPA) architecture, where the antenna positions and rotations are set once deployed. As a result, these FPA systems are unable to fully capitalize on the wireless channel spatial variations within their coverage areas due to their lack of spatial flexibility in antenna position and rotation.
In view of the above concerns and limitations, it is imperative to develop disruptively new and innovative technologies to break through the inherent constraints of conventional FPA technologies and enable the full exploitation of the spatial degrees
of freedom (DoFs), thereby promoting the sustainable capacity growth of future wireless networks without further increasing or even reducing the number of antennas.
\begin{figure*}[!t]
	\centering
	\includegraphics[width=0.70\linewidth]{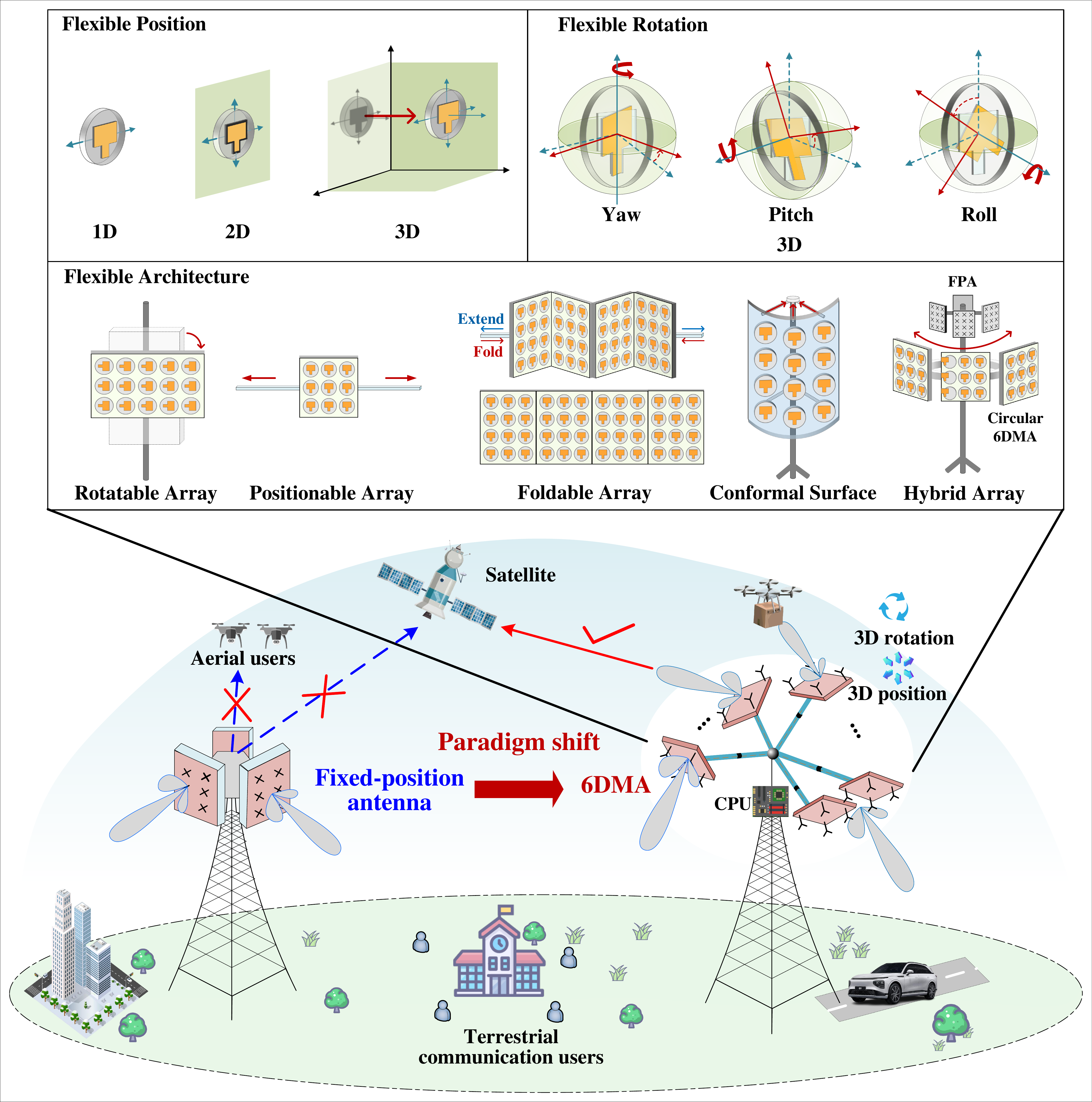}
	\caption{Position/rotation/architecture models and paradigm shifts in wireless network design with 6DMA.}
	\label{6D}
		\vspace{-0.59cm}
\end{figure*} 

\subsection{What Is 6DMA?}
Motivated by the above, six-dimensional movable antenna (6DMA) technology has been recently first proposed by {\emph {Shao et al}}. in \cite{6dmaCon, 6dmaDis, 6dmaSens, 6dmaMag,6dmaChan,near}, 
as an efficient approach to provide the full spatial DoFs and improve MIMO system capacity cost-effectively for future  wireless networks such as 6G, without the need to deploy additional
antennas. 
As shown in Fig.~\ref{function}, the 6DMA-equipped transceivers are deployed with a number of antennas/antenna surfaces, each of which can be independently adjusted in terms of both three-dimensional (3D) position and 3D rotation. By adaptively adjusting the positions and/or rotations (orientations) of the 6DMA surfaces based on the users' spatial distribution and the corresponding long-term/statistical channel state information (CSI), a 6DMA-equipped transceiver can fully exploit the spatial variations of the radio environment and proactively reshape the wireless channels to provide more favorable conditions enhancing system performance without the need for additional antennas/arrays. 6DMA-enabled transceivers are also endowed with more spatial DoFs to reshape the radiation pattern in the angular domain and can adapt to the wireless channel spatial distribution dynamically, and thereby enhance network performance. Here, channel spatial distribution refers to the spatial variations in the channel between the 6DMA-enabled transmitter/receiver and the users, which depend on the 3D spatial distribution of the users as well as the 3D spatial distribution of the scatterers in the environment.
As shown in Fig. \ref{function}, 6DMA is capable of achieving several appealing effects for wireless channel reconfiguration, such as allocating spatial DoFs to adapt to the user spatial distribution (see Fig. \ref{function} (a)), circumventing interference to/from undesired directions (see Fig. \ref{function} (b)), creating new line-of-sight (LoS) links to bypass obstacles between transceivers via flexible array positioning and rotation (see Fig. \ref{function} (c)), and compensating for path loss in rich-scattering propagation by tuning the 6DMA surfaces to cater to the dominant LoS channel path as well as non-negligible non-LoS (NLoS) paths, thereby refining the channel statistics/distribution, e.g., transforming Rayleigh/fast fading to Rician/slow fading facilitating higher rates and improved reliability (see Fig. \ref{function} (d)).

To implement 3D position and 3D rotation reconfiguration capabilities, each 6DMA surface is connected to a central processing unit (CPU) at the transceiver via an extendable and rotatable rod, which contains flexible wires for supplying power to the 6DMA surface and facilitating radio frequency (RF)/control signal exchange between the surface and the CPU.
As shown in Fig.~\ref{function}, the 6DMA-equipped transceiver adjusts the entire antenna surface as a basic unit, which reduces implementation cost and control complexity \cite{6dmaMag}.
Meanwhile, the positions and rotations of the 6DMA surfaces change only when the user spatial distribution in the network varies significantly, i.e., they adapt to the large-scale channel variations of users. Thus, each 6DMA surface can move slowly and infrequently in practice. In addition, a 6DMA surface can employ different array architectures, including single antenna, linear antenna array, planar surface, curved/conformal surface, rotatable/positionable array, foldable array, and so on (see Fig.~\ref{6D}).
Furthermore, 6DMA is compatible with a wide range of multi-antenna technologies, such as intelligent reflecting surface/reconfigurable intelligent surface (IRS/RIS) \cite{proc,ris1,huameng,9844229,yu2021low} and holographic MIMO, and can also be adopted in a hybrid deployment with conventional FPAs at the BS \cite{6dmaPar}.
\begin{figure*}[!t]
	\centering
	\includegraphics[width=0.9\linewidth]{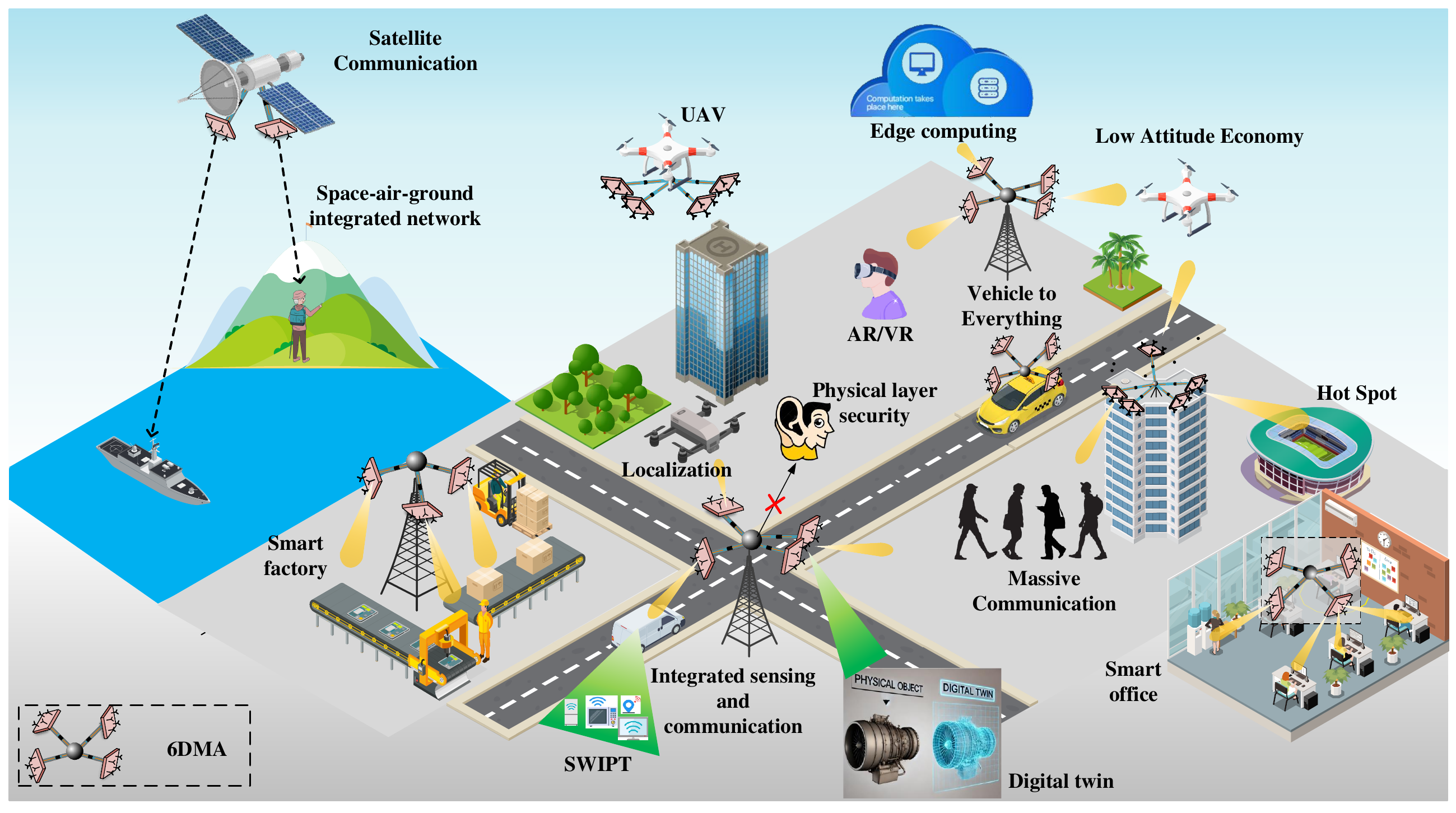}
	\caption{Illustration of 6DMA applications in future wireless network.}
	\label{fig:applications}
	\vspace{-0.59cm}
\end{figure*} 

Due to the above promising advantages, 6DMA has great potential for deployment in future wireless networks, enabling significant improvement of channel conditions for various applications and enhancing communication/sensing performance in a cost-effective manner.
In Fig.~\ref{fig:applications}, we show an envisioned future wireless network empowered by 6DMA with a variety of promising applications. For instance, it is particularly beneficial in scenarios with non-uniform or dynamic user distributions, such as massive machine-type communication (mMTC) networks, where typically only a small portion of devices are active for communication at any time \cite{jadc1,jadc6}. In this context, 6DMA can effectively enhance device activity detection accuracy and efficiency by leveraging its controllable antenna position and rotation, especially when a large number of devices are deployed under diverse propagation conditions. For a cellular network, the BS can adaptively allocate more antenna resources to serve hotspot area by leveraging 6DMA. Unlike conventional down-tilt antennas/arrays that mainly serve terrestrial users on the ground, 6DMA can lift up a portion of its antennas/arrays to extend coverage across a broader 3D space, thus further enhancing connectivity and coverage performance for the growing low-altitude economy.
In this context, by facilitating simultaneous communication between airborne and terrestrial users, 6DMA can serve as a potential enabling technology for space-air-ground integrated networks. In addition to improving communication performance, the unique directional sparsity property of 6DMA channels \cite{6dmaChan} will also elevate the accuracy of sensing and localization \cite{sens1,spoof}, further boosting the performance of integrated sensing and communications (ISAC) \cite{10540249} by flexibly exploiting their joint benefits while avoiding mutual interference in future wireless networks. Moreover, to enhance the efficiency of simultaneous wireless information and power
transfer (SWIPT) from the BS to wireless powered devices in scenarios such as a smart office/home \cite{swi}, the 6DMA can help compensate for the notable power loss over long distances by adjusting its position and rotation to align more favorably with the directions of nearby devices. In outdoor environments, 6DMA can be deployed on building rooftops, lamp posts, advertising boards, and even the surfaces of high-speed vehicles to support various applications \cite{wuwen}, such as ultra-reliable and low latency communication
(URLLC), by effectively compensating for Doppler and delay spread effects.
As a transformative technology, 6DMA has the potential to turn traditional fixed antennas into a flexible and more intelligent system, offering significant benefits to a wide range of vertical industries in 6G/Beyond-6G, including smart factories, smart offices, and smart cities. By integrating digital twin technology with 6DMA \cite{wuwen,twinxinyu,twinxin1,sxf}, a real-time virtual environment can be created to more precisely adapt antenna positions and rotations, thus further enhancing communication and sensing performance.
Notably, to leverage the benefits offered by the spatial flexibility of 6DMA and other non-fixed position/rotation antenna techniques, several relevant industrial activities, prototypes, and projects have been launched to advance the research in this field, see Table~\ref{tab:prototype}.

\subsection{What’s New?}
6DMA is essentially a distributed MIMO architecture, but it also offers the additional capability to collaboratively and adaptively adjust the positions and/or rotations of antennas based on the user spatial distribution. Compared with the traditional large-scale MIMO and antenna selection systems, which mainly rely on a large number of evenly spaced antennas to obtain high spatial diversity/resolution, 6DMA can deliver even more spatial DoFs and comparable performance with significantly fewer antennas. Below, we discuss the new challenges and advances introduced by 6DMA in comparison to state-of-the-art technologies.

\subsubsection {New challenges}
6DMA is endowed with the highest flexibility and compatibility in terms of spatial DoFs by allowing for adjustments in both the 3D positions and 3D rotations of antennas/antenna surfaces (sub-arrays), and it also provides a general model for any types of position and rotation adjustable antennas, as shown in Fig.~\ref{6D}.
The design of 6DMA-enhanced wireless systems and networks presents new and unique challenges from a communication perspective, which are elaborated in the following. Firstly, the positions and rotations of all 6DMAs at each transceiver must be carefully designed to achieve cooperative signal focusing and interference cancellation, while the design of the position and rotation adjustment of 6DMAs is further complicated by various geometric factors and practical constraints \cite{6dmaCon}.
In addition, to serve all users in the network efficiently, the positions and rotations of all 6DMAs must be jointly designed with the transmissions of the BSs and users. This joint design aims to optimize end-to-end communication over the wireless channels reconfigured by the 6DMAs. Secondly, determining the optimal 6DMA positions and rotations requires knowledge of the CSI of the channels between all possible 6DMA candidate positions and rotations and all users. Thus, compared to conventional FPA, channel estimation for 6DMA systems introduces new challenges elaborated as follows. For FPAs, the number of channel coefficients to be estimated is limited to the product of the numbers of transmit and receive antennas. In contrast, for 6DMA systems, CSI must be acquired in a continuous space that includes a vast number of candidate antenna positions and rotations. Therefore, as the numbers of antennas and candidate antenna positions/rotations increase, the computational complexity of channel estimation algorithms and their implementation time and energy costs will increase dramatically. Furthermore, the channels between a user and different candidate antenna positions/rotations in continuous 3D space generally exhibit significantly different distributions (e.g., consider the channels from a user to two
6DMA surfaces whose normal vectors point in the direction
and the opposite direction of the user, respectively), making the CSI estimation/acquisition problem for 6DMA very intricate and challenging to solve. 
Thirdly, since 6DMA surfaces need to be mechanically moved by devices such as rotary motors, the hardware cost, power consumption, and movement/rotation delay should be taken into consideration when designing and deploying 6DMA systems, especially for larger movement regions.
Finally, 6DMA  can also be flexibly customized/simplified to partially position/rotation-adjustable systems, which comprise of both positionable 6DMA (P-6DMA) with fixed antenna rotation and rotatable 6DMA (R-
 6DMA) with fixed antenna position, based on the practical requirements and constraints of
 specific use cases. These variants underscore the need for re-investigating the rotation-only and position-only optimization and application designs and characterizing their performance losses as compared to general 6DMAs. In summary, the efficient integration of 6DMAs into wireless networks introduces significant opportunities while presenting new challenges, which require dedicated and extensive studies.

\subsubsection {New Advances Compared to State-of-the-Art Technologies}
Next, we highlight the key 
differences and advantages of 6DMA as compared to the main competing technologies.
\paragraph {Fluid Antenna System}
\begin{figure*}[!t]
	\centering
	\includegraphics[width=6.3in]{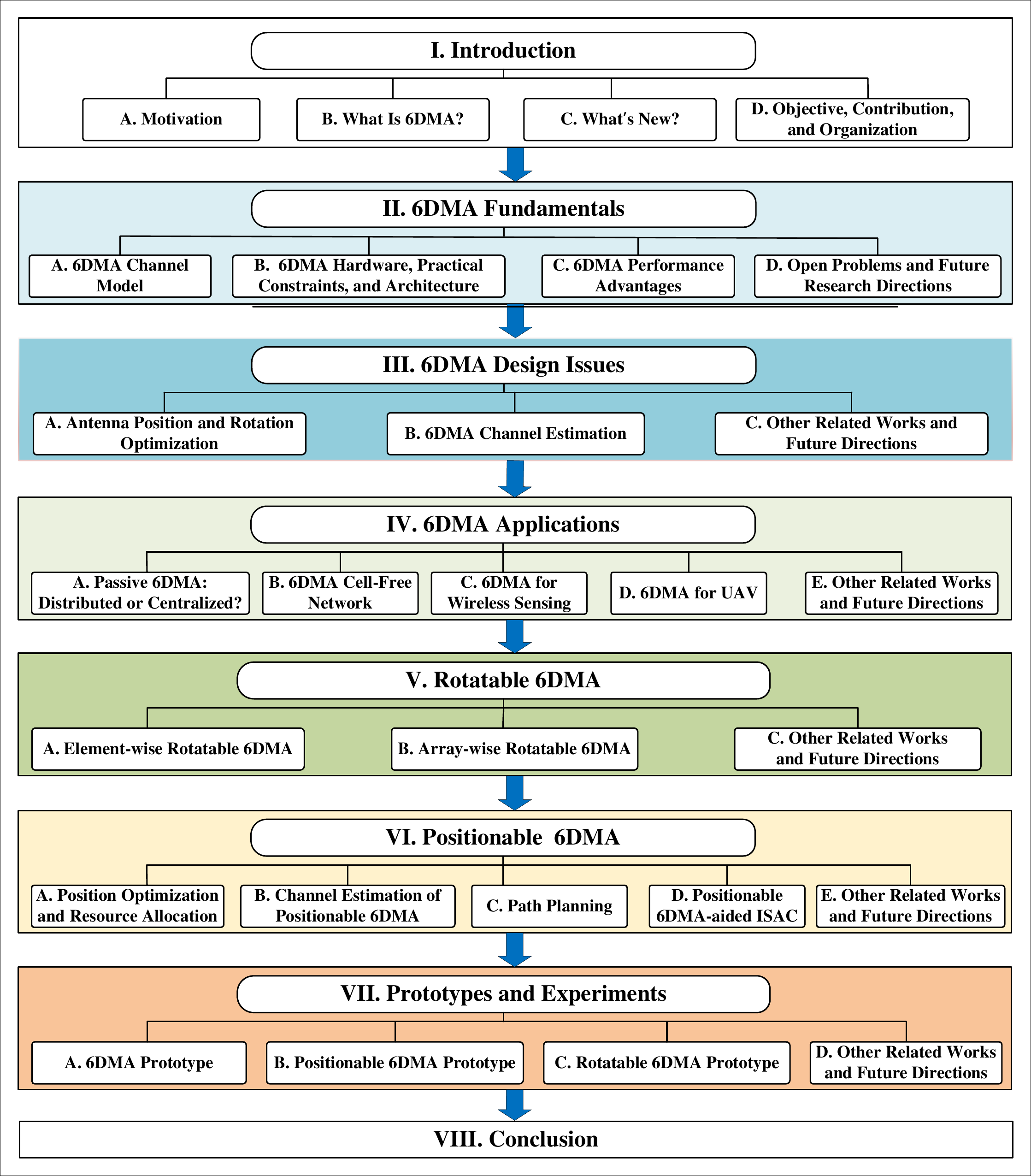}
	\caption{Organization of the tutorial.\label{fig:Organization}}
	\label{Cov_comm}
			\vspace{-0.59cm}
\end{figure*}
The fluid antenna system has emerged as an enhancement method for MIMO by exploiting antenna position flexibility \cite{New2024A,Wong2022Bruce,Wang2024AI, Zhu2024MovableAntennas,Ning2024Movable}. 6DMA systems differ significantly from the existing fluid antenna systems.
Firstly, the existing fluid antenna systems adjust the positions of the antennas on the scale of several wavelengths within a predefined line or surface while keeping the antenna rotations fixed. This results in relatively modest rate  improvements compared to FPA, typically around tens of percent. In contrast, the proposed 6DMA system introduces a new DoF via antenna rotation and allows for adjustment in both the 3D position and 3D rotation of each antenna within the 3D space. Moreover, the antenna position/rotation of 6DMA can be adjusted
on a much larger scale than several wavelengths.
Due to its 6D mobility and enhanced flexibility, 6DMA-aided systems offer significantly greater flexibility in positioning and orienting antenna surfaces to align with the user distribution, as illustrated in Fig. \ref{function}. This capability leads to substantial enhancement in network capacity, with performance improvements exceeding several times that of FPAs \cite{6dmaCon}.
Furthermore, the performance of flexible-position-only antenna systems is practically limited by the response/switching time or the movement speed of the antennas, as they usually require frequent movement (on the order of milliseconds or even microseconds) of individual antennas to quickly adapt to small-scale channel variations, leading to high implementation and energy costs. In contrast, 6DMA adjustments in position and/or orientation are needed much less often because its primary performance advantages come from the adaptive allocation of antenna resources and spatial DoFs, which is determined by the channel's spatial distribution. As the channel distribution changes more gradually in practical scenarios, these adjustments are typically required over much longer intervals, such as hours, days, or even longer \cite{6dmaDis}.

\paragraph{Antenna Selection}
6DMA systems also differ fundamentally from traditional antenna selection systems  \cite{as9,as10,sele}. First, in traditional antenna selection, the antennas are selected for communication from a fixed, predetermined set of antennas at fixed positions. In contrast, 6DMA systems optimize not only the positions but also the rotations of all 6DMA surfaces. Moreover, 6DMA systems require far fewer physically deployed radiating elements and avoid their additional cost and energy consumption. However, on the other hand, the optimal adjustments must adhere to practical constraints on antenna position and rotation, which significantly increases the complexity of solving the resulting optimization problem as compared to antenna selection. Second, unlike traditional antenna systems where radiating elements are typically spaced by half a wavelength, a 6DMA device can precisely configure both the positions and rotations of its radiating elements. This capability enables it to exploit the full spatial variations of the channel, even within small spaces of wavelength size only. Moreover, in multiuser MIMO scenarios, 6DMA offers additional advantages by enabling more distinct spatial signatures with respect to the users to enhance their received signals while simultaneously suppressing multiuser interference. Unlike conventional antenna selection with antenna on/off switching only, 6DMA can reconfigure both the array steering vector and the radiating elements’ positions and rotations to achieve more design flexibility than antenna selection. Finally, by intelligently adjusting the positions and rotations of the radiating elements, 6DMA can achieve a channel hardening effect comparable to, or even stronger than, that of traditional antenna selection methods. 

\paragraph{Extremely Large-Scale MIMO}
Extremely large-scale MIMO (XL-MIMO) deploys a vast number of antennas within a compact area to achieve high spatial resolution and capacity \cite{exl}. Compared to XL-MIMO, the proposed 6DMA system offers similar or superior performance with significantly fewer radiating elements and RF chains. For example, a 6DMA device can operate with just two active 6DMA surfaces, matching the performance of an XL-MIMO setup that requires many more antennas. In addition, 6DMA's 6D movability, encompassing both the 3D positions and 3D rotations of distributed antennas, enables precise adaptation to dynamic channel conditions and user distributions. This flexibility allows 6DMA to optimize the spatial DoF in real-time thereby enhancing spectral efficiency while reducing hardware complexity and energy consumption. In multiuser scenarios, 6DMA can create unique spatial signatures for users more effectively than XL-MIMO, leading to improved signal reception and interference suppression. Overall, 6DMA is an attractive alternative to XL-MIMO, as it combines high performance with reduced implementation costs and offers increased operational flexibility, which makes it well-suited for future dense and dynamic wireless networks.
\begin{table*}[!t]
	\footnotesize
	\vspace{-0.88cm}
	\caption{List of Main Industry Activities, Prototypes, and Projects Related to 6DMA }
	\begin{tabular}{|m{0.1\textwidth}<{\centering}|m{0.05\textwidth}<{\centering}|m{0.78\textwidth}|}  \hline
		\textbf{Reference}      &\textbf{Year} & \multicolumn{1}{c|}{\textbf{Main Objective}}\\ \hline
		\cite{Park2005A} & 2005 & Design a mobile antenna for multimedia communications with a Ku-band geostationary satellite.  \\ \hline
		\cite{Tam2011Electrolytic} & 2008 & Design the first electrolytic fluid antenna.  \\ \hline
		\cite{Fayad2009Mechanically} & 2009 & Demonstrate a heterogeneous dielectric resonator antenna.  \\ \hline
		\cite{Khaleel2013Design} & 2013 & Demonstrate the design, numerical simulation, and fabrication of flexible antennas.  \\ \hline
		\cite{Silva2015Tx} & 2015 & Design a low-cost mobile ground terminal antenna for Ka-band satellite operating in the 20-30 GHz range.  \\ \hline
		\cite{Zhuravlev2015Experimental} & 2015 & Demonstrate a setup for experimental simulation of sparse antenna arrays by independently moving one transmitting antenna and one receiving antenna.  \\ \hline
		\cite{Cosker2016Realization} & 2016 & Design a miniaturized inverted-F antenna operating at 885 MHz, suitable for wearable applications.  \\ \hline
		\cite{Dey2016Microfluidically} & 2016 & Design a microfluidically reconfigured wideband frequency-tunable liquid-metal monopole antenna that operates from 1.29 to 5.17 GHz.  \\ \hline
		\cite{Basbug2017Design} & 2017 & Demonstrate a reconfigurable antenna array design in which the elements can move along a semicircular path.  \\ \hline
		\cite{Lotfi2017Printed} & 2017 & Design an endfire beam-steerable planar printed pixel antenna that can beam steer through 300° in the azimuth plane without using any phase shifters.  \\ \hline
		\cite{Song2019Wideband} & 2019 & Design a wideband frequency reconfigurable patch antenna with switchable slots based on liquid metal manipulation in a 3D printed microfluidic channel.  \\ \hline
		\cite{Teng2019Robust} & 2019 & Design Galinstan-based liquid metal coil wearable antennas within polydimethylsiloxane (PDMS) substrates for wireless power transmission applications.  \\ \hline
		\cite{Shen2021Reconfigurable} & 2021 & Demonstrate a surface wave fluid antenna that realizes beam shaping and spatial diversity for MIMO.  \\ \hline
		\cite{Shen2022Radiation} & 2022 & Design a moving antenna that performs specific motions on a platform during vehicle maneuvers.   \\ \hline
		\cite{Shen2022RadiationPattern} & 2022 & Demonstrate an antenna design that combines surface wave and fluidic reconfigurable techniques with one fluid radiator channel.  \\ \hline
		\cite{Wang2022Continuous} & 2022 & Demonstrate an antenna design that combines surface wave and fluidic reconfigurable techniques with two radiator channels.  \\ \hline
		\cite{Martinez2022Toward} & 2022 & Demonstrate a reconfigurable fluid antenna that utilizes continuous electrowetting techniques for achieving agile radiation patterns.  \\ \hline
		\cite{Li2022Using} & 2022 & Design a surface-wave-based fluid antenna system.  \\ \hline
		\cite{Shen2024Design} & 2024 & Design a dynamically reconfigurable fluid radiator capable of adjusting its position within a predefined space.  \\ \hline
		\cite{Zhang2024A} & 2024 & Demonstrate a pixel-based reconfigurable antenna that is designed to meet the requirements of 2.5 GHz fluid antenna system and the specified switching speed.  \\ \hline
		\cite{Dong2024Movable} & 2024 & Present a hardware architecture that can be separated into a communication module and an antenna positioning module.   \\ \hline
		\cite{Zhu2024MovableAntennas} & 2024 & Design a prototype of a two-dimensional positionable 6DMA communication system operating at 3.5 GHz or 27.5 GHz with ultra-accurate movement control.  \\ \hline
		\cite{Wang2024Movable} & 2024 & Demonstrate a broadband channel measurement system with physical positionable 6DMAs and an extremely high movable resolution reaching 0.02 mm. \\ 
	\end{tabular}
	\begin{tabular}{|m{0.1\textwidth}<{\centering}|m{0.43\textwidth}|m{0.4\textwidth}|}  \hline
		\textbf{Company} & \multicolumn{1}{c|}{\textbf{Main Product}} & \multicolumn{1}{c|}{\textbf{Link}}\\ \hline
		SHXNewStar & 1.8m-4.5m Ka-band satellite communication antenna & \href{https://www.vastantenna.com/earth-station-antenna/}{https://www.vastantenna.com/earth-station-antenna/} \\ \hline
		CPI Comm & GEO/LEO/MEO full motion satellite antenna & \href{https://www.cpii.com/product.cfm/15/112}{https://www.cpii.com/product.cfm/15/112} \\ \hline
		C-COM & Ground auto-acquire VSAT broadband internet antenna & \href{https://www.c-comsat.com/inetvu-mobile-products/?all}{https://www.c-comsat.com/inetvu-mobile-products/?all} \\ \hline
		SAT-LITE & Portable X-Y GEO/LEO/MEO tracking antenna & \href{https://www.sat-litetech.com/antennas/leo-meo-antennas/}{https://www.sat-litetech.com/antennas/leo-meo-antennas/} \\ \hline
		SATPRO & Aircraft/vehicle/ship full motion satellite antenna & \href{https://www.satpro.cn/products.asp?id=1}{https://www.satpro.cn/products.asp?id=1} \\ \hline
		Huawei & AirEngine 8760-X1-PRO smart antenna & \href{https://e.huawei.com/cn/material/networking/wlan/a33b1684cf284bdcb990f3f38dff2672}{https://e.huawei.com/cn/material/networking/wlan} \\ \hline
	\end{tabular}
		\begin{tablenotes}
		\footnotesize
		\item[1] GEO: Geostationary earth orbit; LEO: Low earth orbit; MEO: Medium earth orbit.
	\end{tablenotes}
	\label{tab:prototype}
\end{table*}

\subsection{Objectives, Contributions, and Organization}
The high potential of 6DMA and other related techniques for application in future wireless networks has spurred extensive research recently. A handful of articles have appeared in the literature providing overviews or surveys on various aspects of non-fixed position/rotation antenna techniques, including their practical implementation, channel modeling, applications, and other design issues \cite{6dmaMag, Zhu2024MovableAntennas,Ning2024Movable,Zheng2024Flexible,New2024A,Wong2022Bruce,Wang2024AI,Shojaeifard2022MIMO}, and are summarized in Table~\ref{tab:survey}. However, the aforementioned works have not addressed general models and architectures for both antenna position and rotation adjustments. Compared to all existing works targeting non-fixed position antenna techniques, this paper is the first to offer an in-depth tutorial on the most spatially flexible option, 6DMA, with flexible antenna position and rotation, aiming to provide new and comprehensive guidance to facilitate and inspire future research on 6DMA-empowered wireless networks. To this end, this tutorial covers 6DMA system models, implementation architectures, advantages, design issues, simplified realizations, namely, positionable 6DMA and rotatable 6DMA, and potential applications in various wireless systems.

As depicted in Fig.~\ref{fig:Organization}, the remainder of this paper is structured as follows. Section II introduces the fundamentals of 6DMA, including its channel model and architecture, hardware, practical constraints, and performance
advantages. In Section III, we discuss several critical design issues associated with 6DMA, including channel estimation and antenna position and rotation optimization. In addition, we provide forward-looking solutions and directions for future exploration. Section IV details potential applications of 6DMA. 
In Sections V and VI, we elaborate further on two simplified realizations of 6DMA, i.e., rotatable 6DMA with fixed antenna position and positionable 6DMA with fixed antenna rotation, respectively. In particular, we discuss specific applications and design challenges introduced by antenna flexibility in only rotation and only position, respectively. Section VII provides prototypes of 6DMA, rotatable 6DMA, and positionable 6DMA systems and corresponding
experimental results. Finally, we conclude this paper in Section VIII. 

\emph{Notations}: Here, $(\cdot)^H$ and $(\cdot)^T$ represent the conjugate transpose and transpose operations, respectively, $\mathbb{E}[\cdot]$ denotes the expectation of a random variable, $\left \|\cdot\right \|_2$ refers to the Euclidean norm, $\mathrm{diag}({\bf x})$ generates a diagonal matrix with the elements of vector ${\bf x}$ on its diagonal, $[\mathbf{a}]_j$ indicates the $j$-th component of vector $\mathbf{a}$, $[\mathbf{A}]_{i,j}$ specifies the element in the $i$-th row and $j$-th column of matrix $\mathbf{A}$, $\mathcal{B}/b$ implies that element $b$ is removed from set $\mathcal{B}$, functions $\max\{\cdot\}$ and $\min\{\cdot\}$ select the maximum and minimum values from a given set, respectively, the union of two sets is symbolized by $\cup$, operators \( \Re(\cdot) \) and \( \Im(\cdot) \) extract the real and imaginary parts of a complex number, respectively, ${N}!$ denotes the number of permutations of ${N}$ elements, \(\mathbf{a} \cdot \mathbf{b}\) denotes the dot product of vectors \(\mathbf{a}\) and \(\mathbf{b}\), $|a|$ signifies the magnitude of complex number $a$, $\otimes$ denotes the Kronecker product, $\mathbf{I}_N$ denotes the $N \times N$ identity matrix, \(\mathbf{1}_{N\times 1}\) denotes the all-ones vector with dimensions \(N \times 1\), $\boldsymbol{\Gamma}^{\mathrm{c}}$ denotes the complementary set of set $\boldsymbol{\Gamma}$, $\text{tr}(\cdot)$ denotes the  trace of a matrix, and \(\mathcal{CN}(0, \sigma^2)\) defines a circularly symmetric complex Gaussian (CSCG) distribution with mean 0 and variance \(\sigma^2\).

\begin{table*}[!t]
	\footnotesize
		\vspace{-0.88cm}
	\caption{List of Representative Survey/Overview Papers Related to 6DMA }
	\begin{tabular}{|m{0.1\textwidth}<{\centering}|m{0.85\textwidth}|}  \hline
		\textbf{Reference}      & \multicolumn{1}{c|}{\textbf{Main Contributions}}\\ \hline
		\cite{6dmaMag} & Provides an overview of 6DMA technology, highlighting its potential in wireless networks, including its motivation, competitive advantages, system/channel modeling, and practical implementation. \\ \hline
		\cite{Zhu2024MovableAntennas} & Discusses the applications of movable antennas in wireless communication, including their hardware architecture, channel characteristics, performance benefits over FPA, and design challenges with potential solutions.\\ \hline
		\cite{Zheng2024Flexible} & Introduces the fundamentals of flexible-position MIMO systems, including hardware design, structural design, and potential applications. \\ \hline
		\cite{New2024A} & Provides a comprehensive tutorial on fluid antenna system, covering theoretical foundations, practical implementations, and challenges, and explores their potential to empower other contemporary technologies. \\ \hline
		\cite{Ning2024Movable} & Proposes various positionable 6DMA architectures incorporating ideas from antenna design and mechanical control for implementation methods.  \\ \hline
		\cite{Wong2022Bruce} & Introduces the concept and approach behind fluid antenna system, explores their potential in 6G technology, and outlines six key research topics.  \\ \hline
		\cite{Wang2024AI} & Discusses the inherent challenges of fluid antenna system, emphasizing the role of artificial intelligence (AI) in addressing its complexities. \\ \hline
		\cite{Shojaeifard2022MIMO} & Provides an overview of IRS for free-space and surface-wave communications, and introduces fluid antenna operating principles.  \\ \hline	
	\end{tabular}
	\label{tab:survey}
\end{table*}

\section{6DMA Fundamentals}
In this section, we present the basics pertinent to
6DMA-enhanced wireless networks, where the corresponding fundamental
channel models are first introduced,
followed by architectures, hardware, and practical movement constraints. Then, we detail the
relationship between antenna position/rotation and the configured wireless channel and reveal the sources of the obtained performance gains. Finally, we discuss open issues worthy of 
investigation in future work.
\subsection{6DMA Channel Model} 
For the ease of illustration, we consider downlink multiuser transmission and model the channels from each user to the antennas of all 6DMA surfaces of a 6DMA-BS, as illustrated in Fig. \ref{function} (a). We discuss the fundamental 6DMA channel model, the polarized 6DMA channel model, the rotatable 6DMA channel model, and the positionable 6DMA channel model. This analysis aims to provide readers with the necessary insights to make informed decisions when developing their system models.
\subsubsection{Basic 6DMA Channel Model}
As illustrated in Fig. \ref{6D},  6DMA systems can be flexibly designed by utilizing various DoFs in antenna movement, ranging from one-dimensional (1D) to six-dimensional (6D). Systems with fewer DoFs in antenna movement, such as the  rotatable 6DMA and the positionable 6DMA discussed in Sections V and VI, respectively, can be realized by restricting movement in certain dimensions. The 6DMA system comprises of $B$ antennas/antenna surfaces (subarrays), indexed by set $\mathcal{B} = \{1, 2, \ldots, B\}$. Each 6DMA surface is assumed to be a uniform planar array (UPA) with a given size, which consists of $N\geq 1$ antennas, indexed by set $\mathcal{N} = \{1, 2, \ldots, N\}$. The position and rotation of the $b$-th 6DMA surface center, $b\in\mathcal{B}$, can be characterized by six parameters, i.e., $\mathbf{q}_b$ for the 3D position and $\mathbf{u}_b$ for the 3D rotation, which are given by
\begin{align}
	\mathbf{q}_b=[x_b,y_b,z_b]^T\in\mathcal{C}, ~\mathbf{u}_b=[\alpha_b,\beta_b,\gamma_b]^T,\label{bb1}
\end{align}
where $\mathcal{C}$ denotes the given 3D space at the 6DMA-BS in which the 6DMA surfaces can be flexibly positioned/rotated. We assume that $\mathcal{C}$ is a convex set which has a finite size. In \eqref{bb1}, $x_b$, $y_b$ and $z_b$ represent the coordinates of the $b$-th 6DMA's center in the global Cartesian coordinate system (CCS) $o\text{-}xyz$, with the 6DMA-BS's reference position serving as the origin, $o$; $\alpha_b\in[0,2\pi)$,  $\beta_b\in[0,2\pi)$ and $\gamma_b\in[0,2\pi)$ denote the rotation angles with respect to (w.r.t.) the $x$-axis, $y$-axis and $z$-axis, respectively.
Given $\mathbf{u}_b$, the following rotation matrix can be defined,
\begin{align}\label{R}
	&\!\!\!\mathbf{R}(\mathbf{u}_b)=\begin{bmatrix}
		c_{\beta_b}c_{\gamma_b} & c_{\beta_b}s_{\gamma_b} & -s_{\beta_b} \\
		s_{\beta_b}s_{\alpha_b}c_{\gamma_b}-c_{\alpha_b}s_{\gamma_b} & s_{\beta_b}s_{\alpha_b}s_{\gamma_b}+c_{\alpha_b}c_{\gamma_b} & c_{\beta_b}s_{\alpha_b} \\
		c_{\alpha_b}s_{\beta_b}c_{\gamma_b}+s_{\alpha_b}s_{\gamma_b} & c_{\alpha_b}s_{\beta_b}s_{\gamma_b}-s_{\alpha_b}c_{\gamma_b} &c_{\alpha_b}c_{\beta_b} \\
	\end{bmatrix},\!\!
\end{align}
where $c_{x}=\cos(x)$ and $s_{x}=\sin(x)$.
\begin{figure}[t]
	\centering
	\setlength{\abovecaptionskip}{0.cm}
	\includegraphics[width=3.1in]{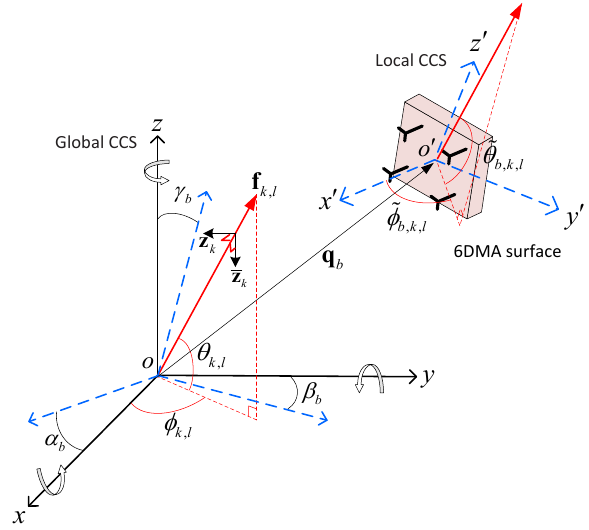}
	\caption{Illustration of the geometry of the 6DMA channel model.}
	\label{system}
			\vspace{-0.59cm}
\end{figure}

As shown in Fig. \ref{system}, each 6DMA surface's local CCS is denoted by $o'\text{-}x'y'z'$, with the surface center serving as the origin, $o'$. 
Let $\bar{\mathbf{r}}_{n}$ denote the position of the $n$-th antenna of the 6DMA surface in its local CCS. Then, the position of the $n$-th antenna of the $b$-th 6DMA surface in the global CCS, denoted by $\mathbf{r}_{b,n}\in \mathbb{R}^3$, can be expressed as
\begin{align}\label{nwq}
	\mathbf{r}_{b,n}(\mathbf{q}_b,\mathbf{u}_b)=\mathbf{q}_b+\mathbf{R}
	(\mathbf{u}_b)\bar{\mathbf{r}}_{n},~n\in\mathcal{N},~b \in\mathcal{B}.
\end{align}

A general multipath channel from each user (assumed to be equipped with a single omnidirectional FPA) to the BS is considered. Let $\mathbf{h}_{k,b}(\mathbf{q}_{b},\mathbf{u}_{b})\in \mathbb{C}^{N\times 1}$ represent the far-field channel from the $k$-th user to all the antennas of the $b$-th 6DMA surface at the BS, which can be expressed as
\begin{align}\label{uk}
	\mathbf{h}_{k,b}(\mathbf{q}_{b},\mathbf{u}_{b})&=
	\sum_{l=1}^{L_k}\eta_{k,l}
	\sqrt{g_{k,l}(\mathbf{u}_{b})}\mathbf{a}_{b,k,l}
	(\mathbf{q}_{b},\mathbf{u}_{b}),
\end{align}
where \( g_{k,l}(\mathbf{u}_{b}) \) represents the effective antenna gain of the \( b \)-th directive 6DMA surface, $L_k$ denotes the number of paths between user $k$ and the BS, and \( \eta_{l, k} \) is the channel coefficient from user \( k \) to the 6DMA-BS over path \( l \). The 6D steering vector \( \mathbf{a}_{b,k,l} (\mathbf{q}_{b},\mathbf{u}_{b}) \) and the effective antenna gain \( g_{k,l}(\mathbf{u}_{b}) \) of the \( b \)-th 6DMA surface are defined in the following.

The 6D steering vector of the $b$-th 6DMA surface for receiving a signal from user $k$ over path $l$ can be expressed as
\begin{align}\label{gen}
	&\mathbf{a}_{b,k,l}(\mathbf{q}_{b},\mathbf{u}_{b})= \!\left[\!e^{-j\frac{2\pi}{\lambda}
		\mathbf{f}_{k,l}^T\mathbf{r}_{b,1}(\!\mathbf{q}_{b},
		\mathbf{u}_{b}\!)},
	\!\cdots,\! e^{-j\frac{2\pi}{\lambda}\mathbf{f}_{k,l}^T
		\mathbf{r}_{b,N}(\!\mathbf{q}_{b},\mathbf{u}_{b}\!)}\!\right]^T,
\end{align}
where $\lambda$ denotes the carrier wavelength, and $\mathbf{f}_{k,l}$ represents the pointing vector corresponding to direction $(\theta_{k,l}, \phi_{k,l})$, which is defined as
\begin{align}\label{KM}
	\!\!\!\mathbf{f}_{k,l}\!=\![\cos(\theta_{k,l})\cos(\phi_{k,l}), \cos(\theta_{k,l})\sin(\phi_{k,l}), \sin(\theta_{k,l})]^T,
\end{align}
where $\phi_{k,l}\in[-\pi,\pi]$ and $\theta_{k,l}\in[-\pi/2,\pi/2]$ denote the azimuth and elevation angles, respectively, for the $l$-th channel path between user $k$ and the 6DMA-BS.
Next, to conveniently define $g_{k,l}(\mathbf{u}_{b})$, we project pointing vector \( \mathbf{f}_{k,l} \) onto the local CCS of the \( b \)-th 6DMA surface, leading to \( \tilde{\mathbf{f}}_{b,k,l}=\mathbf{R}(\mathbf{u}_{b})^{-1}\mathbf{f}_{k,l} \).

Then, we represent $\tilde{\mathbf{f}}_{b,k,l}$ in spherical coordinates as
$\tilde{\mathbf{f}}_{b,k,l}=[\cos(\tilde{\theta}_{b,k,l})\cos(\tilde{\phi}_{b,k,l}), \cos(\tilde{\theta}_{b,k,l})\sin(\tilde{\phi}_{b,k,l}), \sin(\tilde{\theta}_{b,k,l})]^T$, where $\tilde{\theta}_{b,k,l}$ and $\tilde{\phi}_{b,k,l}$ represent the corresponding directions of arrival (DOAs) in the local CCS (see Fig. \ref{system}).
Finally, the effective antenna gain $g_{k,l}(\mathbf{u}_{b})$ of the \(b\)-th directive 6DMA surface along direction \((\tilde{\theta}_{b,k,l}, \tilde{\phi}_{b,k,l})\) in the linear scale can be defined as 
\begin{align}\label{gm}
	g_{k,l}(\mathbf{u}_{b})=10^{\frac{A(\tilde{\theta}_{b,k,l}, \tilde{\phi}_{b,k,l})}{10}},~b\in\mathcal{B}, l\in L_{k}, k\in\mathcal{K},
\end{align}
where \(A(\tilde{\theta}_{b,k,l}, \tilde{\phi}_{b,k,l})\) denotes the effective antenna gain in dBi, which is determined by the radiation pattern of the adopted antenna \cite{6dmaChan}.

\subsubsection{Polarized 6DMA Channel Model}
In the previous two subsections, we simplify the analysis by ignoring the impact of antenna polarization on the channel. However, as each antenna rotates, its polarization properties will also change. To capture this effect, we consider the LoS far-field polarized 6DMA channel and assume that each transmitting user is equipped with a single, fixed-position dual-polarized antenna, and the receiving 6DMA BS is also equipped with multiple, position/rotation-adjustable, dual-polarized antennas. Each antenna consists of two orthogonally oriented linearly polarized elements, with one element for vertical polarization (\(\mathcal{V}\)-element) and the other for horizontal polarization (\(\mathcal{H}\)-element). We assume that the vertical \(\mathcal{V}\)-element is aligned along the positive \(y\)-axis with unit vector \(\mathbf{e}_{\mathrm{v}} = [0, 1, 0]^T\), while the horizontal \(\mathcal{H}\)-element is oriented along the positive \(x\)-axis with unit vector \(\mathbf{e}_{\mathrm{h}} = [1, 0, 0]^T\).

The polarization state of an electromagnetic wave is defined by two orthogonal electric field components on the wavefront. Specifically, these components are represented by the orthogonal unit vectors in the global CCS \cite{heap}, given by \(\mathbf{z}_k = [s_{\theta_k} s_{\phi_k}, -c_{\theta_k}, s_{\theta_k} c_{\phi_k}]^T\) and \(\bar{\mathbf{z}}_k = [c_{\phi_k}, 0, -s_{\phi_k}]^T\), which are perpendicular to the signal's propagation direction (see Fig. \ref{system}) and characterize the polarization state along the LoS path.

The transmit field components of the LoS path are generated by projecting the transmit antenna's time-varying electric fields onto the LoS signal direction. The corresponding transformation is  given by \cite{ipa}
	\begin{align}
		\!\!\!\!\!	\mathbf{P}_{k,b}(\mathbf{u}_b) \!= \!
		\begin{bmatrix}
			(\mathbf{R}(\mathbf{u}_b)\mathbf{e}_\mathrm{v}) \cdot \mathbf{z}_k & (\mathbf{R}(\mathbf{u}_b)\mathbf{e}_\mathrm{h}) \cdot \mathbf{z}_k \\
			(\mathbf{R}(\mathbf{u}_b)\mathbf{e}_\mathrm{v}) \cdot \bar{\mathbf{z}}_k & (\mathbf{R}(\mathbf{u}_b)\mathbf{e}_\mathrm{h}) \cdot \bar{\mathbf{z}}_k
		\end{bmatrix}\in \mathbb{C}^{2\times 2},
	\end{align}
where $\mathbf{R}(\mathbf{u}_b)\mathbf{e}_\mathrm{v}$ and $\mathbf{R}(\mathbf{u}_b)\mathbf{e}_\mathrm{h}$ represent the mapping of the local CCS of the \(\mathcal{V}\)-element and \(\mathcal{H}\)-element of the 6DMA surface at the BS to the global CCS.
Similarly, we define \( \mathbf{u}_{k}^{\mathrm{r}} \) as the rotation angle vector of the \( k \)-th user's local CCS relative to the global CCS \( o\text{-}xyz \).
Then, the receive field components are obtained by projecting the LoS signal direction onto the receive antenna using the projection matrix 
\begin{align}
	\!\!\!\mathbf{Q}_k (\mathbf{u}_{k}^{\mathrm{r}})= 
	\begin{bmatrix}
		\mathbf{z}_k \cdot (\mathbf{R}(\mathbf{u}_{k}^{\mathrm{r}})\mathbf{e}_\mathrm{v}) & \bar{\mathbf{z}}_k \cdot (\mathbf{R}(\mathbf{u}_{k}^{\mathrm{r}})\mathbf{e}_\mathrm{v}) \\
		\mathbf{z}_k \cdot (\mathbf{R}(\mathbf{u}_{k}^{\mathrm{r}})\mathbf{e}_\mathrm{h}) & \bar{\mathbf{z}}_k \cdot (\mathbf{R}(\mathbf{u}_{k}^{\mathrm{r}})\mathbf{e}_\mathrm{h})
	\end{bmatrix}\in \mathbb{C}^{2\times 2},
\end{align}
where $\mathbf{R}(\mathbf{u}_{k}^{\mathrm{r}})\mathbf{e}_\mathrm{v}$ and $\mathbf{R}(\mathbf{u}_{k}^{\mathrm{r}})\mathbf{e}_\mathrm{h}$ denote the transformation of the local CCS of the \(\mathcal{V}\)-element and \(\mathcal{H}\)-element of the user's antenna to the global CCS.
Consequently, the dual-polarized response matrix between the $k$-th user and the dual-polarized antennas on the \( b \)-th 6DMA subarray at the BS is given by  
\begin{align}\label{po3}
	{\mathbf{A}_{k,b}(\mathbf{u}_b,\mathbf{u}_{k}^{\mathrm{r}})} = \mathbf{Q}_k (\mathbf{u}_{k}^{\mathrm{r}})\mathbf{P}_{k,b}(\mathbf{u}_b)\in \mathbb{C}^{2\times 2}.
\end{align}

When the transmission distance is sufficiently large, the phase shift caused by the LoS channel is the same for both the \(\mathcal{V}\)- and \(\mathcal{H}\)-ports
of the antenna, for all pairs of transmit and receive antennas. Therefore, the polarized 6DMA channel from user $k$ to the $b$-th 6DMA subarray at the BS is given by \cite{ipa}:
\begin{align}\label{tu5}
	\!\! \overline{\mathbf{h}}_{k,b}(\mathbf{q}_b,\mathbf{u}_b) = \mathbf{h}_{k,b}^{\mathrm{los}}(\mathbf{q}_b,\mathbf{u}_b) \otimes \mathbf{A}_{k,b}(\mathbf{u}_b,\mathbf{u}_{k}^{\mathrm{r}})\in \mathbb{C}^{2N\times 2},
\end{align}
where $\mathbf{h}_{k,b}^{\mathrm{los}}(\mathbf{q}_b,\mathbf{u}_b)$ is obtained by setting $L_k$ in \eqref{uk} to 1, i.e.,
$
	\mathbf{h}_{k,b}^{\mathrm{los}}(\mathbf{q}_b,\mathbf{u}_b)=\eta_{k,1}
	\sqrt{g_{k,1}(\mathbf{u}_{b})}\mathbf{a}_{k,1}
	(\mathbf{q}_{b},\mathbf{u}_{b})$.
 
\subsubsection{Rotatable 6DMA Channel Model}  
\begin{figure}[t]
	\centering
	\setlength{\abovecaptionskip}{0.cm}
	\includegraphics[width=3.3in]{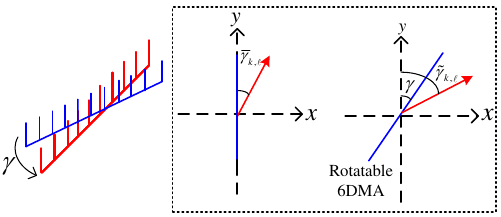}
	\caption{Illustration of the rotatable 6DMA channel model geometry with 1D rotation.}
	\label{hcv}
	\vspace{-0.59cm}
\end{figure}
We discuss both the directive and omnidirectional antenna patterns for the rotatable 6DMA channel model. First, with directive antenna pattern, directive rotatable 6DMAs have the capability to reshape the radiation pattern in the angular domain for achieving flexible beam coverage by adjusting the rotation angles of antennas.
Given a fixed antenna position \(\mathbf{q}_b\), we focus on a scenario with a single 6DMA surface, i.e., $B = 1$ in the
6DMA channel model \eqref{uk}.
The position of the antenna \(\mathbf{q}_b\) is set to \(\mathbf{0}\). By substituting \(\mathbf{q}_{b} = [0, 0, 0]^T\) into \eqref{nwq}, the expression for the antenna location in the global CCS simplifies to
\begin{align}\label{nwqr}
	\mathbf{r}_{b,n}(\mathbf{u}) = \mathbf{R}(\mathbf{u})\bar{\mathbf{r}}_{n},\quad n \in \{1, 2, \cdots, N\},
\end{align}  
where $\mathbf{u}=\mathbf{u}_1$ in \eqref{bb1} as $B = 1$, which represents the rotation vector for the entire array.

The far-field channel from the \(k\)-th user to all antennas of the \(b\)-th rotatable 6DMA surface can then be expressed as  
\begin{align}\label{ukr}
	\mathbf{h}_{k}(\mathbf{u}) =
	\sum_{l=1}^{L_k}\eta_{k,l}
	\sqrt{g_{k,l}(\mathbf{u})}\mathbf{a}_{k,l}(\mathbf{u}),
\end{align}  
where  
$	\mathbf{a}_{k,l}(\mathbf{u}) = \left[e^{-j\frac{2\pi}{\lambda}\mathbf{f}_{k,l}^T\mathbf{R}(\mathbf{u})\bar{\mathbf{r}}_{1}}, \cdots, e^{-j\frac{2\pi}{\lambda}\mathbf{f}_{k,l}^T\mathbf{R}(\mathbf{u})\bar{\mathbf{r}}_{N}}\right]^T$
represents the 6D steering vector of the rotatable 6DMA surface for receiving a signal from user \(k\) over path \(l\).

In contrast, with omnidirectional antenna pattern, omnidirectional rotatable 6DMA (i.e., $g_{k,l}(\mathbf{u})=1$) can only modify the channel phase to maximally exploit small-scale channel variations, thereby mitigating the effect of deep fading. In this scenario, by considering a 1D rotation angle $\gamma$, the channel model of the rotatable 6DMA can be simplified. As illustrated in Fig. \ref{hcv}, \(\bar{\gamma}_{k,\ell}\) represents the reference spatial angle, \(\gamma\) denotes the rotation angle of the antenna array, and the effective spatial angle is defined as \cite{rachang}
\begin{align}
	\tilde{\gamma}_{k,\ell} = \bar{\gamma}_{k,\ell} + \gamma,
\end{align}
for the user $k$ and $\ell \in \{0, 1, \dots, L_k\}$. As such, the steering vector at transmitter is modeled as
\begin{align}
\mathbf{a}_k(\gamma) = 
[			e^{j \frac{2\pi}{\lambda} d_{1} \sin \tilde{\gamma}_{k,\ell}}, \cdots,
e^{j \frac{2\pi}{\lambda} d_{M} \sin \tilde{\gamma}_{k,\ell}}]^T,\label{sra}
\end{align}
where \(d_{n}\) is the distance from the \(n\)-th transmit antenna to the center of the transmitter array. For a uniform linear rotatable 6DMA-array with half-wavelength spacing \( d=\frac{\lambda}{2}\), the distance \(d_{n}\) is given by $
	d_{n} = \frac{(n - 1) \bar{D}}{N - 1} - \frac{\bar{D}}{2}, \quad \forall n \in \{1, 2, \dots, N\}$,
with $\bar{D} = (N - 1) d$ being the aperture of the transmit array.

Subsequently, under the far-field planar wavefront assumption, the channel between the rotatable 6DMA-array and user \(k\) is expressed as
	\begin{align}\label{rago}
		\mathbf{h}_k(\gamma) = \xi_{k,1} \mathbf{a}_k (\gamma)+\sum_{\ell=1}^{L_k-1}  \xi_{k,\ell} \mathbf{a}_k (\gamma),
	\end{align}
	where \(\xi_{k,1}\) and \(\xi_{k,\ell}\) for \(l \in \{2,\cdots, L_k\}\) represent the complex path gains of the LoS and the \(l\)-th NLoS paths for user \(k\), respectively.
	
\begin{figure*}[t!]
	\centering
	\setlength{\abovecaptionskip}{0.cm}
	\includegraphics[width=7.1in]{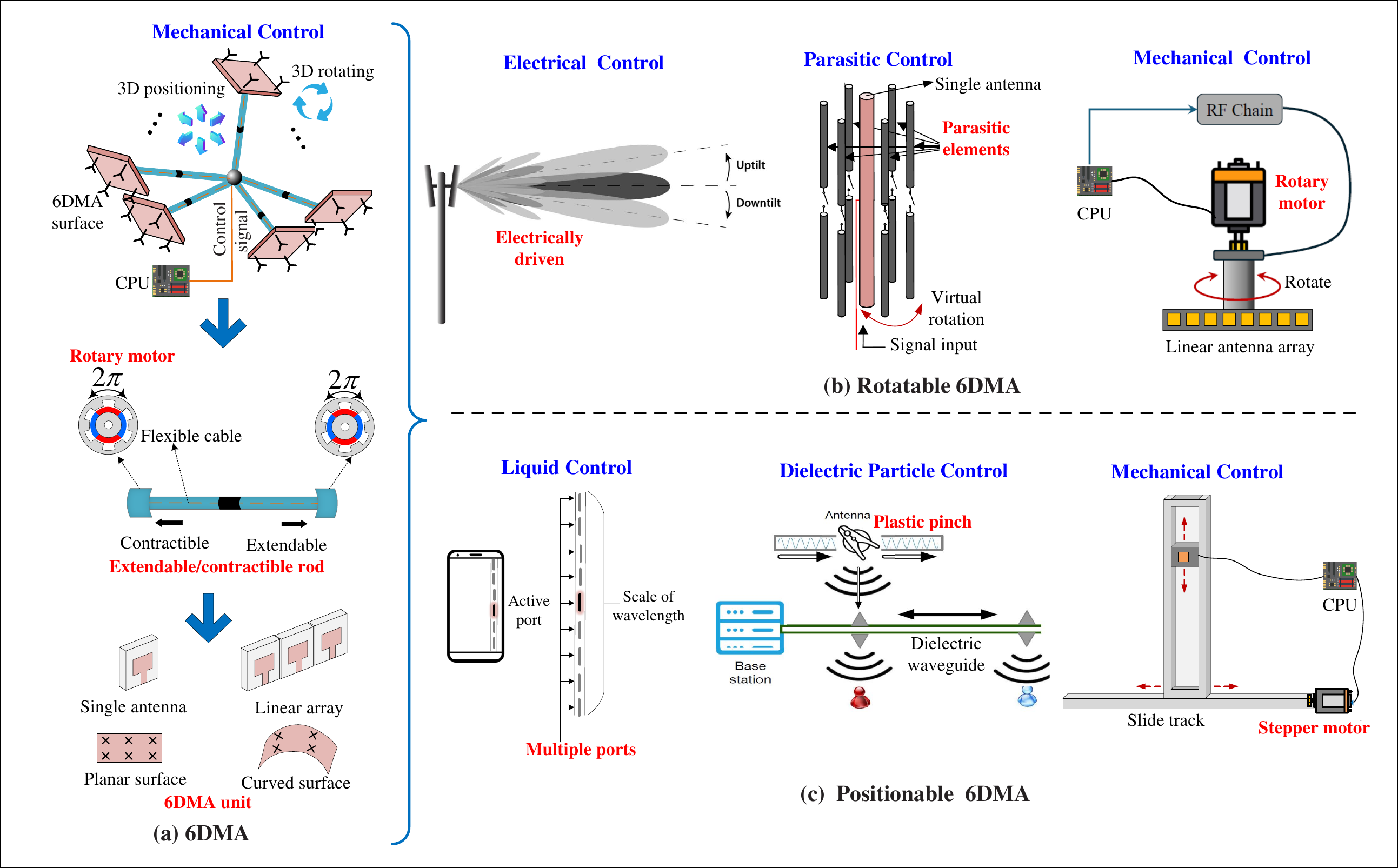}
	\caption{6DMA implementations.}
	\label{hardware}
			\vspace{-0.59cm}
\end{figure*}
\subsubsection{Positionable 6DMA Channel Model}
Given a fixed antenna rotation \(\mathbf{u}_b\), we consider a scenario where each 6DMA surface is equipped with a single antenna, i.e., \( N = 1 \) in the 6DMA channel model \eqref{uk}. The movement region of the BS antenna \(\mathcal{C}\) is defined as a 2D plane. In this case, the location of the antenna in the local CCS simplifies to \(\bar{\mathbf{r}}_{n} = [0, 0, 0]^T\). By substituting \(\bar{\mathbf{r}}_{n} = [0, 0, 0]^T\) into \eqref{nwq}, the expression for the antenna location in the global CCS reduces to 
\begin{align}\label{nwq90}
	\mathbf{r}_{b,n}(\mathbf{q}_b) = \mathbf{q}_b, \quad b \in \{1, 2, \cdots, B\}.
\end{align}
Accordingly, the steering vector for all channel paths and the $b$-th antenna, where $b \in \{1, 2, \cdots, B\}$ at the positionable 6DMA-BS, is given by 
$\bm t_k(\bs{q}_b) =\!\left[e^{-j\frac{2\pi}{\lambda}
		\mathbf{f}_{k,1}^T\bs{q}_b}, e^{-j\frac{2\pi}{\lambda}
		\mathbf{f}_{k,2}^T\bs{q}_b},
	\!\cdots,\! e^{-j\frac{2\pi}{\lambda}\mathbf{f}_{k,L_k}^T
		\bs{q}_b}\!\right]^T\in\mathbb C^{L_k\times1}$. Then, the 6DMA channel model  between user $k$ and the BS, as given in \eqref{uk},  reduces to the following positionable 6DMA channel model:
\begin{align}\label{p6}
	\bm h_k(\bs{q}) =\bm G_k(\bs{q})^H\mathbf {\boldsymbol{\tau}}_k\in\mathbb C^{B\times1},
\end{align}
where $\bs{q}=[\bs{q}_1,\bs{q}_2,\cdots,\bs{q}_{B}]^T$, $\bm G_k(\bs{q}) = \left[\bm t_k(\bs{q}_1), \bm t_k(\bs{q}_2), \cdots, \bm t_k(\bs{q}_{B})\right] \in \mathbb C^{L_{k}\times B}$, and \({\boldsymbol{\tau}}_k=[\tau_{k,1},\tau_{k,2},\cdots,\tau_{k,L_k}]^T\in \mathbb{C}^{L_k \times 1}\) represents the channel coefficient vector with $\tau_{k,l}$ being the complex-valued channel coefficient of the \(l\)-th path from user $k$.

\subsection{6DMA Hardware, Practical Constraints, and Architecture}
\subsubsection{6DMA Hardware}
We provide a comprehensive overview of the state-of-the-art hardware implementations for realizing
antenna movement for 6DMA, positionable 6DMA, and rotatable 6DMA. The 6DMA can be controlled through various methods, including mechanical, liquid, electrical, dielectric particle, and parasitic controls (see Fig. \ref{hardware}).

Mechanically movable components utilize external structures, such as electric motors, precision gears, micro-electro-mechanical-systems (MEMS), or motor-driven shafts, to enable antenna motion. These actuators transform control signals and energy into mechanical movement, allowing for precise positioning or rotation of the antenna elements \cite{moo,Li2022Using}. In the 6DMA system depicted in Fig. \ref{hardware}(a), each rod is equipped with two rotary motors at both ends, and the CPU controls these motors to adjust the position and rotation of each 6DMA surface. The rods can contract or extend to adjust the distance between each 6DMA surface and the CPU. They also contain flexible wires (e.g., coaxial cable) that supply power to the 6DMA surfaces and enable control/RF signal exchange with the CPU. As a result, the transmitter/receiver can flexibly set the 3D positions and 3D rotations of all 6DMA surfaces to enhance wireless network performance. Under this arrangement, the BS can function much like a “transformer,” with the ability to rapidly reconfigure its antenna array into nearly any desired shape to improve wireless network performance.
For the rotatable 6DMA depicted in Fig. \ref{hardware}(b), a single rotary motor can be utilized in the rotary transceiver. Compared to fully-movable 6DMA systems, rotatable 6DMA experiences some performance loss but offers lower deployment, operational, and maintenance costs, as each antenna unit, e.g., a uniform linear array (ULA), only requires one rotary motor to rotate all antennas in it. Finally, for the positionable 6DMA shown in Fig. \ref{hardware}(c), a stepper motor can be employed to control the movement of slide tracks \cite{zhumo, Zhu2024MovableAntennas}. This system consists of an antenna mounted on a high-precision slide track, which can move in both the vertical and horizontal directions, allowing for flexible position (i.e., translation) adjustments.
\begin{figure*}[t!]
	\centering
	\setlength{\abovecaptionskip}{0.cm}
	\includegraphics[width=6.0in]{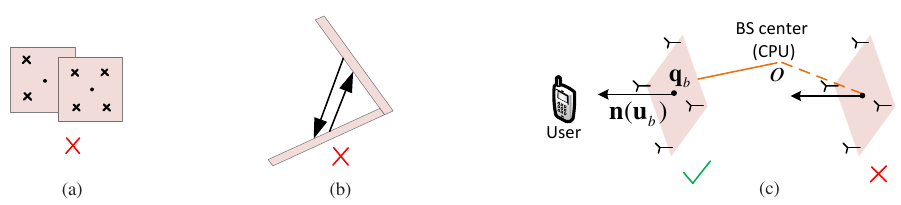}
	\caption{Constraints on the rotations and positions of 6DMA surfaces.}
	\label{rotationangle}
	\vspace{-0.59cm}
\end{figure*}

The key idea of dielectric particle control is illustrated in Fig. \ref{hardware}(c), where positionable 6DMAs, also referred to as pinching antennas, operate as leaky wave antennas by utilizing small dielectric particles, such as plastic pinches, to activate specific points along a waveguide \cite{pin1,pin2}. A positionable 6DMA is assumed to move along a pre-installed track parallel to the waveguide. Due to the low cost of positionable 6DMAs, a large number of them can be pre-deployed on the track, which means that each pinching antenna only needs to cover a small segment of the waveguide and can be quickly moved to the required location. Dielectric particle control enables precise control of radiation locations without requiring additional hardware. This effective approach enables flexible deployment of antennas in optimal positions and guarantees reliable LoS links in challenging environments, thereby significantly reducing path loss while maintaining low deployment costs.

Moreover, electrically controlled antennas can adjust the transmit power in the horizontal/vertical directions by modifying the electrical parameters of the antenna sectors (see the rotatable 6DMA in Fig. \ref{hardware}(b)). In contrast to mechanical tilting, electrical tilting only changes the size of the main lobe and the phase emitted by the sector antenna. Using electrically driven antennas allows remote electrical tilting to be adjusted easily and quickly. Electronically reconfigurable antennas, such as dual-mode patch antennas, can adjust the phase center and radiation pattern to achieve virtual antenna movement. These electronically reconfigurable antennas avoid issues like coupling, wear, and the space required for driving mechanisms, which makes them a promising approach for implementing wireless systems with flexible antenna rotations. However, the displacement of the phase center and the coverage range of the rotation for electronically reconfigurable antennas are usually limited in practice, ranging from $0^\circ$ to $10^\circ$ \cite{Dynamic}, which is less flexible compared to mechanical 6DMAs.

Furthermore, liquid control for positionable 6DMA, also known as fluid antennas \cite{Wong2022Bruce,modu}, is illustrated in Fig. \ref{hardware} (c), where the fluid antenna translates along a line on the scale of a wavelength. Liquid-based elements utilize the flow properties of liquid or fluid materials within a container, which can be driven by a syringe, a nano-pump, or electrowetting. For example, by manually applying pressure to a syringe or digitally controlling a micropump or nanopump, liquid metal can move within the air chamber, thereby changing the position of the antenna within the container. Selecting a suitable fluid material is a major  challenge for fluid antennas, as it must meet various criteria, including cost, safety, physical and chemical stability, melting point, evaporation, viscosity, etc.

Finally, parasitic control for rotatable 6DMA is depicted in  Fig. \ref{hardware} (b). In some cases, the implementation of physically rotating antennas is
not practical. An alternative approach is to employ switched parasitic elements to enable the formation of a directional beam that can rotate within the duration of a symbol period. A particularly practical method for achieving such virtual rotation is to utilize parasitic elements that are either open-circuited or short-circuited. Changing the switch states over time effectively mimics antenna rotation. This process is analogous to sampling and reconstruction using rectangular pulses, which results in effective bandwidth expansion. However, it does not introduce additional DoFs for interference mitigation \cite{vra}. Consequently, while parasitic-controlled rotatable 6DMA offers a simpler implementation, it lacks the spatial flexibility and interference mitigation capabilities of mechanically controlled 6DMA.

Each 6DMA control method provides distinct advantages and limitations. Mechanical control offers precise movement, dielectric particle control enables low-cost precision, electrical control enables rapid adjustments, liquid control adds flexibility, and parasitic control simplifies hardware requirements. Considering the unique advantages and limitations of different
hardware implementations, the choice of it in practice should take into account the specific deployment requirements to achieve an optimal balance between hardware cost and system performance.  It is also possible to combine the above hardware techniques for a more effective hybrid design tailored to specific applications. For example, mechanical adjustment and electrical tuning can be combined to complement each other. While hybrid antenna designs offer superior flexibility for reconfiguration, they also introduce increased design complexity, requiring careful integration of these techniques to maximize their advantages.

\subsubsection{6DMA Position and Rotation Constraints}
Next, we introduce three practical constraints on the  positioning and rotation of 6DMA surfaces.

{\textbf{Minimum-Distance Constraint}}:
We impose a minimum distance, denoted by $d_{\min}$,  between the centers of any pair of 6DMA surfaces to avoid their overlap as well as mutual coupling (see Fig. \ref{rotationangle} (c)). This constraint is expressed as 
\begin{align}\label{jM3}
	\|\mathbf{q}_b-
	\mathbf{q}_{j}\|_2\geq d_{\min},~\forall b ,j \in \mathcal{B}, j\neq b.
\end{align}

{\textbf{Rotation Constraints to Avoid Signal Reflection}}:
6DMA surfaces must meet the following rotation constraints to avoid mutual signal reflections between any two 6DMA surfaces (see Fig. \ref{rotationangle} (a)),
\begin{align}\label{M2}
	\mathbf{n}(\mathbf{u}_b)^T(\mathbf{q}_{j}-\mathbf{q}_b)\leq  0,~\forall b ,j \in \mathcal{B}, j\neq b,
\end{align}
where
$	\mathbf{n}(\mathbf{u}_b)=\mathbf{R}(\mathbf{u}_b)\bar{\mathbf{n}}$
denotes the outward normal vector of the $b$-th 6DMA surface, with $\bar{\mathbf{n}}$ denoting the normal vector of the $b$-th 6DMA surface in the local CCS. To achieve this, each 6DMA surface should not form an acute angle with any of the other 6DMA surfaces.

{\textbf{Rotation Constraints to Avoid Signal Blockage}}:
To prevent each 6DMA surface from rotating towards the CPU of the BS which causes signal blockage, we impose a constraint on the rotation of each 6DMA surface (see Fig. \ref{rotationangle} (b)), which is given by
\begin{align}\label{M3}
	\mathbf{n}(\mathbf{u}_b)^T\mathbf{q}_b\geq 0,~\forall b\in \mathcal{B}.
\end{align}
This is achieved by tuning the normal vector of each 6DMA surface so that it does not point towards the CPU.

\begin{figure*}[t!]
	\centering
	\setlength{\abovecaptionskip}{0.cm}
	\includegraphics[width=6.1in]{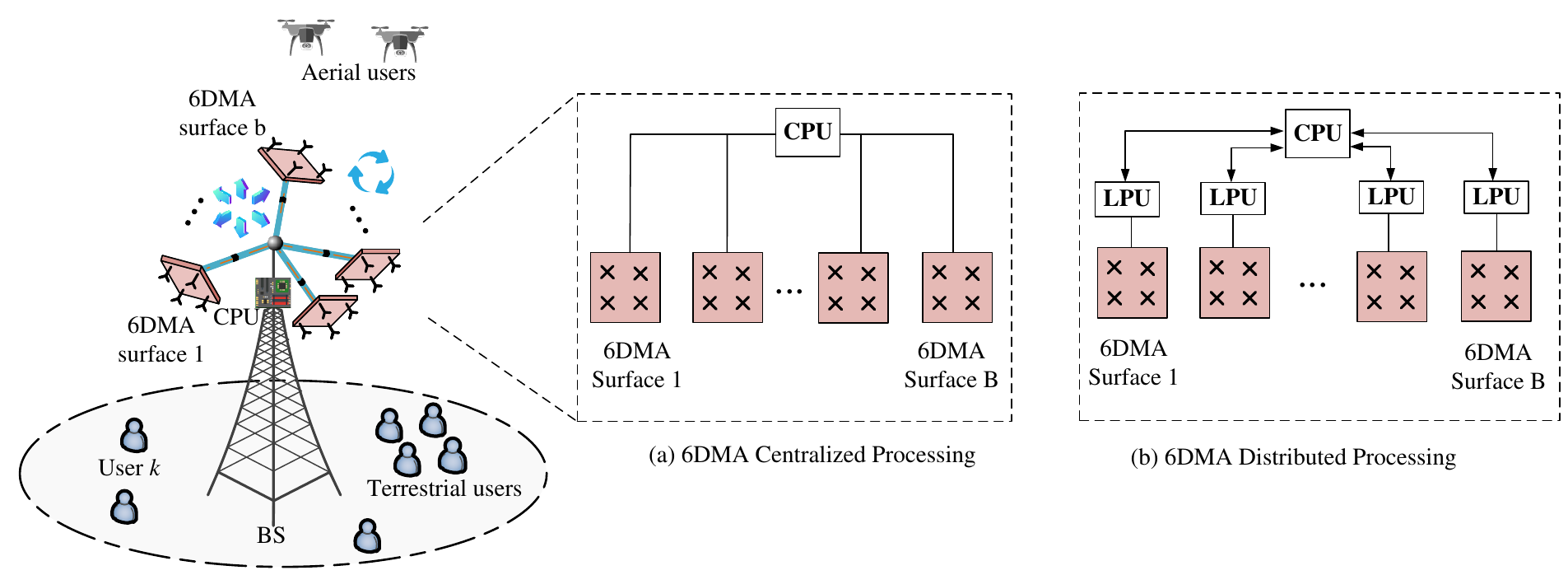}
	\caption{6DMA-equipped BS and different processing architectures.}
	\label{practical_scenario}
	\vspace{-0.59cm}
\end{figure*}
\subsubsection{6DMA Architecture}
The processing architecture is another important factor in implementing practical 6DMA-aided
wireless systems. There are two different architectures for coordinating the signal processing of all 6DMA surfaces: the {\emph{6DMA centralized architecture}} \cite{6dmaCon}, as illustrated in Fig. \ref{practical_scenario} (a), and the {\emph{6DMA distributed architecture}}\cite{6dmaChan}, as illustrated in Fig. \ref{practical_scenario} (b). In the centralized processing architecture, a CPU is connected to all distributed 6DMA surfaces to gather data and execute various tasks such as signal precoding and detection. Generally, centralized processing offers high accuracy in signal precoding and detection because it has access to global data. However, for larger numbers of antennas and candidate antenna positions/rotations, this centralized processing architecture for 6DMA will have to cope with increasingly higher computational cost and complexity as well as the resulting longer processing delay, which may impair the practical deployment of 6DMA in future wireless networks. Therefore, a new distributed 6DMA processing architecture is proposed in \cite{6dmaChan}, as illustrated in Fig. \ref{practical_scenario} (b), to
alleviate the computational complexity of the CPU. Specifically,  
each 6DMA surface is equipped with a local processing unit (LPU), which carries out various signal processing tasks, such as channel estimation and precoding/combining, independently at the same time. All LPUs operate in a decentralized and parallel manner, and they can exchange signals with the CPU for joint signal processing. 
Compared to the conventional centralized 6DMA processing with a single CPU only, the distributed 6DMA processing with both the CPU and LPUs can not only offload computational tasks from the CPU, but also reduce the baseband data transmission rate between the distributed 6DMA surfaces and the CPU by leveraging the local computation and baseband signal processing at the LPUs, thus effectively reducing the overall processing cost and latency. With this distributed 6DMA architecture, advanced signal processing techniques, such as distributed channel estimation and distributed optimization of antenna positions and rotations, are essential for further exploration. Considering the respective advantages and limitations of centralized and distributed 6DMA architectures, it is crucial to select the most
suitable structure based on the specific
system requirements, operating environments, and practical
constraints of a given deployment scenario.

\subsection{6DMA Performance Advantages}
\begin{figure*}[t!]
	\centering
	\setlength{\abovecaptionskip}{0.cm}
	\includegraphics[width=5.5in]{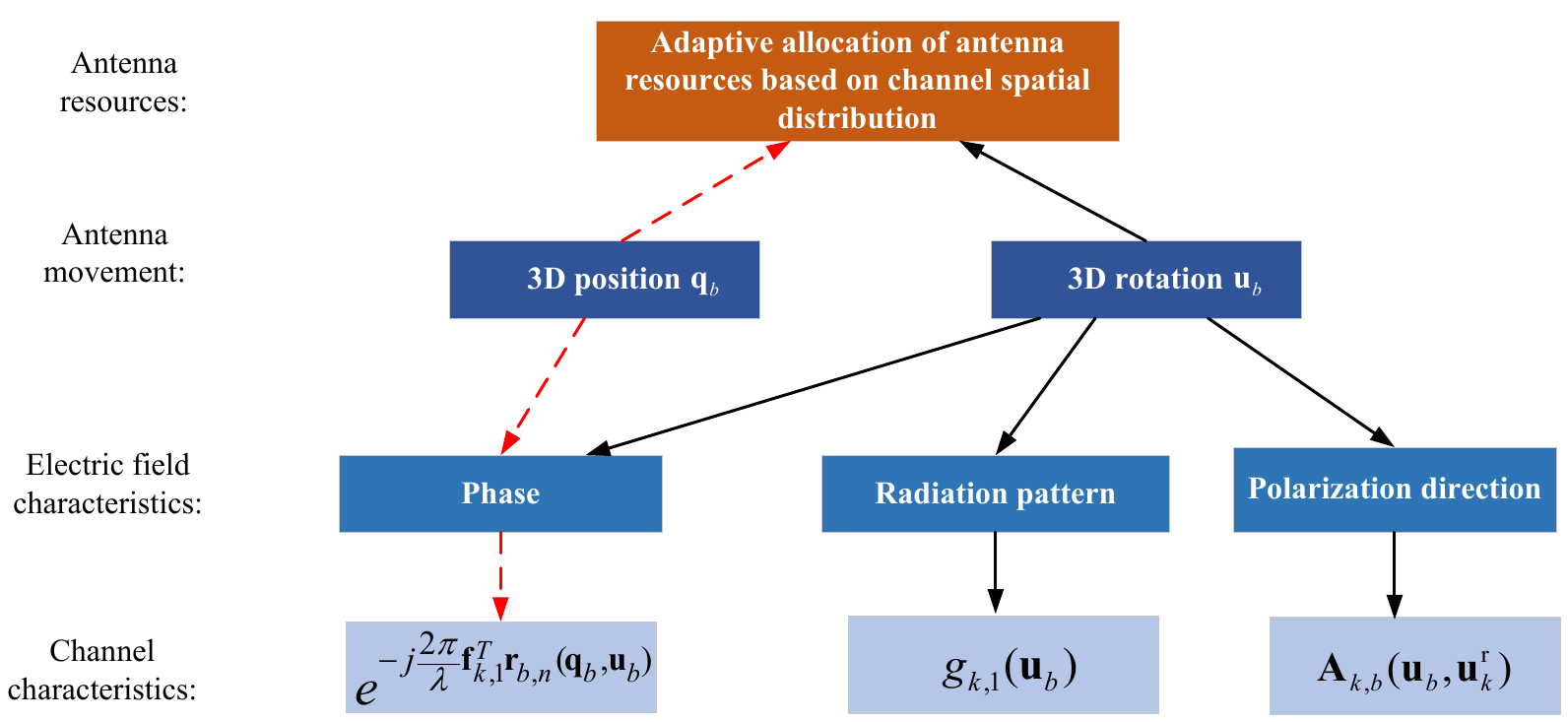}
	\caption{Relationships between antenna movement, antenna resources, electric field, and channel 
		characteristics in 6DMA systems.}
	\label{rela}
	\vspace{-0.59cm}
\end{figure*}

As shown in Fig. \ref{rela}, based on the polarized 6DMA channel model  $\overline{\mathbf{h}}_{k,b}(\mathbf{q}_b,\mathbf{u}_b)$ in \eqref{tu5}, this subsection analyzes how antenna positioning only (i.e., translation), rotation only, and the combination thereof influence channel quality, providing insights into the key characteristics of 6DMA and identifying the sources of its performance gains.

First, we present Remark 1 to explain how the antenna position affects the 6DMA channel.

{\emph{\textbf{Remark 1}:  When the antenna position (i.e., translation) \(\mathbf{q}_b\) is adjusted on the scale of a wavelength, it primarily results in a phase change in the channel gain from any path. This indicates that the impact of antenna position movement or translation, with fixed antenna rotation, can not only exploit small-scale channel variations to mitigate deep fading in the multipath channel but also enable flexible beamforming in a pure-LoS channel.		
Moreover, when the antenna position is adjusted on a much larger scale, which significantly changes the distances between it and users, it can also have an impact on the average strength of the links and even their LoS availability.  
		}}

From Fig. \ref{rela}, we see that the relative orientation of the electric field vectors between the transmitter and receiver remains unchanged regardless of an antenna position change on the wavelength scale, i.e., translation. This implies that the antenna position does not affect the amplitude or polarization alignment of the received signal from any path, and only the phase is influenced through the term \(e^{-j\frac{2\pi}{\lambda} \mathbf{f}_{k,l}^T\mathbf{r}_{b,n}(\mathbf{q}_b, \mathbf{u}_b)}\) in \eqref{gen}, which accounts for path length differences, thereby leading to Remark 1. Thus, for a 6DMA system with both flexible antenna positioning only (e.g., fluid antennas), by adjusting the positions of active elements on a finite two-dimensional (2D) surface or 1D line without considering antenna rotation, the individual antenna positions \(\mathbf{q}_b\) can be optimized to fully exploit small-scale channel variations and mitigate deep fading in the multipath channel.

Next, we present Remark 2 to explain how antenna rotation affects the 6DMA channel.

{\emph{\textbf{Remark 2}: The rotation of the antenna alters the amplitude of the 6DMA channel gain from any path by modifying both the antenna's radiation pattern and polarization. 
}}

When the antenna rotates, it changes the impinging angle of the electromagnetic wave. This alteration affects the antenna's radiation pattern, i.e., the distribution of the electric field with respect to the local CCS through the effective antenna gain $g_{k,l}(\mathbf{u}_{b})$ in \eqref{gm}, which in turn influences the received electric field intensity. 
Furthermore, the rotation of the transmitting/receiving antenna also modifies its polarization orientation through the polarization response matrix ${\mathbf{A}_{k,b}(\mathbf{u}_b,\mathbf{u}_{k}^{\mathrm{r}})}$ in \eqref{po3}, thereby changing the 6DMA channel gain.

Finally, we present Remark 3 to highlight the combined impact of flexible antenna positions and rotations. 

{\emph{\textbf{Remark 3}: The main performance gain of 6DMA with both flexible antenna positions and rotations stems from the adaptive allocation of antenna resources and spatial DoFs based on the channel spatial distribution\footnote{The channel spatial distribution characterizes the large-scale variations in the channel between a 6DMA-enabled transmitter and receiver, which depend on the 3D spatial distribution of devices/users/scatterers in the environment.}, which varies slowly in practice.
This gain is primarily achieved through large-scale antenna position adjustments and rotations, rather than fine-scale position changes at the wavelength level.
}}

As shown in Fig. \ref{gain}, a transmitter/receiver equipped with  6DMAs can adaptively allocate antenna resources in space to match the spatial distribution of users. For instance, if users are at similar distances from the transmitter and a specific region contains a large number of users (see Fig. \ref{gain}(a)) or dense scatterers (see Fig. \ref{gain}(b)), the 6DMA system assigns more antennas to that area to improve communication and sensing performance. By allocating antennas to different communication/sensing regions, the system achieves balanced power distribution across users/scatterers, thus improving the overall system performance.

More specifically, the 6DMA system offers an improved array gain, spatial multiplexing gain, and geometric gain while effectively suppressing interference, which leads to significant enhancement in wireless network performance. They are further elaborated in the following.

\subsubsection{Array Gain}
A 6DMA system achieves superior array gain compared to FPA arrays by adjusting the position and orientation of the array towards the desired direction. This fully utilizes the directive radiation pattern of each antenna element on the 6DMA surface. As illustrated in Fig. \ref{gain}, aligning the position and rotation of each 6DMA surface with the direction of the desired signal effectively compensates for path loss. This advantage is especially valuable in propagation environments with rich scattering and significant multi-path fading.

\subsubsection{Spatial Multiplexing Gain}
Employing 6DMAs at the transmitter and/or receiver enhances the spatial multiplexing gain of MIMO systems. Traditional MIMO systems often face challenges in fully achieving the spatial multiplexing gain needed for maximizing data rates due to issues like antenna mutual coupling or insufficient multi-path scattering, which lead to channel rank deficiency. In contrast, the flexibility of 6DMA surfaces allows them to be strategically positioned and rotated based on the propagation channel between the transmitter and receiver. This optimizes the singular values of the MIMO channel matrix, which leads to a more balanced singular value distribution and increased system capacity, as shown in Fig. \ref{gain}.

\subsubsection{Interference Suppression}
In multi-user scenarios, 6DMAs enhance signal power towards desired directions while effectively suppressing interference from undesired ones. In addition, by strategically adjusting the positions and orientations of 6DMA surfaces at the BS according to the users' long-term channel characteristics, the transmit and receive beamforming based on instantaneous channels becomes significantly more effective in mitigating interference.

\subsubsection{Geometric Gain}
Beyond enhancing communication performance, 6DMA systems improve sensing accuracy by leveraging the geometric flexibility of position- and rotation-adjustable 6DMA surfaces. For instance, for localization tasks performed by a 6DMA-BS, the accuracy is influenced by the relative geometry between the target and the transmitting antennas. By optimizing antenna position/rotation at the BS, a geometric gain for sensing can be achieved \cite{6dmaMag, 6dmaSens}.
\begin{figure*}[t!]
	\centering
	\setlength{\abovecaptionskip}{0.cm}
	\includegraphics[width=5.5in]{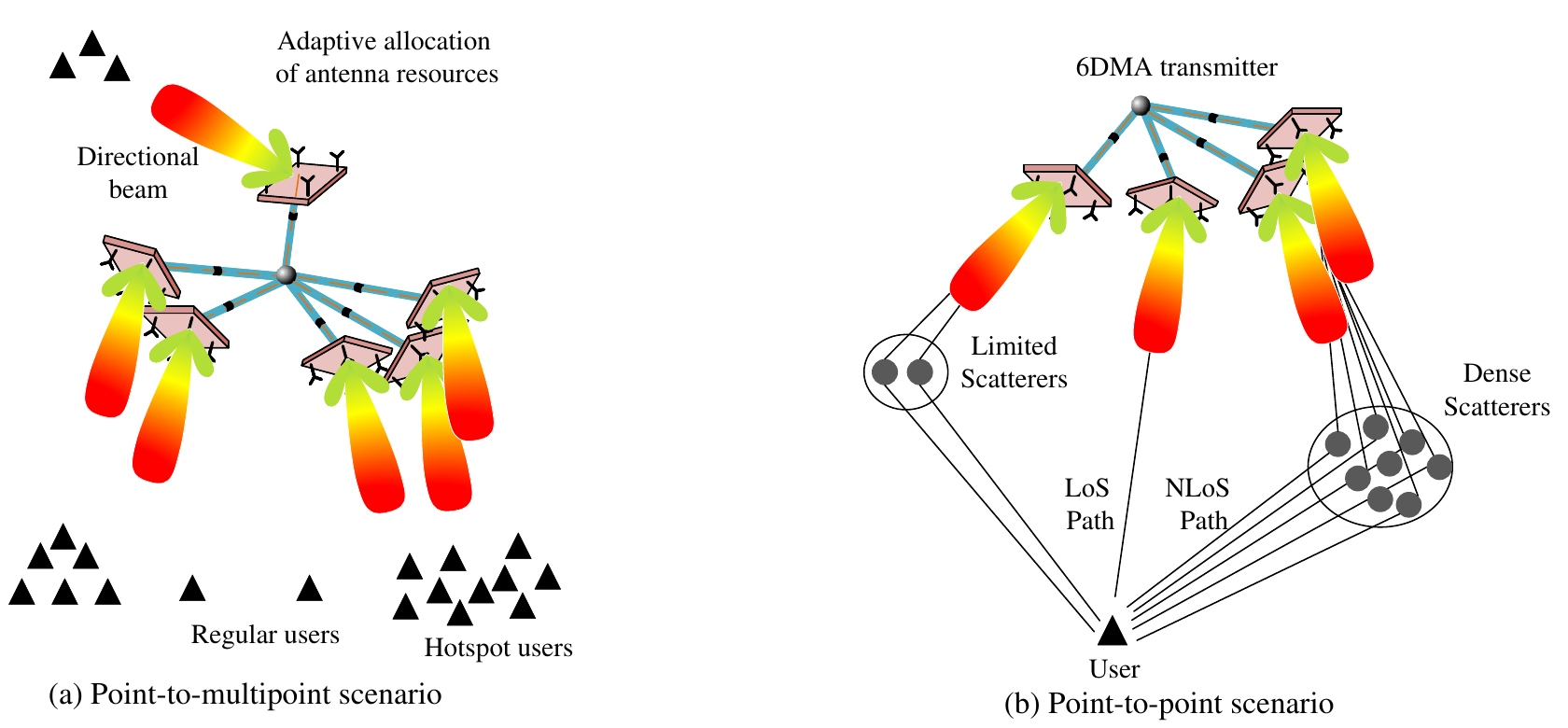}
	\caption{6DMA performance enhancement.}
	\label{gain}
	\vspace{-0.59cm}
\end{figure*}
\subsection{Open Problems and Future Research Directions}
Modeling the 6DMA channel, designing practical hardware, and accounting for position/rotation constraints are essential for optimizing the positions and rotations of 6DMAs and realizing their potential performance gains in 6DMA-assisted wireless systems. However, research in these areas is still in its infancy and encounters numerous critical challenges that require further exploration. Below, we highlight several promising directions to inspire future studies.
One key assumption in the 6DMA channel model discussed in Section II-A is that signal coupling among adjacent antenna elements can be neglected. In reality, increasing the number of possible positions and rotations for 6DMA surfaces within a fixed movement range typically enhances movement precision, which translates into improved performance. However, this approach reduces the spacing between antenna elements, potentially making mutual coupling effects more pronounced and no longer negligible. Such coupling arises from the interaction between adjacent elements through circuit coupling, which results in coupled channel coefficients. Consequently, the channel model for 6DMA, as outlined in Section II-A, may become inaccurate. Strategies
such as leveraging circuit and antenna theory, which
involve matching networks and employing isolation techniques
in antenna design, can help mitigate mutual coupling effects \cite{cou1,cou2}. Addressing this issue may also require more complex, nonlinear models to characterize mutual coupling effects accurately \cite{wcx1, wcx2,wcx5,shao2019complementary}, which presents an intriguing avenue for future research.  

Developing an empirical 6DMA channel model necessitates extensive channel measurements, a process highly dependent on the advancement of 6DMA devices. However, the current development of 6DMA technology is still in an early stage, with existing prototypes not yet suitable for practical deployment, despite significant efforts in this area. Overcoming this challenge requires collaboration among physicists, electromagnetic specialists, and antenna engineers to develop reliable 6DMA prototypes and perform extensive channel measurements across diverse environments. 
The measurement data collected can then be utilized to construct empirical channel models tailored to various scenarios. In addition, while this tutorial primarily emphasizes the spatial correlation of the candidate antenna positions and rotations, it is essential to acknowledge that correlations also exist in the time and frequency domains. The development of theoretical channel models should further consider additional factors such as atmospheric conditions, weather variations, and other environmental influences, particularly for applications in THz communications \cite{thz}.

In Section II-B, three position and two rotation constraints are considered. However, depending on the deployment scenario, additional constraints may need to be introduced. For example, within the same 6DMA-BS, it is necessary to determine how many 6DMA surfaces should be responsible for receiving signals and how many for transmitting signals at the same frequency. Thus, the number of transmitting and receiving surfaces should be constrained to ensure feasible resource allocation and achievable performance. Balancing these roles among the available 6DMA surfaces would require new optimization frameworks that account for channel conditions, traffic demands, and hardware capabilities. Investigating such constraints is essential to further enhance 6DMA system's efficiency and practicality in various deployment scenarios.

\section{6DMA Design Issues}
In this section, we discuss the main design issues of
6DMA-aided wireless systems and present some corresponding promising
approaches to tackle them. In particular, we investigate antenna position and rotation optimization considering both continuous and discrete antenna movement. Then, we examine 6DMA channel estimation techniques, addressing both statistical channel estimation and instantaneous channel estimation. In this section, we consider multiuser communication between $K$ users with FPAs and a 6DMA-BS, as illustrated in Fig. \ref{practical_scenario}.
\begin{figure*}[!t]
	\centering   
	\subfigure[Simulation setup for 6DMA system.] 
	{
		\begin{minipage}{6cm}
			\centering         
			\includegraphics[scale=0.36]{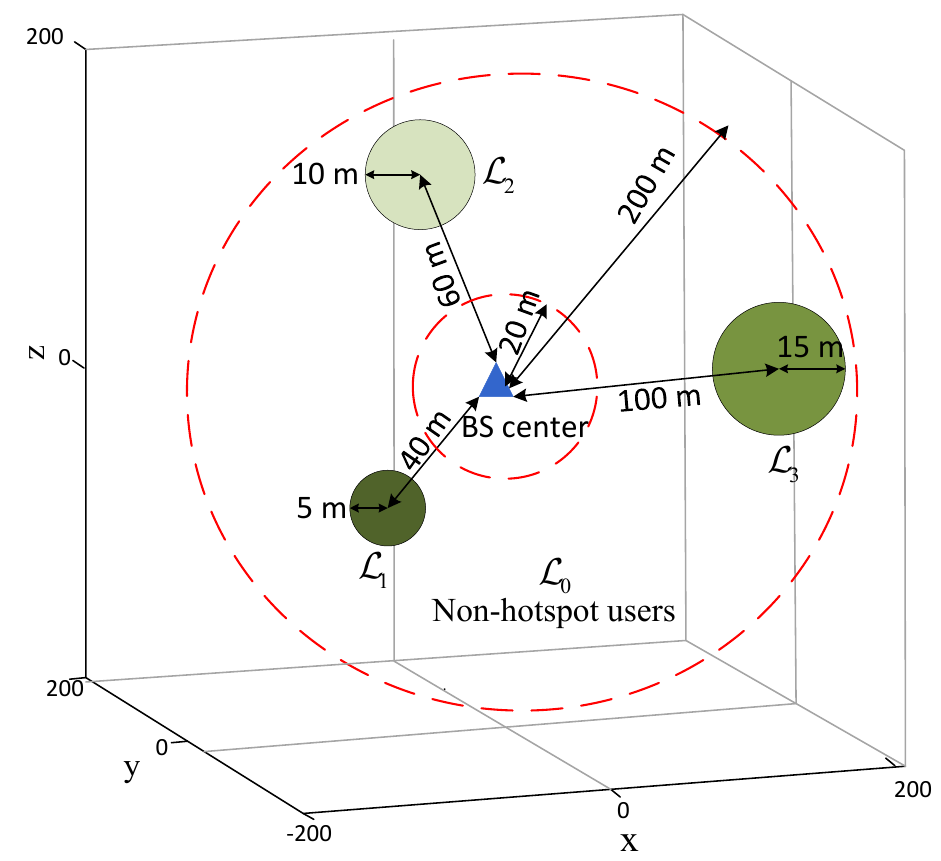}   
		\end{minipage}
	}
	\hspace{0.9cm}
	\subfigure[The optimized positions/rotations of 6DMA surfaces.] 
	{
		\begin{minipage}{8cm}
			\centering      
			\includegraphics[scale=0.36]{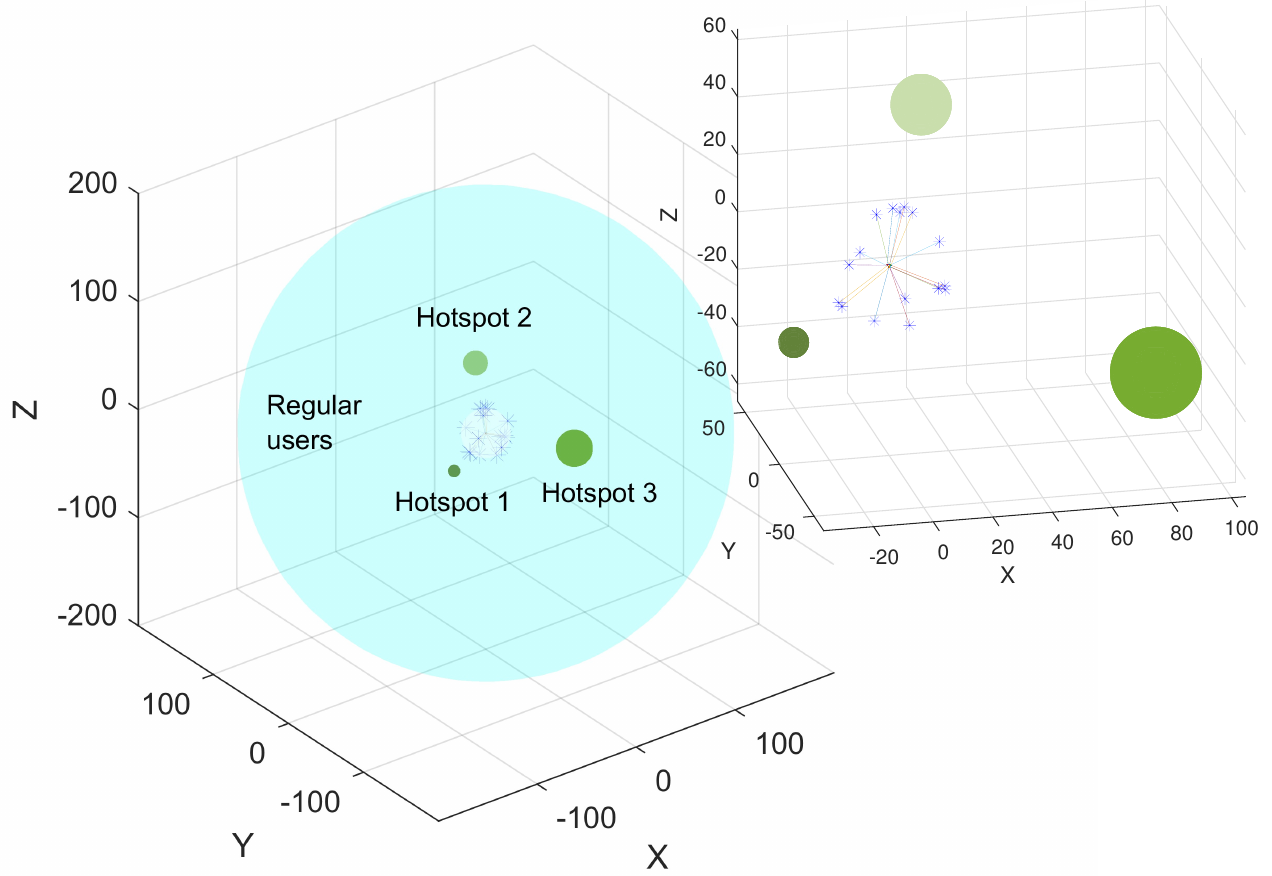}  
		\end{minipage}
	}
	\caption{Simulation setup of the 6DMA system and the optimized positions/rotations of the 6DMAs using the proposed algorithm (each blue snowflake represents the normal vector of its corresponding 6DMA surface, and darker hotspots indicate higher user density).} 
	\label{USERS}  
	\vspace{-0.59cm}
\end{figure*}
\subsection{Antenna Position and Rotation Optimization}
The general optimization problem for a 6DMA system with flexible antenna positioning and rotation can be formulated as follows
	\begin{align}	\label{gene_op}
		~&~\mathop {\max }\limits_{\mathbf{q}, \mathbf{u}, \mathbf{m}} U\left( \mathbf{q}, \mathbf{u}, \mathbf{m} \right) \\
		\text{s.t.}~&~ f_i(\mathbf{q}, \mathbf{u}, \mathbf{m})\geq 0, i = 1, 2, \dots \text{(Movement constraints)}  \nonumber\\
		~&~ \bar{f}_i(\mathbf{q}, \mathbf{u}, \mathbf{m}) \geq 0, i = 1, 2, \dots \text{(Resource constraints)}  \nonumber\\
		~&~ 	\tilde{f}_i(\mathbf{q}, \mathbf{u}, \mathbf{m})\geq 0, i = 1, 2, \dots \text{(Performance constraints)}  \nonumber
		\end{align}
where $\mathbf{m}$ represents communication or/and sensing resources, such as power, bandwidth, and beamforming/waveform design. The utility function, \( U\left( \mathbf{q}, \mathbf{u}, \mathbf{m} \right) \), comprises system performance metrics, such as achievable rate, secrecy rate, energy efficiency, and sensing accuracy.  
The constraints \( f_i(\mathbf{q}, \mathbf{u}, \mathbf{m}) \) impose limitations on antenna movement, including finite antenna movable regions, antenna position, antenna rotation, minimum antenna spacing, maximum moving speed, and so on. The constraints \( \bar{f}_i(\mathbf{q}, \mathbf{u}, \mathbf{m}) \) pertain to communication or/and sensing resources, such as maximum transmit power, system bandwidth, and the maximum number of associated users. Lastly, the constraints \( \tilde{f}_i(\mathbf{q}, \mathbf{u}, \mathbf{m}) \) ensure communication and sensing performance requirements, such as maximum interference leakage, minimum signal-to-interference-plus-noise ratio (SINR), and required  Cramer-Rao Bound (CRB).

In this subsection, we consider the 6DMA centralized processing architecture in Fig. \ref{practical_scenario}(a). 
\subsubsection{Continuous Antenna Position and Rotation Optimization} 
We first consider the uplink transmission from $K$ users, each equipped with a single FPA, to a 6DMA-BS. Let $\mathbf{H}(\mathbf{q},\mathbf{u})\in \mathbb{C}^{NB\times K}$ denote
the multiple-access channel from all $K$ users to all 6DMA surfaces at the BS, modeled using the basic 6DMA channel model in \eqref{uk}. The average network capacity of a 6DMA-enabled system, accounting for random channel realizations, is expressed as
\begin{align}
	R_{\mathrm{avg}}(\mathbf{q},\mathbf{u})=\mathbb{E}_{\mathbf{H}}\left[\log_2 \det \left(\mathbf{I}_{NB}+\frac{p}{\sigma^2}\mathbf{H}(\mathbf{q},\mathbf{u})
	\mathbf{H}(\mathbf{q},\mathbf{u})^H\right)\right], \label{pc0}
\end{align}
where $\sigma^2$ denotes the noise variance, and $p$ represents the transmit power for each user. To maximize \(R_{\mathrm{avg}}(\mathbf{q}, \mathbf{u})\) by jointly optimizing the 3D positions \(\mathbf{q}\) and 3D rotations \(\mathbf{u}\) of all 6DMA surfaces at the BS, subject to the practical constraints outlined in Section II-B, the optimization problem is formulated as
\begin{subequations}
	\label{MG3}
	\begin{align}
			\text{(P1-CO)}:~~&~\mathop{\max}\limits_{\mathbf{q},\mathbf{u}}~~
		R_{\mathrm{avg}}(\mathbf{q},
		\mathbf{u})\\
		\text {s.t.}~&~\mathbf{q}_b\in\mathcal{C}, ~\forall b \in \mathcal{B}, \label{M1}\\
		~&~ \eqref{M2}, \eqref{M3}, \eqref{jM3},
	\end{align}
\end{subequations}
where constraint \eqref{M1} ensures that each 6DMA surface's center is situated within the 3D site space of the BS, $\mathcal{C}$. As discussed in Section II-A, \eqref{M2}--\eqref{jM3} are antenna position and rotation constraints. Note that since the expectation in \eqref{pc0} is difficult to compute analytically, the Monte Carlo method can be used to approximate \( 	R_{\mathrm{avg}}(\mathbf{q},\mathbf{u}) \) \cite{6dmaCon}.

Unfortunately, problem $\text{(P1-CO)}$ is inherently non-convex due to the non-concave nature of the objective function, the coupling of position and rotation variables, and the non-convexity of the constraints in \eqref{M2}--\eqref{jM3}, which makes the joint optimization highly challenging.
A practical solution to address this issue is to apply alternating optimization (AO), which iteratively optimizes the position and rotation of each 6DMA surface while keeping those of the others fixed \cite{6dmaCon}. 
In particular, during each AO iteration, the first step is to optimize \(\mathbf{q}_b\), with \(\{\mathbf{q}_j\}_{j \in \mathcal{B}/b}\) and \(\{\mathbf{u}_j\}_{j \in \mathcal{B}}\) fixed. The resulting optimization problem for \(\mathbf{q}_b\) can be expressed as 
\begin{subequations}
	\label{mm}
	\begin{align}
		\text{(P2-CO-b)}:~~&~\mathop{\max}\limits_{\mathbf{q}_b}
		~\tilde{R}_{\mathrm{avg}}(\mathbf{q}_b,\mathbf{u}_b) \\
		\text {s.t.}~&~ \mathbf{q}_b\in\mathcal{C}, \label{MM1}\\
		~&~ \|\mathbf{q}_{b}-
		\mathbf{q}_{j}\|_2\geq d_{\min},~\forall j \in \mathcal{B}/b,\label{MM2}\\
		~&~ \mathbf{n}(\mathbf{u}_b)^T(\mathbf{q}_{j}-\mathbf{q}_b)\leq  0, ~j\in \mathcal{B}/b,\label{MM3}\\
		~&~ \mathbf{n}(\mathbf{u}_j)^T(\mathbf{q}_{b}-\mathbf{q}_j)\leq  0,~ j\in \mathcal{B}/b,\label{MM33}\\
		~&~\mathbf{n}(\mathbf{u}_b)^T\mathbf{q}_b\geq 0, \label{oMM33}
	\end{align}
\end{subequations}
where $\tilde{R}_{\mathrm{avg}}(\mathbf{q}_b,\mathbf{u}_b)$ is a re-expression of $R_{\mathrm{avg}}(\mathbf{q},\mathbf{u})$ in terms of the position $\mathbf{q}_b$ (for $b \in \mathcal{B}$) and the rotation $\mathbf{u}_b$ (for $b \in \mathcal{B}$) of a single 6DMA surface, with the positions and rotations of the other surfaces held constant.
To efficiently solve \(\text{(P2-CO-b)}\), non-convex constraint \eqref{MM2} is first reformulated into a convex form with respect to \(\mathbf{q}_b\). After this transformation, the feasible direction method \cite{linear} is applied to solve the problem \cite{6dmaCon}.

In the second step, \(\mathbf{u}_b\) is optimized for fixed \(\{\mathbf{u}_j\}_{j \in \mathcal{B}/b}\) and \(\{\mathbf{q}_j\}_{j \in \mathcal{B}}\). The optimization problem for \(\mathbf{u}_b\) is expressed as
\begin{subequations}
	\label{um}
	\begin{align}
		\text{(P3-CO-b)}:~~&~\mathop{\max}\limits_{\mathbf{u}_b}
		~	\tilde{R}_{\mathrm{avg}}(\mathbf{q}_b,\mathbf{u}_b),\\
		\text {s.t.}~&~ \mathbf{n}(\mathbf{u}_b)^T(\mathbf{q}_j-\mathbf{q}_{b})\leq  0,~ j\in \mathcal{B}/b,\label{uuM3}\\
		~&~\mathbf{n}(\mathbf{u}_b)^T\mathbf{q}_b\geq 0. \label{wgw}
	\end{align}
\end{subequations}
The objective function in \(\text{(P3-CO-b)}\) is non-concave, and constraints \eqref{uuM3} and \eqref{wgw} are non-convex, which poses significant difficulties in obtaining an optimal solution. To address these challenges, we can first approximate the rotation matrix using a linearization technique for small rotation angle changes \cite{6dmaCon}, thereby converting the non-convex rotation constraints into convex ones. Subsequently, the feasible direction method can again be applied to solve \(\text{(P3-CO-b)}\).

\begin{figure}[t!]
	\centering
	\setlength{\abovecaptionskip}{0.cm}
	\includegraphics[width=3.3in]{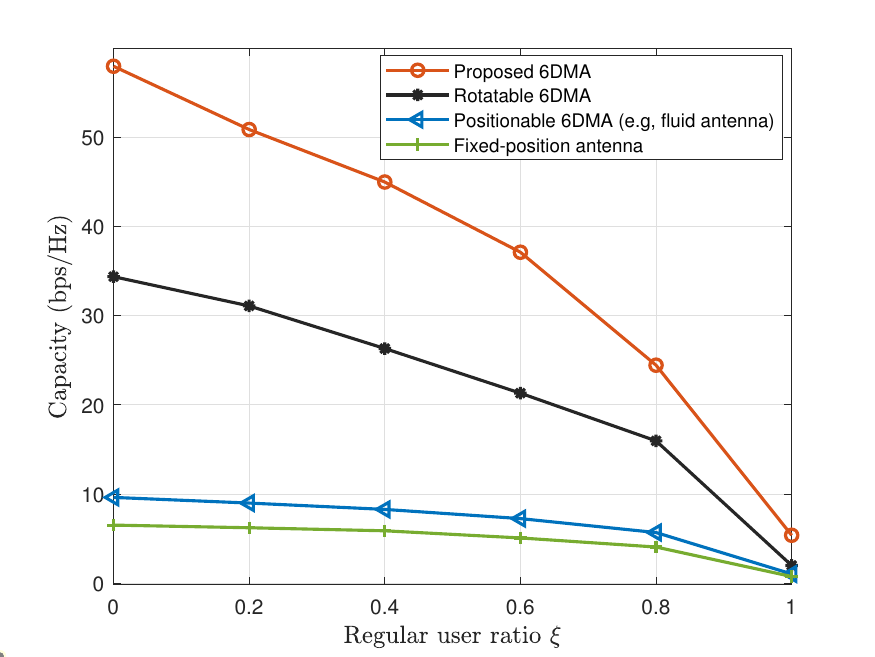}
	\caption{The sum rate versus the non-hotspot user ratio \(\xi\) for continuous antenna position and rotation optimization.}
	\label{back}
			\vspace{-0.59cm}
\end{figure}

To demonstrate the network capacity achieved through AO-based 6DMA position and rotation optimization, we consider
the setup shown in Fig. \ref{USERS} (a). According to the 3GPP standard, the directive antenna gain \( A(\tilde{\theta}_{b,k,l}, \tilde{\phi}_{b,k,l}) \) in dBi is given by
$A(\tilde{\theta}_{b,k,l}, \tilde{\phi}_{b,k,l}) = G_{\max} - \min\left\{ -\left[A_{\mathrm{H}}(\tilde{\phi}_{b,k,l}) + A_{\mathrm{V}}(\tilde{\theta}_{b,k,l})\right], G_s \right\}$,
where \( A_{\mathrm{H}}(\tilde{\phi}_{b,k,l}) = -\min\left\{ 12\left(\frac{\tilde{\phi}_{b,k,l}}{\phi_{\mathrm{3dB}}}\right)^2, G_s \right\} \) and \( A_{\mathrm{V}}(\tilde{\theta}_{b,k,l}) = -\min\left\{ 12\left(\frac{\tilde{\theta}_{b,k,l}}{\theta_{\mathrm{3dB}}}\right)^2, G_v \right\} \). Here, \( \theta_{\mathrm{3dB}} = \phi_{\mathrm{3dB}} = 25^\circ \), \( G_{\max} = 8 \) dBi, and \( G_s = G_v = 25 \) dB \cite{6dmaChan}.
Fig. \ref{USERS}(b) illustrates the optimized positions and rotations of all 6DMA surfaces using the proposed algorithm. The results show that, in a network containing both hotspot and non-hotspot users (with the proportion of non-hotspot users denoted by \(\xi = 0.6\)), the 6DMA surfaces adjust their 6D positions and rotations to accommodate the distribution of both user types. The allocation of 6DMA surfaces to the different hotspots depends primarily on the BS-to-user distance and the user density in each hotspot. In Fig. \ref{back}, we evaluate the effect of \(\xi\) on the sum rate for different schemes. Specifically, we consider three benchmark schemes based on a three-sector BS, with each sector covering approximately $120^\circ$. These schemes are: FPA \cite{exl}, rotatable 6DMA with fixed position \cite{6dmaCon}, and positionable 6DMA with fixed rotation (i.e., fluid antenna) \cite{9650760, flu0, flu1, flu2}. The proposed general 6DMA system consistently outperforms all other schemes for all values of \(\xi\), as general 6DMA provides higher spatial DoFs and can allocate antenna resources more effectively to match the channel's spatial distribution. However, the performance gain of 6DMA reduces as \(\xi\) approaches one (i.e., with non-hotspot users only). These results demonstrate that the proposed 6DMA design is especially beneficial when the user spatial distribution is non-uniform and exhibits clustering in hotspots.

On the other hand, the antenna position and rotation of
6DMA generally need to be jointly designed with the transmit
beamforming of the transceiver in networks such as the BS or
access point (AP) to optimize network performance. Similar to \(\text{(P1-CO)}\), the rate maximization problem in the multi-user downlink MIMO case can be equivalently formulated as
\begin{subequations}
	\label{MGII3}
	\begin{align}
		\text{(P-JT)}:~~&~\mathop{\max}\limits_{
			\mathbf{W},\mathbf{q},
			\mathbf{u}}~~
		R(\mathbf{W},\mathbf{q},\mathbf{u})\\
		\text {s.t.}~&~\sum_{k=1}^{K}\|\mathbf{w}_k\|_2^2\leq P_{\max}, \label{M00}\\
		~&~\eqref{M1}, \eqref{M2}, \eqref{M3}, \eqref{jM3},
	\end{align}
\end{subequations}
where $\mathbf{w}_k$ denotes the transmit beamforming vector for user $k$ at the BS, $\boldsymbol{W}=\{\boldsymbol{w}_k\}$ denote the set of beamforming vectors, $P_{\max}$ is the maximum transmit power, and $R(\mathbf{W},\mathbf{q},\mathbf{u})$ is the achievable rate of all users, given by
\begin{align}\label{r}
	&R(\mathbf{W},\mathbf{q},\mathbf{u})\!=\!\sum_{k=1}^{K}\log_2\left(1+
	\frac{|(\mathbf{h}_k(\mathbf{q},\mathbf{u}))^H
		\mathbf{w}_k|^2}{\sum_{j\neq k}|(\mathbf{h}_k(\mathbf{q},\mathbf{u}))^H
		\mathbf{w}_j|^2\!+\!\sigma^2}\right). 
\end{align}

For the sake of simplicity, we assume that the CSI is known in \eqref{r}. The main challenge in this joint optimization lies in the coupling between the transmit beamforming and the antenna position and rotation optimization variables. A possible approach to address this problem is deep learning \cite{DLS,DLS1,gao2022robust}. Unlike the AO technique discussed earlier, which iteratively optimizes the antenna positions and rotations, the low-complexity deep reinforcement learning (DRL)-based algorithm proposed in \cite{near} jointly designs the 6DMA positions, rotations, and beamforming as the output of a DRL neural network, which leads to more stable and faster convergence.

\subsubsection{Discrete Antenna Position and Rotation Optimization}
In the preceding subsections, we primarily considered continuous positions and rotations of 6DMA surfaces. While continuously tuning the position and/or rotation of each 6DMA surface provides the greatest flexibility and achieves the highest performance gains over FPAs, it is challenging to implement in practice. This is because adjusting 6DMA surfaces using mechanical systems such as rotary motors typically allows only discrete changes in position and rotation. In addition, implementing continuous adjustments increases hardware costs, power consumption, and movement time, especially when a large number of 6DMA surfaces are required to cover a wide movement range at the transmitter or receiver. As such, a more practical and cost-effective approach is to use a finite set of discrete positions and rotation levels for 6DMA surfaces to strike a balance between performance gain and implementation cost. We consider the uplink transmission from \(K\) users, each equipped with a FPA, to a 6DMA-BS, where the basic 6DMA channel model in \eqref{uk} is employed. There are a total of \( M \) possible positions, and each position has \( L_{\mathrm{d}} \) possible rotations. We denote \( \breve{\mathbf{q}}_m \in \mathbb{R}^3\) as the \( m \)-th discrete position and \( \breve{\mathbf{u}}_{l}^{(m)} \in \mathbb{R}^3\) as the \( l \)-th discrete rotation at the \( m \)-th position. 

A major challenge in designing discrete positions and rotations for 6DMA surfaces arises from the large number of possible combinations of discrete positions and rotations across all surfaces, which makes it computationally expensive to find the optimal solution. Solving this problem optimally requires an exhaustive search over all combinations, which becomes impractical when the number of positions, rotations, and antennas are large.
To solve this problem, we consider a modified version of problem \text{(P1-CO)} by replacing the continuous  position and rotation constraints in \eqref{MG3} with their discrete counterparts. As a result,
the achievable rate maximization
problem with discrete antenna positions and rotations is formulated as follows
\begin{subequations}
	\label{DMG3}
	\begin{align}
		\text{(P-DIS)}:~~&~\mathop{\max}\limits_{\mathbf{z}}~~R_{\mathrm{avg}}(\mathbf{z})
		\\
		\text {s.t.}
		~&~(\mathbf{s}_v^T-\mathbf{s}_b^T)
		\mathbf{U}\mathbf{g}_b\leq 0, b\neq v,\forall b,v\in\mathcal{B},\label{zh1}\\
		~&~\mathbf{s}_b^T
		\mathbf{U}\mathbf{g}_b\geq 0,~\forall b\in\mathcal{B},\label{zh}\\
		~&~\mathbf{s}_b^T
		\mathbf{D}\mathbf{s}_v\geq d_{\min},~b\neq v,\forall b,v\in\mathcal{B},\label{M20}\\
		~&~\mathbf{1}^T\mathbf{s}_b =1, ~\forall b\in\mathcal{B}, \label{MD1}\\
		~&~\mathbf{1}^T\mathbf{g}_b =1, ~\forall b\in\mathcal{B}, \label{M10}\\
		~&~ [\mathbf{s}_b]_m\in\{0,1\},~\forall m\in\mathcal{M}, \forall b\in\mathcal{B}, \label{MD2}\\
		~&~ [\mathbf{g}_b]_l\in\{0,1\},~\forall l\in\mathcal{L}, \forall b\in\mathcal{B}, \label{M8}
	\end{align}
\end{subequations}
where $\mathcal{M}=\{1,2,\cdots, M\}$, $R_{\mathrm{avg}}(\mathbf{z})$ is the average network capacity, $[\mathbf{D}]_{m,m'}=\| \breve{\mathbf{q}}_m- \breve{\mathbf{q}}_{m'}\|_2$ denotes the distance between the $m$-th discrete position and the $m'$-th discrete position, $[\mathbf{U}]_{m,l}=\mathbf{n}( \breve{\mathbf{u}}_l^{(m)})^T \breve{\mathbf{q}}_m$, and  $\mathbf{z}=[\mathbf{s}_1^T,\mathbf{s}_2^T,\cdots,\mathbf{s}_B^T,
\mathbf{g}_1^T,\mathbf{g}_2^T,\cdots,\mathbf{g}_B^T]^T$ denotes the position and rotation indicator vector. Here, 
$\mathbf{s}_b\in\mathbb{R}^{M}$ and $\mathbf{g}_b\in\mathbb{R}^{L}$, $b\in\mathcal{B}$, are the position and rotation indicator vectors  respectively, which are given by
\begin{align}\label{3b}
[\mathbf{s}_b]_m &= \left\{\begin{matrix}
		1, & \text{if}~b=m,\\
		0, &\mathrm{otherwise},
	\end{matrix}\right.~~[\mathbf{g}_b]_l&= \left\{\begin{matrix}
		1, & \text{if}~j_{b}=l,\\
		0, &\mathrm{otherwise}.
	\end{matrix}\right.
\end{align}
Constraints \eqref{zh1}, \eqref{zh}, and \eqref{M20} in \(\text{(P-DIS)}\) are the discrete versions of \eqref{M2}, \eqref{M3}, and \eqref{jM3} in \(\text{(P1-CO)}\), respectively. Constraints \eqref{MD2} and \eqref{M8} ensure that the position and rotation indicator vectors are binary (0 or 1), indicating whether a specific position or rotation is chosen by the \(b\)-th 6DMA surface. Furthermore, constraints \eqref{MD1} and \eqref{MD2} guarantee that no position is chosen by more than one 6DMA surface, while constraints \eqref{M10} and \eqref{M8} ensure that each 6DMA surface selects only one rotation for each discrete position. 

The non-convex nature of the objective function and constraints, combined with the binary variables, makes \(\text{(P-DIS)}\) a challenging non-convex integer programming problem. Moreover, the objective function of problem \(\text{(P-DIS)}\) requires statistical information of the multiple-access channel matrix, which is not always readily available in practice.
An offline optimization approach was proposed in \cite{6dmaDis}, which approximated the average network capacity through Monte Carlo simulations, where the channel statistics of users were assumed to be known a priori. The discrete position and rotation variables were transformed into continuous variables, and the modified problem was addressed using effective optimization techniques. The final discrete solutions were obtained by quantizing the results to the nearest values within the discrete sets \cite{6dmaDis}.
Furthermore, for scenarios without prior statistical channel knowledge, a low-complexity conditional sample mean (CSM)-based online optimization method was introduced in \cite{6dmaDis} to address \(\text{(P-DIS)}\). This method randomly generated various sets of discrete position and rotation combinations for all 6DMA surfaces and evaluated the actual achievable sum rate for each set. The collected data was subsequently utilized to directly optimize the 3D positions and rotations of all available 6DMA surfaces to maximize the average sum rate of the users without requiring explicit computation of statistical channel information.

\begin{figure}[t!]
	\centering
	\setlength{\abovecaptionskip}{0.cm}
	\includegraphics[width=3.69in]{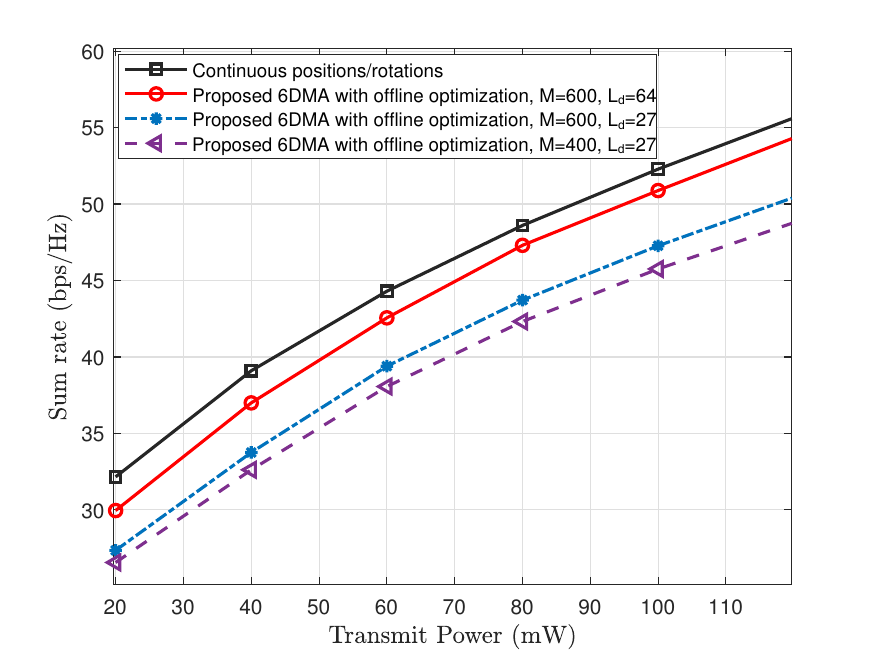}
	\caption{Sum rate versus transmit power for discrete antenna position and rotation optimization.}
	\label{alg_user}
			\vspace{-0.59cm}
\end{figure}

To assess the benefits of discrete antenna position and rotation optimization in improving system performance, Fig. \ref{alg_user} presents the sum rate achieved by the offline optimization algorithm from \cite{6dmaDis} as a function of the users' transmit power. The simulation setup is the same as in Fig. \ref{USERS}(a). It is observed that increasing the number of discrete positions, \(M\), or rotations, \(L_{\mathrm{d}}\), enhances the sum rate due to higher spatial DoFs, which enables better adaptation to user distributions and improves both array and spatial multiplexing gains.
Furthermore, one can observe that the offline algorithm with discrete positions/rotations suffers a performance loss compared to the
AO algorithm with continuous positions and rotations \cite{6dmaCon}. This is because discrete position/rotation adjustments of 6DMA surfaces will limit their spatial DoFs for adaptation.

\subsection{6DMA Channel Estimation}
This subsection discusses channel estimation for the 6DMA system based on the basic 6DMA channel model in \eqref{uk}.
\subsubsection{Problem Description and Challenges}
To obtain optimal 6DMA positions and rotations, the CSI between all 6DMA candidate positions/rotations and all users is essential, which, however,
is practically challenging to acquire. Specifically, we consider the 6DMA centralized processing architecture in \ref{practical_scenario}(a) for uplink multi-user MIMO communication in a narrowband 6DMA system. The received signal at the \(NB\)-antenna 6DMA BS from \(K\) users (each equipped with a single omnidirectional FPA) can be expressed as follows
\begin{align}\label{gene_channel}  
	\mathbf{Y}=\mathbf{X}\overline{\mathbf{H}}^H(\bar{\mathbf{q}},\bar{\mathbf{u}})+\mathbf{N},  
\end{align}  
where $\overline{\mathbf{H}}(\bar{\mathbf{q}},\bar{\mathbf{u}})\in\mathbb{C}^{NM\times K}$ denotes the multiple-access channel from the $K$ users to a 6DMA surface across all $M$ possible positions $\bar{\mathbf{q}}\in\mathbb{C}^{3M\times 1}$ and rotations $\bar{\mathbf{u}}\in\mathbb{C}^{3M\times 1}$ in the specified region, with $M$ being the number of possible position-rotation pairs, which can be large in practice. \(\mathbf{X} \in \mathbb{C}^{L \times K}\) is the pilot matrix, where each entry is independent and identically distributed (i.i.d.) according to a complex Gaussian distribution with zero mean and unit variance, and $\mathbf{N}$ denotes the complex additive white Gaussian noise (AWGN) matrix at the BS, with $L$ being the pilot sequence length. For 6DMA systems, channel estimation aims to obtain the instantaneous or statistical CSI of $\overline{\mathbf{H}}(\bar{\mathbf{q}},\bar{\mathbf{u}})$ for all available 6DMA positions and rotations within a specified transmitter/receiver region, to optimize the 6DMA's position and rotation.  

Compared to conventional FPAs, 6DMA channel estimation introduces new challenges. First, unlike conventional MIMO systems, where the number of channels is limited due to the fixed antenna positions, 6DMA involves a much larger number of channels as the antennas can move to arbitrary positions and rotations within a specified region. This high-dimensional nature causes a significant increase in the pilot overhead for channel estimation.  
In addition to the larger number of 6DMA-induced channel coefficients compared to FPA systems, another challenge arises from the uneven channel power distribution of the 6DMA system. Specifically, for a 6DMA system with flexible position and rotation, the channel power between a user and candidate antenna can vary significantly for different positions/rotations in continuous 3D space. For example, the channels between a user and two 6DMA surfaces with opposing normal vectors will exhibit very different distributions. This significantly complicates CSI estimation for 6DMA compared to traditional MIMO systems.  

By exploiting angular-domain channel sparsity, sparse signal recovery methods like compressed sensing can be applied to estimate the instantaneous CSI based on measurements from only a limited number for antenna positions and rotations, thereby reducing training overhead. However, the assumption of angular-domain sparsity does not always hold in practice. Existing literature primarily focuses on two approaches for 6DMA channel estimation, namely statistical 6DMA channel estimation and instantaneous 6DMA channel estimation. In the following, we discuss a protocol for  CSI acquisition and data transmission in 6DMA systems, the unique directional sparsity of 6DMA channels, two types of channel estimation \cite{6dmaChan}, and key open issues for future research.  

\subsubsection{6DMA Directional Sparsity and Protocol Design}
In existing wireless networks, the BS is usually installed at a high altitude. Thus, there are only a limited number of scatterers around the BS, while there may be rich scattering near each of the users. Due to the rotatability, positionability, and antenna directivity of 6DMA, the channels between a given user and different candidate 6DMA positions and rotations in a continuous 3D space generally exhibit drastically different distributions.
 For example, in Fig. \ref{practical_scenario}, user $k$ on the ground 
	can establish a significant channel with 6DMA surface 1, but its
	channel with 6DMA surface $b$, which is directed towards the sky, is much weaker and thus can  be
	assumed to be approximately zero. We define this property as {\textit{directional sparsity}} as follows \cite{6dmaChan}.

{\emph{\textbf{Definition 1 (Directional Sparsity)}: In the considered 6DMA system, user $k$ is assumed to have non-zero channel gains to only a subset of 6DMA position-rotation pairs indexed by the set $\mathcal{W}_k\subseteq \mathcal{M}=\{1,2,\cdots,M\}$, 
(i.e., antenna gain $g_{k,l}(\mathbf{u}_{m})\neq 0, \forall m\in \mathcal{W}_k$ in \eqref{gm}); while for the remaining 6DMA position-rotation pairs $m\in \mathcal{W}_k^c$, where $
\mathcal{W}_k \cup \mathcal{W}_k^{\mathrm{c}} = \mathcal{M}$ and $ \mathcal{W}_k \cap \mathcal{W}_k^{\mathrm{c}} = \emptyset
$, the channels to user $k$ are assumed to be zero (i.e.,  $g_{k,l}(\mathbf{u}_{m})=0, \forall m\in \mathcal{W}_k^{\mathrm{c}}$ in \eqref{gm}). }}
\begin{figure}[t!]
	\centering
	\setlength{\abovecaptionskip}{0.cm}
	\includegraphics[width=3.1in]{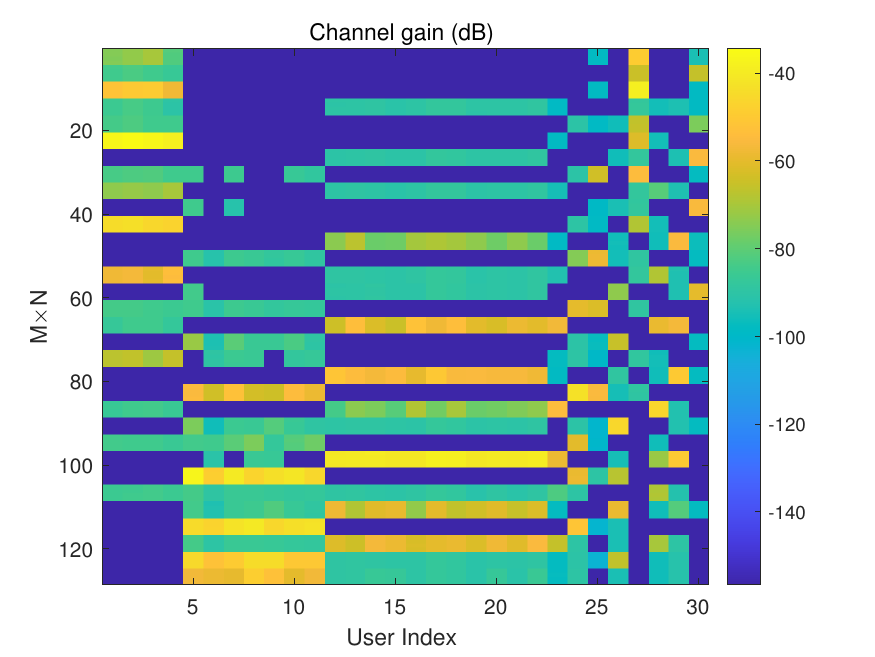}
	\caption{Illustration of the directional sparsity pattern of 6DMA channel $\overline{\mathbf{H}}(\bar{\mathbf{q}},\bar{\mathbf{u}})$.}
	\label{gain_N}
	\vspace{-0.59cm}
\end{figure}

The directional sparsity of the 6DMA channels \(\overline{\mathbf{H}}(\bar{\mathbf{q}},\bar{\mathbf{u}})\) is illustrated in Fig. \ref{gain_N}. The deep blue regions correspond to channel gains approximately  equal to zero. To characterize the directional sparsity of 6DMA channels, we define $\mathbf{Z}\in \mathbb{R}^{M \times K}$ as the directional sparsity indicator matrix as follows: 
\begin{align}
	[\mathbf{Z}]_{m,k} = \begin{cases} 
		1, & \text{if the channel between the}~ k \text{-th user} \\
		&\text{and the}~ m \text{-th candidate 6DMA}\\ &  \text{ position-rotation is non-zero}, \\
		0, & \text{otherwise}.
	\end{cases}
\end{align}

\begin{figure}[t!]
	\centering
	\setlength{\abovecaptionskip}{0.cm}
	\includegraphics[width=3.6in]{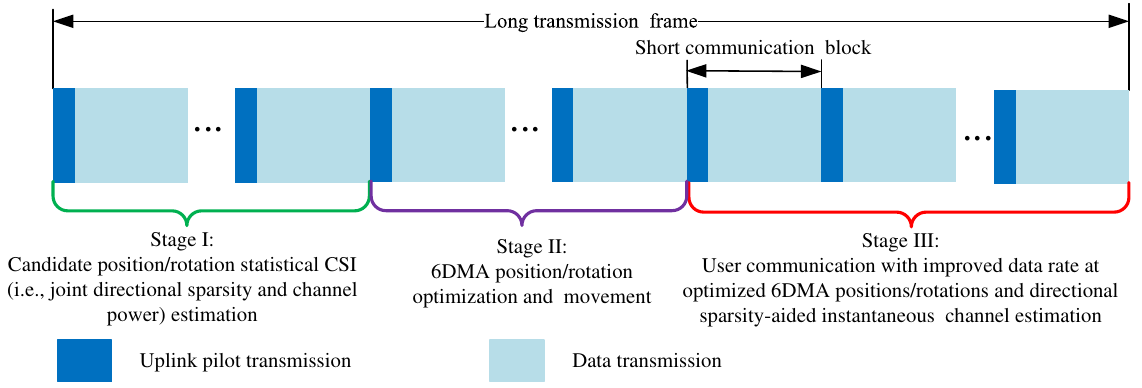}
	\caption{The proposed three-stage protocol for 6DMA systems.}
	\label{timescale}
			\vspace{-0.59cm}
\end{figure}

By leveraging the unique directional sparsity of 6DMA, a practical protocol for the operation of the 6DMA system was proposed in \cite{6dmaChan}. As illustrated in Fig. \ref{timescale}, the protocol is divided into three stages within a long transmission frame. Note that user data transmission continues throughout all three stages, with enhanced data rates achieved in the third stage after the 6DMA surfaces have been positioned and rotated optimally. The protocol is detailed as follows:
\begin{itemize}
	\item Stage I (Candidate position/rotation statistical CSI estimation): In the long transmission frame, we assume that the statistical CSI between all 6DMA candidate position-rotation pairs and all users remains constant. Thus, we can estimate the statistical CSI in terms of the directional sparsity matrix $\mathbf{Z}$ and the average channel power matrix \(\mathbf{P} \)  in Stage I. Specifically, the 6DMA surface moves over $\overline{M} \ll M$ different sampling position-rotation pairs to collect data for estimating the corresponding statistical CSI.
    Based on the estimated channel, we can reconstruct the statistical CSI for all $M$ possible position-rotation pairs.
	
	\item  Stage II (6DMA position/rotation optimization and movement):  With the global statistical CSI, the 6DMA-BS then optimizes the positions and rotations of all $B$ 6DMA surfaces in this stage to maximize the ergodic sum rate. Once the optimized positions and rotations of the 6DMA surfaces have been determined, the 6DMA surfaces are gradually moved over consecutive short communication  blocks to the optimized positions and rotations.

	\item  Stage III (User communication with improved data rate at optimized 6DMA positions/rotations and instantaneous channel estimation): With all 6DMA surfaces located at their optimized positions and rotations obtained in Stage II, the users can communicate with the BS with an improved sum rate in Stage III, where the instantaneous channels between the users and each 6DMA surface located at its optimized  position and rotation can be estimated at the BS by leveraging the directional sparsity determined in Stage I.
\end{itemize}
		
In this protocol, the candidate position/rotation channels are high-dimensional but statistically constant during the long transmission frame. Thus, the candidate position/rotation statistical CSI can be estimated much less frequently than the candidate position/rotation instantaneous CSI, and thus the average pilot overhead for the former is low in the long term. Furthermore, the directional sparsity of the 6DMA channel can be leveraged to further reduce the pilot overhead for both the estimation of the statistical CSI for the candidate positions/rotations and the estimation of instantaneous CSI for the optimized positions/rotations. Next, we discuss specific methods for estimating the statistical and instantaneous CSI based on this protocol, respectively.

\subsubsection{Statistical Channel Estimation}
Assuming that the channels of different users are statistically independent, the average network capacity $	R_{\mathrm{avg}}(\mathbf{q},\mathbf{u})$ in \eqref{pc0} can be upper-bounded by 
	\begin{align} 
		R_{\mathrm{avg}}(\mathbf{q},\mathbf{u})&\leq 
		\log_2 \det \left(\mathbf{I}_{K}+\frac{p}{\sigma^2}\mathbb{E}\left[{\mathbf{H}}(\mathbf{q},\mathbf{u})^H
		{\mathbf{H}}(\mathbf{q},\mathbf{u})\right]\right)
		\nonumber\\
				&=\sum_{k=1}^K\log_2 \left( 1+\frac{p}{\sigma^2}\sum_{b=1}^{B}[\mathbf{P}]_{k,b}\right),\label{o1}
	\end{align}
where \(\mathbf{P} \in \mathbb{R}^{B \times K}\) denotes the average channel power matrix, with its \((b,k)\)-th element given by \([\mathbf{P}]_{b,k} = \sum_{j=N(b-1)+1}^{Nb} \mathbb{E}\left[\left|[{\mathbf{H}}(\mathbf{q},\mathbf{u})]_{j,k}\right|^2\right]\). This element corresponds to the average power of the channels between user \(k\) and all antennas on the \(b\)-th 6DMA surface.  
From \eqref{o1}, it is observed that $R_{\mathrm{avg}}(\mathbf{q},\mathbf{u})$ depends on the statistical CSI (i.e., \(\mathbf{P}\)), which is determined by the antenna positions and rotations. Therefore, the statistical CSI for all candidate positions and rotations must be estimated to optimize the 6DMA surfaces' positions and rotations for maximizing the ergodic sum rate.

Next, we present the statistical channel estimation method proposed in \cite{6dmaChan, icc} to efficiently estimate the average channel power for 6DMAs with affordable complexity. This scheme leverages the directional sparsity characteristic of 6DMA channels and involves two main steps. First, all $B$ 6DMA surfaces are repositioned and rotated across a total of $\overline{M} > B$ distinct position-rotation pairs to gather channel measurements for estimating the channel power. Second, using the estimated channel power, the multi-path average power and the DOA vector are calculated to reconstruct the average channel power for all potential 6DMA positions and rotations.
\subsubsection*{\textbf{Step 1: Covariance-Based Average Channel Power Estimation}}
\begin{figure}[t!]
	\centering
	\setlength{\abovecaptionskip}{0.cm}
	\includegraphics[width=3.3in]{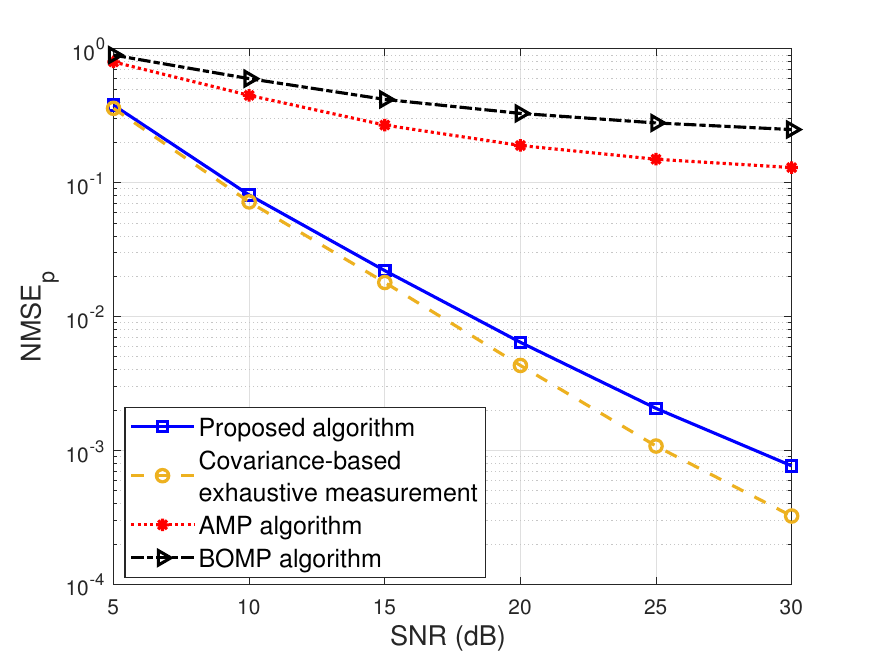}
	\caption{The NMSE of channel power estimation versus SNR with $K=50$.}
	\label{snr}
	\vspace{-0.65cm}
\end{figure}
To enable average channel power estimation in Step I, the 6DMA surfaces at the BS move over $\overline{M}$ different position-rotation pairs, and the BS collects $L$ pilot symbols from each user. The received signal at the \(m\)-th (\(m \in \{1,2,\cdots, \overline{M}\}\)) candidate position-rotation pair can be expressed as  
	\begin{align}
		\mathbf{Y}_m=\mathbf{X}\text{diag}([\bar{\mathbf{Z}}]_{m,:})\overline{\mathbf{H}}_m^T+\mathbf{N}_m,\label{aps1}
	\end{align}
where ${\bar{\mathbf{Z}}}\in \mathbb{R}^{\overline{M} \times K}$ represents the directional sparsity indicator matrix for the $\overline{M}$ position-rotation pairs, $\overline{\mathbf{H}}_m\in \mathbb{C}^{N\times K}$ denotes the channel matrix from all $K$ users to all the antennas of the $m$-th 6DMA candidate position-rotation pair, and $\mathbf{N}_m$ is the AWGN matrix. 
Next, we introduce the power state vector \(\boldsymbol{\eta}_m\in\mathbb{R}^{K} \), which is defined as $
\boldsymbol{\eta}_m\!=\!\left[[\bar{\mathbf{P}}]_{m,1}[\bar{\mathbf{Z}}]_{m,1}, [\bar{\mathbf{P}}]_{m,2}[\bar{\mathbf{Z}}]_{m,2}, \!\cdots,\! [\bar{\mathbf{P}}]_{m,K}[\bar{\mathbf{Z}}]_{m,K}\right]^T$,
where $\bar{\mathbf{P}}\in \mathbb{R}^{\overline{M} \times K}$ specifies the average channel power for the \( \overline{M} \) sampling position-rotation pairs. Our goal in Stage I is to estimate $\bar{\mathbf{P}}$ and ${\bar{\mathbf{Z}}}$, which will then be used as partial statistical CSI for reconstructing the full statistical CSI corresponding to all possible $M \gg \overline{M}$ 6DMA positions and rotations. This process requires estimating the power state vector $\boldsymbol{\eta}_m$ from the noisy observations $\mathbf{Y}_m$, based on the known pilot matrix $\mathbf{X}$. In general, this estimation can be formulated as a maximum likelihood estimation (MLE) problem \cite{zhilin}.

Specifically, we define
$\boldsymbol{\Sigma}_m=\mathbf{X}\text{diag}(\boldsymbol{\eta}_m)\mathbf{X}^H+\sigma^2\mathbf{I}_L$ and let $\hat{\boldsymbol{\Sigma}}_{m}=\frac{1}{N}\mathbf{Y}_m\mathbf{Y}_m^H$ denote the sample covariance matrix of the received signal for the $m$-th candidate position/rotation pair averaged over all $N$ antennas of the 6DMA surface. Then, the MLE problem can be formulated as follows
\begin{align} \label{co}
	\arg \min_{\boldsymbol{\eta}_m\in\mathbb{R}_+}\ln\text{det}(\boldsymbol{\Sigma}_m)+\text{tr}(\boldsymbol{\Sigma}_m^{-1}\hat{\boldsymbol{\Sigma}}_{m}).
\end{align}
To solve problem \eqref{co}, a coordinate descent method can be used \cite{6dmaChan}. Once \(\boldsymbol{\eta}_m\) has been obtained, the average channel power matrix can be expressed as \(\bar{\mathbf{P}} = [\boldsymbol{\eta}_1, \ldots, \boldsymbol{\eta}_M]^T\).
Furthermore, we set the directional sparsity matrix \([ \bar{\mathbf{Z}}]_{m,k} = 1\) if \([\boldsymbol{\eta}_m]_k\) exceeds a given threshold \(\epsilon > 0\); otherwise, \([ \bar{\mathbf{Z}}]_{m,k} = 0\). 
\begin{figure}[t!]
	\centering
	\setlength{\abovecaptionskip}{0.cm}
	\includegraphics[width=3.3in]{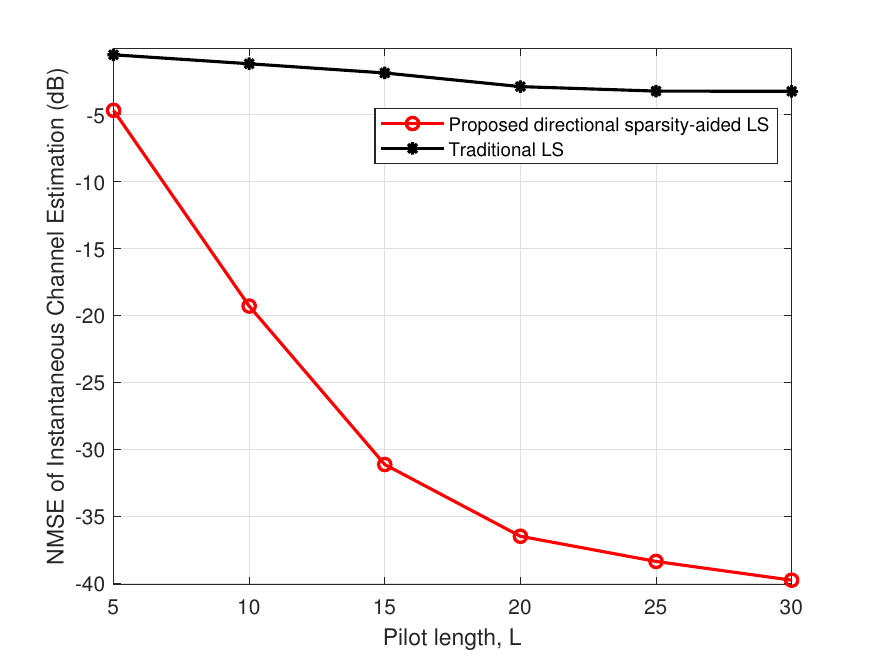}
	\caption{The NMSE of instantaneous channel estimation versus  pilot length with $K=30$ users.}
	\label{actual}
	\vspace{-0.59cm}
\end{figure}
\subsubsection*{\textbf{Step 2: Multi-Path Average Power and DOA Vector Estimation}}
Once the statistical CSI for a small number of \( \overline{M} \) position-rotation pairs, i.e., $\bar{\mathbf{P}}\in \mathbb{R}^{\overline{M} \times K}$, is estimated, the statistical CSI for all $M$ possible position-rotation pairs in the BS site region \(\mathcal{C}\) can be reconstructed at the BS \cite{icc}. In particular, the entries of \(\bar{\mathbf{P}}\) can be expressed as follows:
	\begin{align}
		[\bar{\mathbf{P}}]_{k,m}
		=N g_{k}(\mathbf{u}_{m},\mathbf{f}_{k}) s_k, \label{mm}
	\end{align}
where channel \(\mathbf{h}_{k,m}(\mathbf{q}_{m}, \mathbf{u}_{m})\) is given in \eqref{uk}, $s_k=\mathbb{E}\left[\left| \sum_{l=1}^{L_k}\sqrt{\mu_{l, k}}e^{-j \bar{\varphi}_{l, k}} \right|^2\right]$ represents the multi-path average power from the $k$-th user to the 6DMA-BS, and $\bar{\varphi}_{l, k}={\varphi}_{l, k}+\frac{2\pi}{\lambda}
(\mathbf{f}_{l, k}-\mathbf{f}_{k})^T\mathbf{r}_{m,n}(\!\mathbf{q}_{m},
\mathbf{u}_{m})$, which is modeled as an independent and uniformly distributed random variable in \([0, 2\pi)\). Eq. \eqref{mm} holds because 
$[\mathbf{h}_{k,m}(\mathbf{q}_{m},
\mathbf{u}_{m})]_n
\approx \sqrt{g_{k}(\mathbf{u}_{m},\mathbf{f}_{k})}
e^{-j\frac{2\pi}{\lambda}
	\mathbf{f}_{k}^T\mathbf{r}_{m,n}(\!\mathbf{q}_{m},
	\mathbf{u}_{m}\!)}\sum_{l=1}^{L_k}\sqrt{\mu_{l, k}}e^{-j \bar{\varphi}_{l, k}}$, 
where $g_{k, l}(\mathbf{u}_{m},\mathbf{f}_{k, l})$  is assumed to remain constant for all $l$, i.e., $g_{k, l}(\mathbf{u}_{m},\mathbf{f}_{k, l}) = g_{k}(\mathbf{u}_{m},\mathbf{f}_{k}), \forall l \in \{1, 2, \cdots, L_k\}$, with $\mathbf{f}_{k}$ denoting the unit DOA vector corresponding to the signal arriving at the BS from the center of the scattering cluster of user $k$.	

\begin{table*}[!t]
	\footnotesize
	\vspace{-0.88cm}
	\centering
	\caption{An Overview of Representative Works on 6DMA Position and Rotation Optimization}
	\label{6DMApr}
	\begin{tabular}{|>{\centering\arraybackslash}m{1.8cm}|p{1.6cm}|p{2cm}|p{2.5cm}|p{3.2cm}|p{2.2cm}|p{1.8cm}|}
		\hline
		\textbf{6DMA Configuration} & \textbf{Reference} & \textbf{System Setup} & \textbf{Design Objective} & \textbf{Optimization Techniques} & \textbf{CSI Assumption} & \textbf{Continuous or Discrete Movement} \\ \hline
		
		\multirow{15}{1.8cm}{6DMA \\ \vspace{5pt} (Joint position and rotation optimization)}
		& \cite{6dmaCon}  & MIMO communication & Sum-rate maximization         & AO and conditional
		gradient algorithm     & Statistical CSI       & Continuous \\ \cline{2-7}
		& \cite{6dmaDis}  & MIMO communication & Sum-rate maximization         & CSM and conditional
		gradient algorithm     & Statistical CSI  and no CSI     & Discrete \\ \cline{2-7}
		& \cite{6dmaChan}  & MIMO communication & Sum-rate maximization         & PSO      & Statistical CSI       & Discrete \\ \cline{2-7}
		& \cite{6dmaSens}  & Sensing, multiple-targets & CRB minimization        & PSO
		& Statistical CSI       & Continuous \\ \cline{2-7}
		& \cite{near}  & Hybrid-field 6DMA communication & Sum-rate maximization         & Deep reinforcement learning      & Statistical CSI       & Continuous \\ \cline{2-7}
		& \cite{passive, weidong6d}  & Passive 6DMA (6D-IRS) & Sum-rate maximization         & AO      &  Instantaneous CSI        & Continuous \\ \cline{2-7}
		& \cite{free6DMA}  & Cell-free communication & Sum-rate maximization         &  Bayesian
		optimization-based algorithm     & Statistical CSI       & Continuous \\ \cline{2-7}
		& \cite{lll}  & MIMO communication & Average sum logarithmic rate maximization         &  Low-complexity sequential optimization     & Statistical CSI       & Continuous \\ \cline{2-7}
				&\cite{6dmaPar}  & MIMO communication & Sum-rate maximization         &  Adaptive Markov Chain Monte Carlo based method     & Statistical CSI       & Continuous \\ \cline{2-7} \hline
		\multirow{6}{1.8cm}{Rotatable 6DMA \\ \vspace{5pt} (Rotation optimization)} 
		& \cite{rot1}  & High-speed communication &  Channel gain rank maximization         &  Mobility-aware beam training      & Instantaneous CSI       & Continuous \\ \cline{2-7} 
		& \cite{rachang}  & ISAC & Maximize the sum-rate and minimize the CRB        &  Block coordinate descent      & Instantaneous CSI       & Continuous \\ \cline{2-7} 
		& \cite{rot6}  & ISAC & Optimize sensing beampattern        &  AO      & Instantaneous CSI       & Continuous \\ \cline{2-7} 
		& \cite{rot2}  & MIMO communication &  Achievable spectral
		efficiency maximization         &  PSO    & Instantaneous CSI       & Continuous \\ \cline{2-7} 
		\hline	
		\multirow{17}{1.8cm}{Positionable 6DMA \\\vspace{5pt} (Position optimization)}
		& \cite{p94}  & MISO communication& Power maximization         & Discrete sampling and graph-based method      & Instantaneous CSI       & Discrete  \\ \cline{2-7}
		& \cite{p75}  & MIMO  communication & MIMO capacity maximization & AO and SCA                                    & Statistical CSI         & Continuous \\ \cline{2-7}
		& \cite{p98}  & MISO  communication& Transmit power minimization & Gradient descent algorithm                    & AoD/AoA, LoS only       & Continuous \\ \cline{2-7}
		& \cite{p49}  & MISO  communication& Sum-rate maximization       & Regularized LS-based OMP                     & Instantaneous CSI       & Continuous \\ \cline{2-7}
		& \cite{pwd}  & Secure communication & Secrecy rate maximization  & Partial enumeration algorithm  & Instantaneous and AoD/AoA & Discrete \\ \cline{2-7}
		& \cite{p60,p101,p102} & ISAC, single-target & Sum-rate maximization       & PSO and projected gradient                             & Instantaneous and AoD/AoA  & Continuous \\ \cline{2-7}
		& \cite{p61}  &ISAC, single-target & Spatial correlation         & Deep reinforcement learning                   & AoD/AoA                & Continuous \\ \cline{2-7}
		& \cite{p59}  &ISAC, single-target & Communication rate maximization & Gradient projection method                 & Instantaneous, AoD/AoA & Continuous \\ \cline{2-7}
		& \cite{jing, p24,p104}   & Sensing and ISAC & CRB minimization & AO and SCA          & Instantaneous and AoD/AoA        & Continuous \\ \cline{2-7} \hline
	\end{tabular}
	\begin{tablenotes}
		\footnotesize
		\item[1] AO: alternating optimization; MISO: multiple-input single-output; PSO: particle swarm optimization; AoD: angle of departure; AoA: angle of arrival; SCA: successive convex optimization; LS: least-square; OMP: orthogonal matching pursuit.
	\end{tablenotes}
	\vspace{-0.5cm}
\end{table*}

From \eqref{mm}, it is clear that, by determining \(\mathbf{f}_{k}\) and \(s_k\) based on the estimated \(\bar{\mathbf{P}}\) and \(\bar{\mathbf{Z}}\), the average channel power between the users and all feasible 6DMA positions and rotations at the BS can be reconstructed. Specifically, we utilize the channel directional sparsity matrix $\bar{\mathbf{Z}}$, estimated in Step I, to decrease the complexity of parameter estimation in Step II.
Specifically, we construct the support set \(\hat{\mathcal{I}}_k \in \mathbb{R}^{M_k}\) for \([\bar{\mathbf{Z}}]_{:,k}\), which includes the indices of the non-zero elements of \([\bar{\mathbf{Z}}]_{:,k}\), where \(M_k < \overline{M}\) denotes the number of non-zero elements.
We define $\bar{\mathbf{p}}_k=[\bar{\mathbf{P}}]_{:,k}\in \mathbb{R}^{M}$ and $\mathbf{v}_k =[ g_{k}(\mathbf{u}_{1},\mathbf{f}_{k}),\cdots, g_{k}(\mathbf{u}_{M},\mathbf{f}_{k})]^T\in \mathbb{R}^{M}$. 
Then, let \(\bar{\mathbf{p}}_{k,\hat{\mathcal{I}}_k} \in \mathbb{R}^{M_k}\) and \(\mathbf{v}_{k,\hat{\mathcal{I}}_k} \in \mathbb{R}^{M_k}\) denote the new vectors formed by the non-zero elements of \(\bar{\mathbf{p}}_k\) and \(\mathbf{v}_k\) that correspond to the indices stored in \(\hat{\mathcal{I}}_k\). 
Consequently, we optimize 
$\mathbf{f}_{k}$ and $s_k$ by minimizing the reconstruction error:
\begin{subequations}
	\begin{align}\label{ac}
		~&~ \min_{ s_k, \mathbf{f}_{k} } \left\|\bar{\mathbf{p}}_{k,\hat{\mathcal{I}}_k} - N\mathbf{v}_{k,\hat{\mathcal{I}}_k} s_k  \right\|^2_2,\\
		~&~ \text{s.t.}~~~s_{k}\ge 0.
	\end{align}
\end{subequations}
Problem \eqref{ac} can be reformulated as a compressed sensing problem by uniformly discretizing $\mathbf{f}_{k}$ into $G\geq 0$ grid
points, and can be efficiently solved using standard compressed sensing algorithms, such as non-negative OMP \cite{icc, 6dmaChan, let}. Once the estimates of \( s_k \) and \( \mathbf{f}_{k} \) are obtained, the channel power for all possible 6DMA positions and rotations, represented as \( \hat{\mathbf{P}} \in \mathbb{R}^{M \times K} \), can be reconstructed according to \eqref{mm}. Additionally, the directional sparsity for all possible 6DMA positions and rotations, represented as \( \hat{\mathbf{Z}} \in \mathbb{R}^{M \times K} \), can be reconstructed by setting \( [\hat{\mathbf{Z}}]_{m,k} = 1 \) if \( [\hat{\mathbf{P}}]_{m,k} \) exceeds a specified threshold \( \epsilon > 0 \).
\subsubsection{Instantaneous Channel Estimation}
With the estimated directional sparsity from Stage I and the optimized 6DMA positions/rotations from Stage II of the protocol, we proposed a directional sparsity-aided LS algorithm in \cite{6dmaChan} to estimate the instantaneous CSI from the users to each 6DMA surface at the optimized position/rotation in Stage III. Unlike statistical channels, instantaneous channels vary rapidly and require a high estimation frequency. This leads to prohibitively high processing costs and latency when using the centralized 6DMA processing architecture shown in \ref{practical_scenario}(a). Therefore, in this subsection, we consider the 6DMA distributed processing architecture shown in Fig. \ref{practical_scenario}(b), where all LPUs can simultaneously and in parallel estimate the instantaneous channels between the users and their respective 6DMA surfaces. Moreover, we leverage the knowledge of channel directional sparsity matrix $\hat{\mathbf{Z}}$ estimated during the statistical channel estimation stage to improve the instantaneous channel estimation accuracy. Specifically, we construct the support set of $\mathbf{h}_{b}=\text{vec}(\overline{\mathbf{H}}_b^T)\in\mathbb{C}^{NK\times 1}$ as  \(\bar{\mathcal{I}}_b=\mathrm{support}\left([\hat{\mathbf{Z}}]_{b,:}^T\otimes  \mathbf{1}_{N \times 1}\right)\in \mathbb{C}^{\bar{K}\times 1}\), which identifies the indices of non-zero elements in $[\hat{\mathbf{Z}}]_{b,:}^T\otimes  \mathbf{1}_{N \times 1}$. Here, \( \bar{K} < NK \) denotes the number of indexed elements in \( \bar{\mathcal{I}}_b \).
Let $\mathbf{A}_{\bar{\mathcal{I}}_b}$ denote the matrix composed of the corresponding columns of $\bar{\mathcal{I}}_b$ in matrix $\mathbf{A}=\mathbf{I}_{K}\otimes{\mathbf{X}}\in\mathbb{C}^{NL\times KN}$, and $\mathbf{h}_{b,\bar{\mathcal{I}}_b}$ represent the vector consisting of the rows of $\bar{\mathcal{I}}_b$ from vector ${\mathbf{h}}_{b}$. Consequently, we derive the support-restricted estimates at the $b$-th LPU ($b\in \mathcal{B}$) using the following LS-based approach: 
\begin{align}
	& \mathbf{h}_{b,\bar{\mathcal{I}}_b}\leftarrow\arg \min_{\mathbf{h}} \|\mathbf{A}_{\bar{\mathcal{I}}_b}\mathbf{h}-
	\mathbf{y}_{b}\|_2^2,\nonumber\\
	& \text{and}~~\mathbf{h}_{b,\bar{\mathcal{I}}_b^{\mathrm{c}}}\leftarrow 0, \label{ops}
\end{align}
where $
\bar{\mathcal{I}}_b \cup \bar{\mathcal{I}}_b^{\mathrm{c}} = \{1,2,\cdots,NK\}$ and $ \bar{\mathcal{I}}_b \cap \bar{\mathcal{I}}_b^{\mathrm{c}} = \emptyset
$.
\begin{table*}[!t]
	\footnotesize
	\centering
	\caption{Summary of Representative Works on 6DMA Channel Estimation}
	\label{tab:6dma_summary}
	\begin{tabular}{|p{2.1cm}|p{1.2cm}|p{3.4cm}|p{9.5cm}|} % 新增第二列宽度可根据需要调整
		\hline
		\textbf{6DMA Configuration} & \textbf{Reference} & \textbf{System Setup} & \textbf{Main Contribution} \\ \hline
		\multirow{4}{2.1cm}{{6DMA}}
		& \cite{icc} 
		& Uplink multi-user MIMO               
		& Statistical channel estimation leveraging directional sparsity of 6DMA Channel. \\ \cline{2-4}
		& \cite{6dmaChan} 
		& Distributed processing architecture          
		& Distributed statistical and instantaneous channel estimation using directional sparsity.\\ \cline{2-4}
		& \cite{htz6} 
		& Uplink THz communication              
		& Instantaneous channel estimation leveraging uneven channel power distribution properties through the concept of directional sparsity of the 6DMA channel. \\
		 \hline
		\multirow{5}{2.2cm}{{Rotatable 6DMA}}
		& \cite{15} 
		& Rotational uniform linear array       
		& Achieves high-resolution DOA estimation for under-determined cases and handles spatial noise covariance. \\ \cline{2-4}
		& \cite{lie} 
		& Single rotational antenna                    
		& Enables high-resolution DOA estimation using a single rotating directional antenna. \\ \cline{2-4}
		& \cite{rot9} 
		& Dual rotating antennas     
		& Proposes space–time-ambiguity decomposition method for DOA estimation to boost antispoofing with rotating dual antenna. \\ \hline
		\multirow{10}{2.2cm}{{Positionable 6DMA}}
		& \cite{123} 
		& Multi-cell homogeneous network                 
		& Proposes linear minimum mean-squared error (LMMSE)-based channel estimation to minimize overhead by leveraging observable positions. \\ \cline{2-4}
		& \cite{147} 
		& Multiuser uplink mmWave MIMO             
		& Proposes a low-sample-size sparse channel reconstruction method that leverages the sparse propagation paths of mmWave channels. \\ \cline{2-4}
		& \cite{148} 
		& Point-to-point mmWave system              
		& Estimates channel parameters using least squares regression for a linear fluid antenna system. \\ \cline{2-4}
		& \cite{xiao2024channel, ma2023compressed,h00} 
		& Point-to-point system                
		& Estimates channel parameters using OMP via successive or joint estimation and studies measurement setups. \\ \cline{2-4}
		& \cite{linglong} 
		& Point-to-point system with linear positionable 6DMA        
		& Proposes successive Bayesian reconstruction for robust channel estimation across antennas. \\ \hline
	\end{tabular}
	\vspace{-0.5cm}
\end{table*}

To illustrate the performance of our proposed 6DMA channel estimation algorithm, Fig. \ref{snr} presents the normalized mean square error (NMSE) of statistical channel estimation as a function of the signal-to-noise ratio (SNR) for the proposed algorithm and the following benchmark methods: 1) Covariance-based exhaustive measurement, which uses the covariance-based algorithm in Step I to estimate the channel powers from all users to all ${M}$ possible position-rotation pairs without performing parameter estimation in Step II; 2) Approximate message passing (AMP) algorithm \cite{vamp}; and 3) Block orthogonal matching pursuit (BOMP) algorithm \cite{bomp}. For both AMP and BOMP, Step I involves estimating the row-sparse channel \(\tilde{\mathbf{H}}_m=\text{diag}([\bar{\mathbf{Z}}]_{m,:})\overline{\mathbf{H}}_m^T\) from \( \mathbf{Y}_m \) as defined in \eqref{aps1}. Once \(\tilde{{\mathbf{H}}}_m\) is recovered, the channel power is subsequently determined based on it, while Step II remains consistent with the proposed algorithm. 
The results demonstrate that the proposed algorithm outperforms both the AMP and BOMP algorithms because the benchmark methods are required to estimate a significantly larger number of unknown parameters to accurately recover the specific entries of \(\tilde{\mathbf{H}}_m\). In addition, the proposed statistical channel reconstruction algorithm exhibits a slightly higher NMSE compared to the covariance-based exhaustive measurement approach. However, it only samples $\overline{M}=32$ sampled position-rotation pairs, which is considerably fewer than the ${M}=350$ required by the covariance-based exhaustive measurement method. Furthermore, Fig. \ref{actual} shows the NMSE for instantaneous CSI estimation as a function of the pilot sequence length $L$. It is observed that the proposed directional sparsity-aided LS algorithm achieves significantly higher estimation accuracy compared to the traditional LS algorithm. This is because the proposed LS algorithm utilizes the directional sparsity from Stage I, thereby requiring fewer measurements than the traditional LS method, which does not exploit this sparsity.

\subsection{Other Related Works and Future Directions}
In Table \ref{6DMApr}, we summarize representative works on 6DMA position and rotation optimization, focusing on the considered system setups, adopted optimization techniques, and key findings.  
The antenna position and rotation optimization works discussed above primarily rely on the 6DMA centralized processing architecture shown in Fig. \ref{practical_scenario}(a). However, there is an urgent need to design distributed or parallel position and rotation optimization algorithms \cite{dis3} by effectively combining the signal processing capabilities of both LPUs and the CPU of the distributed processing architecture shown in Fig. \ref{practical_scenario}(b). From an optimization perspective, it is also essential to develop more advanced and computationally efficient algorithms for 6DMA position and rotation design, particularly for systems with practically large antenna arrays. A valuable direction is the development of stochastic geometry frameworks that accurately incorporate accurate spatial correlation, mutual coupling, and optimization effects, which would allow researchers to evaluate the impact of large-scale 6DMA deployment in future 6G networks. In addition, the impact of polarization changes due to antenna positioning and rotation has been largely overlooked. Polarization is a fundamental property of electromagnetic waves, describing the orientation of the electric field vector as the wave propagates. Future position/rotation optimization research should consider this factor to further enhance the performance of 6DMA systems.  Furthermore, the analysis of 6DMA performance is highly challenging in many cases, which makes it essential to explore novel methods for effective evaluation. One possible approach is to simplify the 6DMA model, such as by applying the circular 6DMA discussed in Section IV-B, which makes performance analysis more manageable. 

In Table \ref{tab:6dma_summary}, we summarize representative works on 6DMA channel estimation based on the considered system setups and 6DMA configuration. The application of directional sparsity in 6DMA systems requires further exploration. For example, the estimated directional sparsity matrix identifies the subsets of candidate 6DMA positions and rotations providing significant channel power for each user. This directional information can be utilized to design an efficient codebook that prioritizes beamforming directions aligned with the dominant scattering paths of the users. By focusing on a reduced set of significant directions, the codebook size can be optimized to balance coverage and computational complexity. For 6DMA channel estimation, in addition to the above-discussed methods, other signal processing techniques from 6DMA-assisted systems can also be incorporated. Examples include the sparse Bayesian algorithm \cite{baye}, the tensor method \cite{xinran}, the estimation of signal parameters via rotational invariance technique (ESPRIT) \cite{esp}, machine learning, and the space-alternating generalized expectation maximization scheme \cite{maxm}. These methods vary in terms of estimation accuracy and computational complexity, offering different trade-offs depending on system requirements. Future research should integrate these advanced signal processing techniques with existing/new  6DMA frameworks to further improve channel estimation efficiency and accuracy. However, the minimum number of observable sampling positions/rotations needed to reliably recover the CSI for different 6DMA surface configurations, including variations in movement region size and position/rotation dimensions, remains an open question. One possible approach to address this issue is to apply the Nyquist-Shannon sampling theorem, which defines the conditions for accurately reconstructing a continuous signal from its discrete samples without information loss. Another approach is to investigate the functional DoFs from electromagnetic information theory \cite{239f}. In this context, the functional DoFs refer to the minimum number of samples required to reconstruct a given electromagnetic field \cite{240f}.
\section{6DMA Applications}
\begin{figure}[!t]
	\centering
	\includegraphics[width=0.39\textwidth]{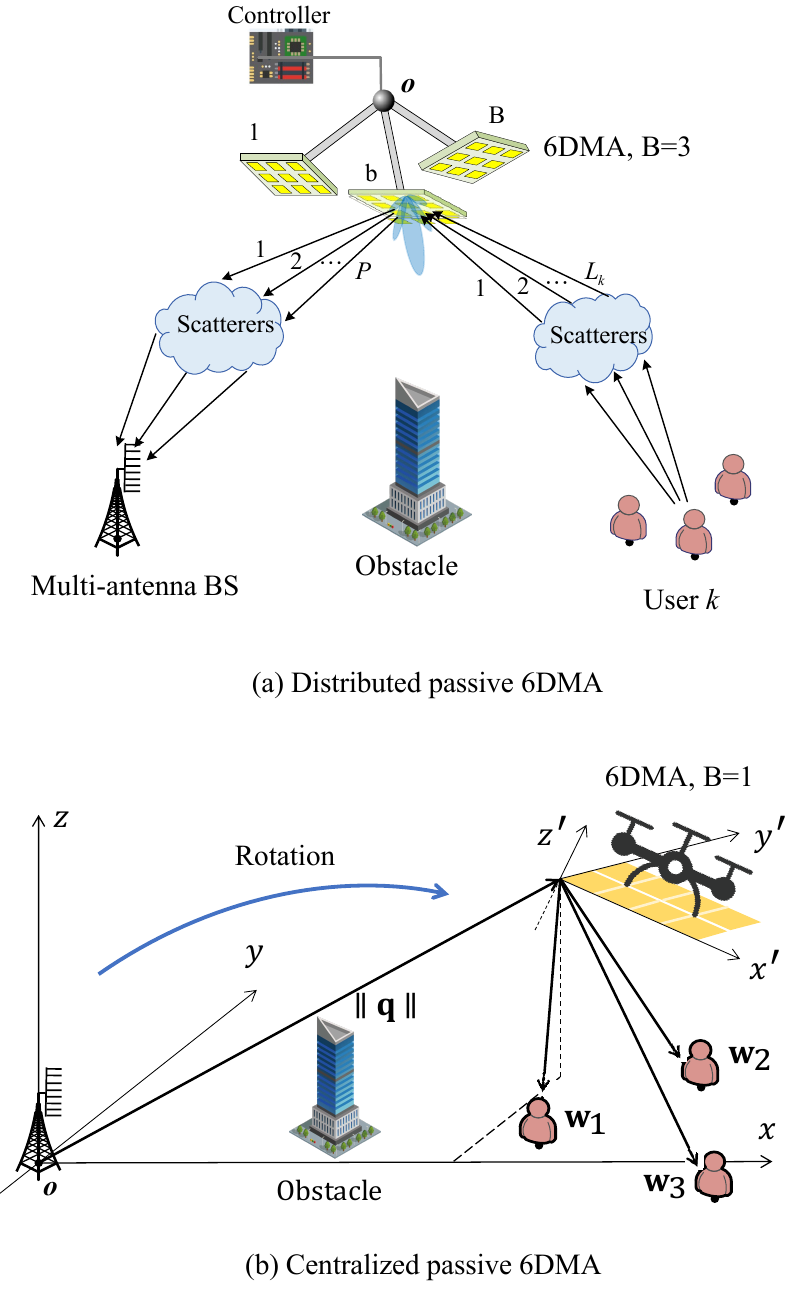}
	\caption{Passive 6DMA-assisted multiuser uplink communication.}
	\label{Fig_main}
			\vspace{-0.59cm}
\end{figure}

Benefiting from the efficacy of 6DMA in enhancing communication performance through optimized antenna positions and rotations (orientations), as demonstrated in the previous sections, this section discusses typical applications of 6DMA in wireless networks, including passive 6DMA, 6DMA cell-free networks, 6DMA for wireless sensing, 6DMA for unmanned aerial vehicles (UAVs), and other pertinent applications.

\subsection{Passive 6DMA: Distributed or Centralized?}
Previous studies have primarily focused on 6DMAs composed of active antenna elements, referred to as active 6DMA. 6DMA surfaces can also be made of passive reflecting elements (via, e.g., IRS/RIS) \cite{ris2, ris3,ris4,ris5} to reconfigure the wireless signal propagation in a more cost-effective manner as compared to active 6DMA. Furthermore, compared to traditional IRS/RIS with fixed-position elements, passive 6DMA can enhance the wireless system performance by optimizing the positions and rotations of their elements. In this subsection, we discuss a passive 6DMA system consisting of passive IRSs that are adjustable in both 3D position and 3D rotation. 
\begin{figure*}[!t] 
	\centering  
	\vspace{-0.35cm} 
	\subfigtopskip=2pt 
	\subfigbottomskip=2pt 
	\subfigcapskip=-5pt
	\subfigure[End-to-end path gain]{
		\label{level.sub.1}
		\includegraphics[width=0.43\linewidth]{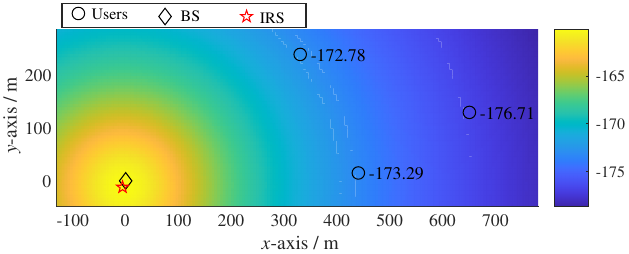}}
	\quad 
	\subfigure[Effective aperture gain]{
		\label{level.sub.2}
		\includegraphics[width=0.43\linewidth]{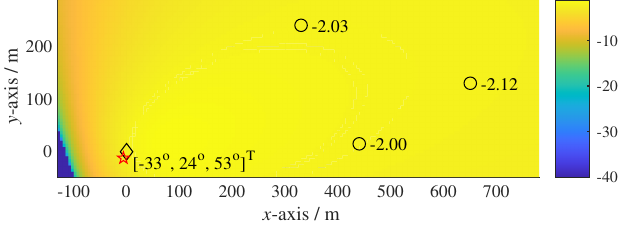}}
	\subfigure[Passive beamforming gain]{
		\label{level.sub.3}
		\includegraphics[width=0.43\linewidth]{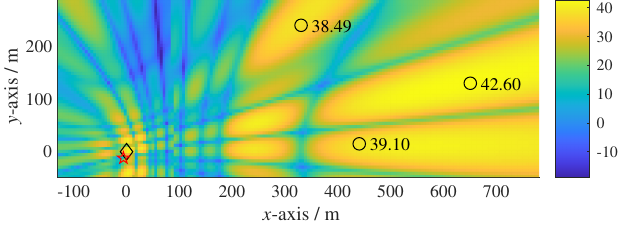}}
	\quad
	\subfigure[Max-min user SNR]{
		\label{level.sub.4}
		\includegraphics[width=0.43\linewidth]{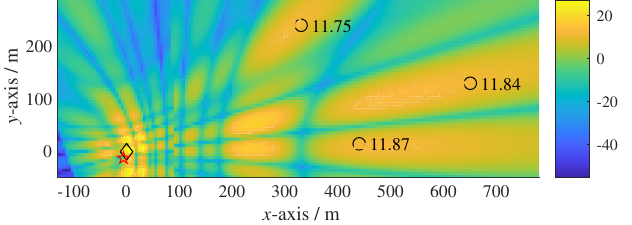}}
	\caption{Distribution of the optimized path gain, effective aperture gain, passive beamforming gain, and max-min user SNR over the passive 6DMA's movement region.}
	\label{level}
	\vspace{-0.59cm}
\end{figure*}

In this context, passive 6DMA can be deployed in two ways: distributed passive 6DMAs, where the reflecting elements form $B>1$ distributed IRSs, each capable of independently rotating and adjusting its position (illustrated in Fig. \ref{Fig_main}(a)); or a centralized passive 6DMA, where all reflecting elements form a single (i.e., $B=1$) large IRS that can freely rotate and adjust its position (e.g., via  a UAV as illustrated in Fig. \ref{Fig_main}(b)) \cite{passive}. It is worth noting that, in a single-user LoS scenario, these two passive 6DMA strategies are equivalent since both of them can yield the maximum received signal power at the user/BS. However, for multi-user scenarios (Fig. \ref{gain} (a)) or single-user scenarios with multipath channels (Fig. \ref{gain} (b)), the two passive 6DMA strategies generally result in distinct channel characteristics between the users and the BS.

In the following, we first consider the distributed passive 6DMA system proposed in \cite{passive}, where $K$ single omnidirectional-FPA users transmit data to a BS equipped with $Q$ antennas, with the assistance of a distributed passive 6DMA (see Fig. \ref{Fig_main} (a)).
We assume that the direct link between each user and the BS is severely blocked by obstacles and thus can be ignored. There are in total $B$ passive 6DMA/IRS surfaces, and each surface is equipped with $N$ IRS elements. The location and rotation of the $b$-th IRS are represented by $\mathbf{q}_b$ and $\mathbf{u}_b$, respectively. Let $\boldsymbol{\theta}_b\triangleq\left[\theta_{b,1},\cdots\theta_{b,N}\right]^T$ denote the reflection coefficients of passive 6DMA surface $b$, where the reflection amplitude is set to one to maximize the signal reflection power and thus $\left|\theta_{b,n}\right|=1$, $\forall n=1\cdots N$, $b\in \mathcal{B}$. 
Then, the cascaded channel between user $k$ and the BS via all passive 6DMA surfaces can be expressed as \cite{passive}

\begin{align}\label{pcr}
	\boldsymbol{h}_k\left(\mathbf{q},\mathbf{u},\boldsymbol{\theta}\right)=\boldsymbol{V}\left(\mathbf{q},\mathbf{u}\right)\text{diag}\left(\boldsymbol{\theta}\right)\boldsymbol{\chi}_k\left(\mathbf{q},\mathbf{u}\right),
\end{align}
with $\mathbf{q}$ and $\mathbf{u}$ being defined in \eqref{bb1},
$	\boldsymbol{V}\left(\mathbf{q},\mathbf{u}\right)=\left[\boldsymbol{V}\left(\mathbf{q}_1,\mathbf{u}_1\right),\cdots,\boldsymbol{V}\left(\mathbf{q}_B,\mathbf{u}_B\right)\right]$ and $
\boldsymbol{\chi}_k\left(\mathbf{q},\mathbf{u}\right)=\left[\boldsymbol{\chi}_k\left(\mathbf{q}_1,\mathbf{u}_1\right)^T,\cdots,\boldsymbol{\chi}_k\left(\mathbf{q}_B,\mathbf{u}_B\right)^T\right]^T$, where $\boldsymbol{\chi}_k\left(\mathbf{q}_b, \mathbf{u}_b\right)=\sum_{l=1}^{L_k}a_{k,l}\sqrt{G_{k,l}^{I}\left( \mathbf{u}_b\right)}\boldsymbol{t}_{k,l}\left(\mathbf{q}_b, \mathbf{u}_b \right)\in\mathbb{C}^{N\times1}$ and $\boldsymbol{V}\left(\mathbf{q}_b, \mathbf{u}_b\right)=\sum_{p=1}^{P}v_{p}\sqrt{G^{R}_{p}\left( \mathbf{u}_b\right)}\mathbf{z}_p\bar{\mathbf{e}}^{H}_{p}\left(\mathbf{q}_b, \mathbf{u}_b \right)\in\mathbb{C}^{Q\times N}$ denote the baseband equivalent channels for the user $k$$\rightarrow$passive 6DMA surface $b$ and passive 6DMA surface $b$$\rightarrow$BS links w.r.t. surface $b$'s position $\mathbf{q}_b$ and rotation $\mathbf{u}_b$, respectively. Here, $L_k$ and $P$ denote the number of propagation paths between user $k$ and passive 6DMA surface $b$ and that between surface $b$ and the BS, respectively, $a_{k,l}$ and  $v_{p}$ are the complex-valued gains of the corresponding channels, \(\mathbf{z}_p \in \mathbb{C}^{M \times 1}\) represents the steering vector of the BS, and $G_{k,l}^{I}\left( \mathbf{u}_b\right)$ $\left(G^{R}_{p}\left( \mathbf{u}_b\right)\right)$ denotes the incident (reflective) radiation pattern of each reflecting element in surface $b$ corresponding to the $l$-th path from user $k$ (the $p$-th path to the BS). $\boldsymbol{t}_{k,l}\left(\mathbf{q}_b, \mathbf{u}_b \right)=\left[e^{j\frac{2\pi}{\lambda}\boldsymbol{f}_{k,l}^T\boldsymbol{r}_{b,1}\left(\mathbf{q}_b, \mathbf{u}_b \right)},\cdots,e^{j\frac{2\pi}{\lambda}\boldsymbol{f}_{k,l}^T\boldsymbol{r}_{b,N}\left(\mathbf{q}_b, \mathbf{u}_b \right)}\right]^T$ and 
$\bar{\mathbf{e}}_p\left(\mathbf{q}_b, \mathbf{u}_b\right) = \left[e^{j\frac{2\pi}{\lambda}\boldsymbol{s}_{p}^T\boldsymbol{r}_{b,1}\left(\mathbf{q}_b, \mathbf{u}_b \right)},\cdots,e^{j\frac{2\pi}{\lambda}\boldsymbol{s}_{p}^T\boldsymbol{r}_{b,N}\left(\mathbf{q}_b, \mathbf{u}_b \right)}\right]^T$
denote the array response vectors of passive 6DMA surface $b$ from the $l$-th transmit path of user $k$ and towards the $p$-th reflective path to the BS, respectively, where $\boldsymbol{r}_{b,n}\left(\mathbf{q}_b, \mathbf{u}_b \right)$ denotes the position of the $n$-th antenna of the $b$-th passive 6DMA surface, as defined in \eqref{nwq}, and $\boldsymbol{f}_{k,l}$ and $\boldsymbol{s}_{p}$ are the direction of arrival/departure vectors with unit-norm for the $l$-th path from user $k$ and the $p$-th path to the BS, both w.r.t. the reference position $\boldsymbol{o}$ in Fig. \ref{Fig_main} (a).

To maximize the achievable rate of all users, denoted by \( R(\mathbf{q}, \mathbf{u}, \boldsymbol{W}, \boldsymbol{\theta}) \), the IRS phase shifts need to be jointly optimized with the surfaces' positions and rotations as well as the receive beamforming matrix \( \boldsymbol{W} \). The resulting optimization problem is formulated as \cite{passive}
\begin{subequations}\label{prob_original2}
	\begin{align}
		\text{(P-PAD)}:~& \mathop{\max}\limits_{\mathbf{q}, \mathbf{u}, \boldsymbol{W}, \boldsymbol{\theta}} ~ R(\mathbf{q}, \mathbf{u}, \boldsymbol{W}, \boldsymbol{\theta})
		\label{ooP2a}\\
		\mathrm{s.t.} ~ & \eqref{M2}, \eqref{M3}, \eqref{jM3},\eqref{M1}\\
		~ & |\theta_{n}|= 1,  n=1,\cdots,NB, \label{ooP2aaaa}
	\end{align}
\end{subequations}
where (\ref{ooP2aaaa}) is the unit-modulus constraint for each passive reflecting element. It is important to note that the phase shift of the IRS makes the resulting optimization problem, \((\text{P-PAD})\), more challenging to solve compared to problem \((\text{P1-CO})\) in the active 6DMA case. To address this non-convex optimization problem, an AO algorithm was employed in \cite{passive} to decompose the problem into three subproblems, which were then solved alternately in an iterative manner.

On the other hand, to reduce hardware costs and signaling overhead among multiple distributed passive 6DMAs, centralized passive 6DMA is an alternative option. As depicted in Fig. \ref{Fig_main} (b), the authors in \cite{liu2024joint} and \cite{weidong6d} focused on a new UAV-enabled passive 6DMA-aided downlink communication system, where an IRS with $N$ passive reflecting elements and  $B=1$ surface is mounted on a UAV to assist in the wireless transmission from a $Q$-antenna BS to $K$ remote users, each equipped with a single omnidirectional FPA.

For convenience, we establish a global CCS in Fig.~\ref{Fig_main} (b) and assume that the BS is located at the origin, denoted as ${\boldsymbol{o}}=[0,0,0]^T$. Denote by $\mathbf{q}$ and $\mathbf{J}_k$ the coordinates of the UAV/IRS and user $k$, respectively. Furthermore, let the rotation (or orientation) of the IRS be represented by $\mathbf{u}$. Assuming LoS BS-UAV and UAV-user links as well as blocked BS-user links, let $\phi_1(\mathbf{q}, \mathbf{u})$ and $\phi_{2,k}(\mathbf{q}, \mathbf{u})$ denote the incident angle and reflected angle at the IRS from the BS to user $k$ given the location and rotation of the IRS. Hence, the coordinates of user \( k \in {\cal K},\) and the BS in the local CCS are given respectively by
\begin{subequations}
\begin{align} \label{wklocal}
	&\mathbf{J}_k^{\text{local}} = \mathbf{R}^T(\mathbf{u})(\mathbf{J}_k-\mathbf{q})=[w_{kx}^{\text{local}},w_{ky}^{\text{local}}, w_{kz}^{\text{local}}]^T\!, \\
	&{\boldsymbol{o}}^{\text{local}}=\mathbf{R}^T(\mathbf{u})({\boldsymbol{o}}-\mathbf{q}) =[b_{x}^{\text{local}},b_{y}^{\text{local}},b_{z}^{\text{local}}]^T. \label{bzlocal}
\end{align}
\end{subequations}
Based on the above, we have
$\phi_1(\mathbf{q},\mathbf{u}) =\arccos \frac{-b_{z}^{\text{local}}}{\|\mathbf{q}\|}, 
	\phi_{2,k}(\mathbf{q}$, $\mathbf{u}) =\arccos \frac{-w_{kz}^{\text{local}}}{\|\mathbf{q}-\mathbf{J}_k\|}$,
and $F_k(\mathbf{q},\mathbf{u})=\cos\phi_1(\mathbf{q}, \mathbf{u})\cos\phi_{2,k}(\mathbf{q}, \mathbf{u})$ denotes the effective aperture gain under the angle-dependent model.
Moreover, we define the AoD from the BS to the IRS, the elevation/azimuth AoA at the IRS from the BS, the elevation/azimuth AoD from the IRS to user $k$, and the reference path gain as $\phi_{BI}$, $\vartheta_{IB}^{(e)}$, $\vartheta_{IB}^{(a)}$, $\phi_{Ik}^{(e)}$, $\phi_{Ik}^{(a)}$, and $\beta_0$, respectively. The channel from the BS to the IRS is given by
\begin{equation} \label{G}
	\mathbf{H}_{BI} = \frac{\sqrt{\beta_0}}{\|\mathbf{q}\|}e^{-j\frac{2\pi \|\mathbf{q}\|}{\lambda}}\mathfrak{\mathbf{a} }_I (\vartheta_{IB}^{(e)},\vartheta_{IB}^{(a)})\mathfrak{\mathbf{a} }^H_B({\phi}_{BI}),
\end{equation}
where $\lambda$ represents the wavelength, and $\mathbf{a}_{I} \in \mathbb{C}^{N\times1}$ along with $\mathbf{a}_{B} \in \mathbb{C}^{Q \times 1}$ denote the 
receive array response vector at the IRS and the transmit array response vector at the BS, respectively \cite{weidong6d}. Similarly, the channel from the IRS to user $k$ can be given as 
\begin{equation} 
	\mathbf{h}_{k}^H = \frac{\sqrt{\beta_0}}{\|\mathbf{q}-\mathbf{J}_k\|} e^{-j\frac{2\pi \|\mathbf{q}-\mathbf{J}_k\|}{\lambda}}\mathfrak{\mathbf{a} } ^H_k(\phi_{Ik}^{(e)}, \phi_{Ik}^{(a)}),\;k \in \mathcal{K}, \label{h}
\end{equation}
where $\mathfrak{\mathbf{a}}_k \in \mathbb{C}^{N \times 1}, k \in \cal K$ denotes the transmit array response vector from the IRS to user $k$ \cite{weidong6d}.

To fully exploit all 6D DoFs available for centralized passive 6DMA, we aim to  maximize the minimum received SNR, denoted by $r_k(\mathbf{q},\mathbf{u},\boldsymbol{\theta})=
	\frac{P_B Q \beta_0^2 F_{k}(\mathbf{q}, \mathbf{u})G_k(\mathbf{q}, \mathbf{u},\boldsymbol{\theta})}{\sigma_k^2\|\mathbf{q}\|^2\|\mathbf{q}-\mathbf{J}_k\|^2}$, among all users, by jointly optimizing the IRS's reflection coefficients \(\boldsymbol{\theta}\), position vector \(\mathbf{q}\), and rotation angle vector \(\mathbf{u}\). Here, $\sigma_k^2$ denotes the noise power at user $k$ and $G_k(\mathbf{q}, \mathbf{u},\boldsymbol{\theta}) = \lvert \mathfrak{\mathbf{a}}^H_k(\phi_{Ik}^{(e)}, \phi_{Ik}^{(a)})\text{diag}(\boldsymbol{\theta})\mathfrak{\mathbf{a}}_I (\vartheta_{IB}^{(e)},\vartheta_{IB}^{(a)})\rvert^2$ denotes the IRS passive beamforming gain. 
	The associated design problem can be formulated as follows
\begin{subequations}
\begin{align} 
	\text{(P-PAC)}: \max_{\mathbf{q},\mathbf{u},\boldsymbol{\theta}}\;&\;\min_{k \in \mathcal{K}}\;\;r_k(\mathbf{q},\mathbf{u},\boldsymbol{\theta}) \label{UAV_Pas6DMA} \\
	\mathrm{s.t.} \;
	\;&\mathbf{q} \in {\cal C}, \\
	\;&F_k(\mathbf{q},\mathbf{u}) \ge 0, \\
	\;&|\theta_{n}|= 1, n=1,2,\cdots,N. 
\end{align} 
\end{subequations}

The authors in \cite{weidong6d} proposed an AO algorithm to solve $\text{(P-PAC)}$, where the IRS's position, rotation, and reflection coefficients were alternately optimized until convergence. To avoid undesirable local optima, a Gibbs sampling method was also introduced in \cite{weidong6d} between two consecutive AO iterations for improved solution exploration.
\begin{figure}[!t]
	\centering
	\includegraphics[width=0.43\textwidth]{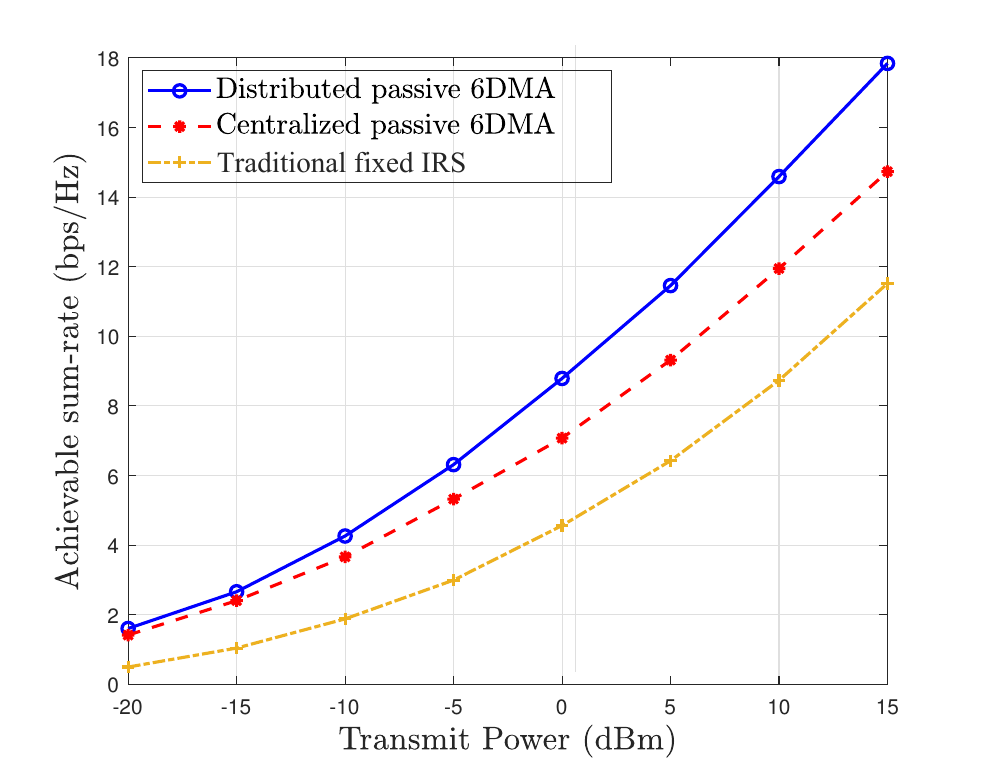}
	\caption{Sum rate versus transmit power for reflecting elements with directive radiation patterns.}
	\label{Fig2_directive}
	\vspace{-0.59cm}
\end{figure}

To evaluate the effectiveness of centralized passive 6DMA, Fig. \ref{level} illustrates the distributions of the end-to-end path gain, effective aperture gain, passive beamforming gain, and max-min user SNR (in dB) across the movement region using the Gibbs sampling-aided AO algorithm proposed in \cite{weidong6d}. In the simulation, the number of IRS elements was set to \(N=256\) and the number of users to \(K=3\). As shown in Figs. \ref{level}(a)-\ref{level}(c), the proposed AO algorithm with Gibbs sampling effectively balances the path gain, effective aperture gain, and beamforming gain among the three users, thereby ensuring the max-min SNR performance depicted in Fig. \ref{level}(d). It can be shown that the passive beamforming gain exhibits significant variations, with a maximum value of \(10 \log_{10}(N^2) = 48.2~\text{dB}\) (with \(N = 256\)) for each user, while all users achieve gains around 40~dB in Fig. \ref{level}(c). This indicates that the proposed algorithm effectively generates multiple high-gain passive beams aligned to all users.

Furthermore, to compare centralized and distributed passive 6DMA systems, Fig. \ref{Fig2_directive} presents the sum rate versus user transmit power for various schemes for reflecting elements with directive radiation patterns \cite{passive}. We consider the setup shown in Fig. \ref{Fig_main}(a)) and set \( Q = 6 \), \( K = 6 \), \( L_k = 2 \), and \( P = 6 \). It is observed that the proposed passive 6DMA schemes (either distributed or centralized) outperform the fixed IRS setup, where the performance gaps become  more substantial as the transmit power increases due to the additional DoFs enabled by position and rotation adjustments. Although the total number of reflecting elements is the same, the passive 6DMA with distributed surfaces performs better than the benchmark scheme with a centralized surface, thanks to its greater flexibility in position and rotation adjustments, which also helps mitigating inter-user interference. However, the distributed system also incurs higher hardware cost and higher complexity for controlling the movement of the individual 6DMA surfaces.

\subsection{6DMA Cell-Free Network}
\begin{figure}[t!]
	\centering
	\setlength{\abovecaptionskip}{0.cm}
	\includegraphics[width=3.6in]{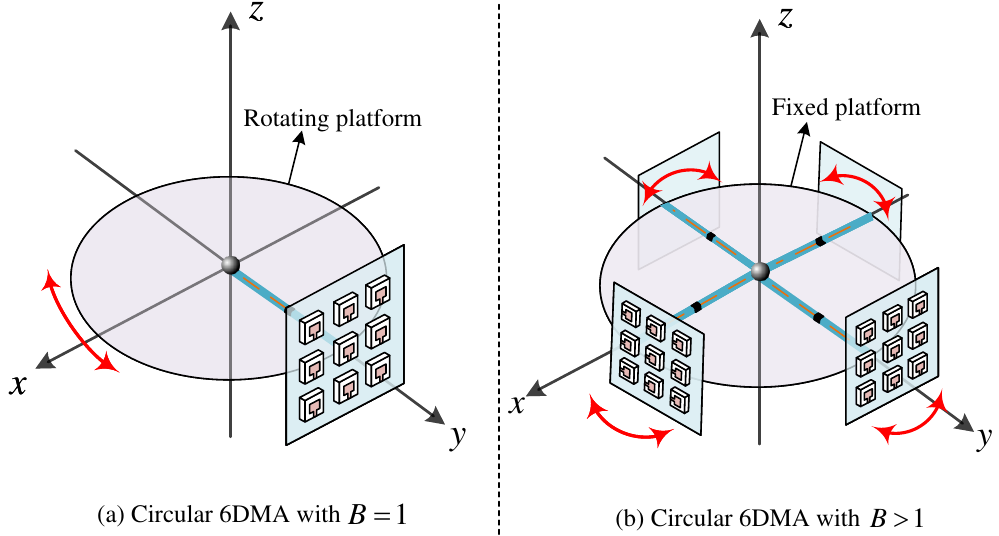}
	\caption{Circular 6DMA architecture.}
	\label{ring}
	\vspace{-0.59cm}
\end{figure}

In Sections III and IV-A, we have considered a single-cell communication system with a single BS and/or AP. To further achieve massive connectivity and seamless coverage in 6G wireless networks, cell-free networks have emerged as a promising technology \cite{encellfree,encellfree1,encellfree2}. To fully exploit the spatial variation of the wireless channel at the APs and user terminals, this subsection investigates a 6DMA-aided cell-free MIMO system, where 6DMAs are deployed at multiple APs to jointly decode signals from multiple users by dynamically adjusting antenna rotations, thereby enhancing the performance of cell-free MIMO communications.
For simplicity of implementation, this subsection examines two circular 6DMA architectures for each AP: circular 6DMA with \( B = 1 \) and circular 6DMA with \( B > 1 \) (see Fig. \ref{ring}).  

In the case of circular 6DMA with \( B = 1 \), as illustrated in Fig. \ref{ring}(a), one 6DMA surface is positioned at the edge of a rotating platform, and the entire 6DMA array rotates along a circular track. The track radius corresponds to the distance between the array center and the center of rotation. By rotating the platform, the circular 6DMA has the capability to scan the surrounding environment, enabling full-view sensing, localization, and imaging. 
In addition, this system can acquire rich spatial frequency data, which can be leveraged to enhance spatial resolution and construct an enlarged virtual array aperture. Note that both the full array's position \(\mathbf{q}\) and rotation \(\mathbf{u}\) are adjustable for circular 6DMA. This type of circular 6DMA with $B=1$ has been widely used in wireless/radar imaging  \cite{zhao2023efficient,zhang2017portable}. For example, the authors in~\cite{zhao2023efficient} proposed a rotating synthetic aperture radar (ROSAR) system for generating $360^\circ$ images. The idea was further employed in~\cite{zhang2017portable} for realizing 3D imaging. 
Specifically, one important performance metric for imaging is image entropy, where a lower image entropy indicates a better quality-of-image (QoI)~\cite{jiang2017fast}. For circular 6DMA, the synthetic aperture of the array is fixed; thus $\mathbf{u}$ is a key factor for the QoI that determines the integral side-lobe ratio (ISLR) and impulse response width (IRW). 
Therefore, the choice of $\mathbf{u}$ should ensure that ISLR is below a specified threshold for satisfactory focusing performance, while IRW is sufficiently narrow to achieve higher spatial resolution. Additionally, the angular velocity of rotation is another crucial factor, which has an impact on the azimuth and elevation resolution~\cite{zhang2017portable}.
\begin{figure}[t]
	\centering
	\includegraphics[width=3.1in]{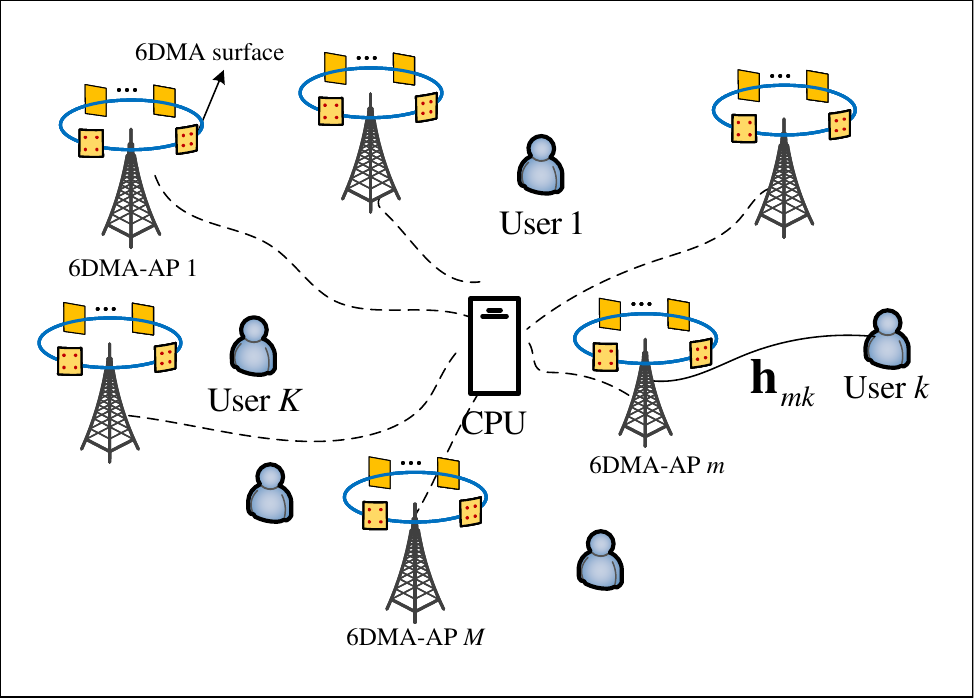}
	\caption{6DMA-aided cell-free network \cite{free6DMA}.} 
	\label{cellfree}
	\vspace{-0.59cm}
\end{figure}

However, a circular 6DMA with \( B = 1 \) cannot freely allocate antenna resources based on the user distribution. Therefore, a circular 6DMA architecture with \( B > 1 \) was proposed in \cite{6dmaPar}, where all 6DMA surfaces at each AP can move freely along a fixed circular track (see Fig. \ref{ring}(b)). By employing circular 6DMAs with \( B > 1 \), we further propose a 6DMA-aided cell-free network in \cite{free6DMA}, as illustrated in Fig. \ref{cellfree}. Specifically, the cell-free network consists of \( K \) users, each equipped with a single isotropic FPA, and \( M_{\mathrm{a}} \) 6DMA-enabled APs, represented by the set \( \mathcal{M}_{\mathrm{a}} = \{1, 2, \dots, M_{\mathrm{a}}\} \). Each 6DMA-AP comprises \( B \) 6DMA surfaces, with each surface containing \( N = N_h \times N_v \) antennas, where \( N_h \) and \( N_v \) represent the number of antennas in the horizontal and vertical directions, respectively.
The rotation angle of 6DMA surface $b$ at AP $m$ is defined as $\gamma_{mb} \in [0,2\pi)$. 
The rotation angle vector is denoted as \({\boldsymbol \gamma }_m = \left[\gamma_{m1}, \dots, \gamma_{mB}\right]^T\). We denote $\phi_{mk}$ and $\theta_{mk}$ as the azimuth and elevation angles, respectively, of the signal from user $k$ to the reference position of AP, $m$.
Let $\mathbf{q}_{mb}$ denotes the position and rotation of the center of 6DMA surface $b$ at AP $m$ in its local CCS.
In this case, the position of any 6DMA surface, $\mathbf{q}_{mb}$, can be uniquely represented by rotation angle $\gamma_{mb}$, i.e., 
\begin{align}
	\mathbf{q}_{mb}(\gamma_{mb}) = 
	\begin{bmatrix}
		\varsigma  \cos\gamma_{mb}, & \varsigma  \sin\gamma_{mb}, & 0
	\end{bmatrix}.
\end{align}
where $\varsigma $ represents the radius of the circular ring.  Thus, although both the position and rotation of the 6DMA surfaces change in the circular 6DMA architecture, its position is determined solely by the rotation angle. This significantly simplifies the system model and reduces the complexity of position/rotation design.

To maximize the average achievable sum-rate of the 6DMA-aided cell-free network, denoted as \( R_{\mathrm{avg}}({\boldsymbol \gamma}) \), we formulate the following optimization problem by jointly optimizing the rotation angles of all 6DMA surfaces at \( M_{\mathrm{a}} \) APs:  
\begin{subequations}
\begin{align}
	\text{(P-FR)}:~&	\underset{{  {\boldsymbol \gamma }}}{\text{max}}~R_{\mathrm{avg}}({\boldsymbol \gamma})  \label{Problem}\\
	\text {s.t.}&~ {{\gamma _{m(b+1)}} \!-\! {\gamma _{mb}}}  \geqslant  \varepsilon ,\forall m \in \mathcal{M}_{\mathrm{a}},b \in \mathcal{B}, \label{Problema}\\
	&~{{\gamma _{m1}} + 2\pi - {\gamma _{mB}}}  \geqslant  \varepsilon ,\forall m \in \mathcal{M}_{\mathrm{a}}, \label{Problemb}
\end{align}
\end{subequations}
where ${\boldsymbol \gamma}=\left[{\boldsymbol \gamma}_1^T,\cdots,{\boldsymbol \gamma}_{M_{\mathrm{a}}}^T\right]^T\in\mathbb{R}^{BM_{\mathrm{a}}\times1}$ denotes the collection of all rotation angle vectors.
Constraints (\ref{Problema}) and (\ref{Problemb}) prevent overlapping and coupling among 6DMA surfaces at the same AP by enforcing a minimum rotation angle difference, denoted as \( \varepsilon \). Compared to problem \((\text{P1-CO})\), problem \((\text{P-FR})\) is easier to solve since the optimization variables include only angular information and not position information. To handle this problem, an efficient Bayesian optimization-based algorithm was proposed in \cite{free6DMA}, which iteratively refines a Gaussian process-based surrogate model and selects new observation points using an expected improvement-based acquisition function.
\begin{figure}[t]
	\centering
	\includegraphics[width=3.5in]{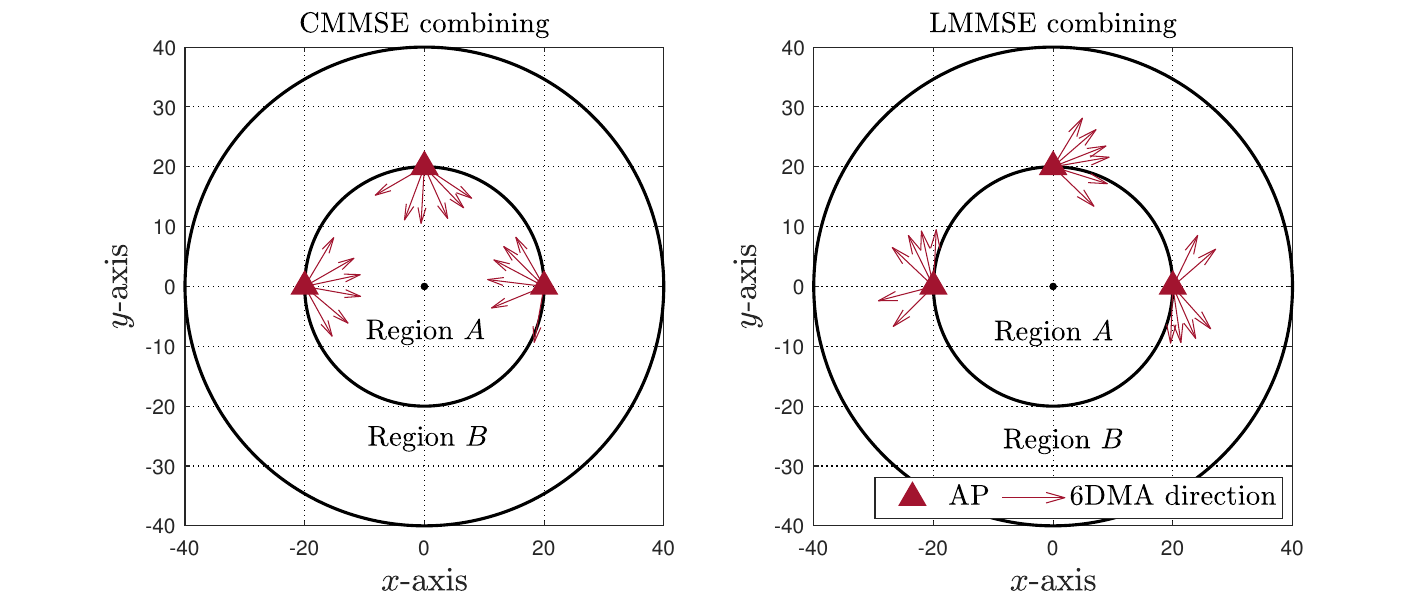}
	\caption{Optimization of 6DMA surface rotations in a cell-free system with CMMSE and LMMSE combining.} 
	\label{result1}
	\vspace{-0.59cm}
\end{figure}

\begin{figure}[t]
	\centering
	\includegraphics[width=2.7in]{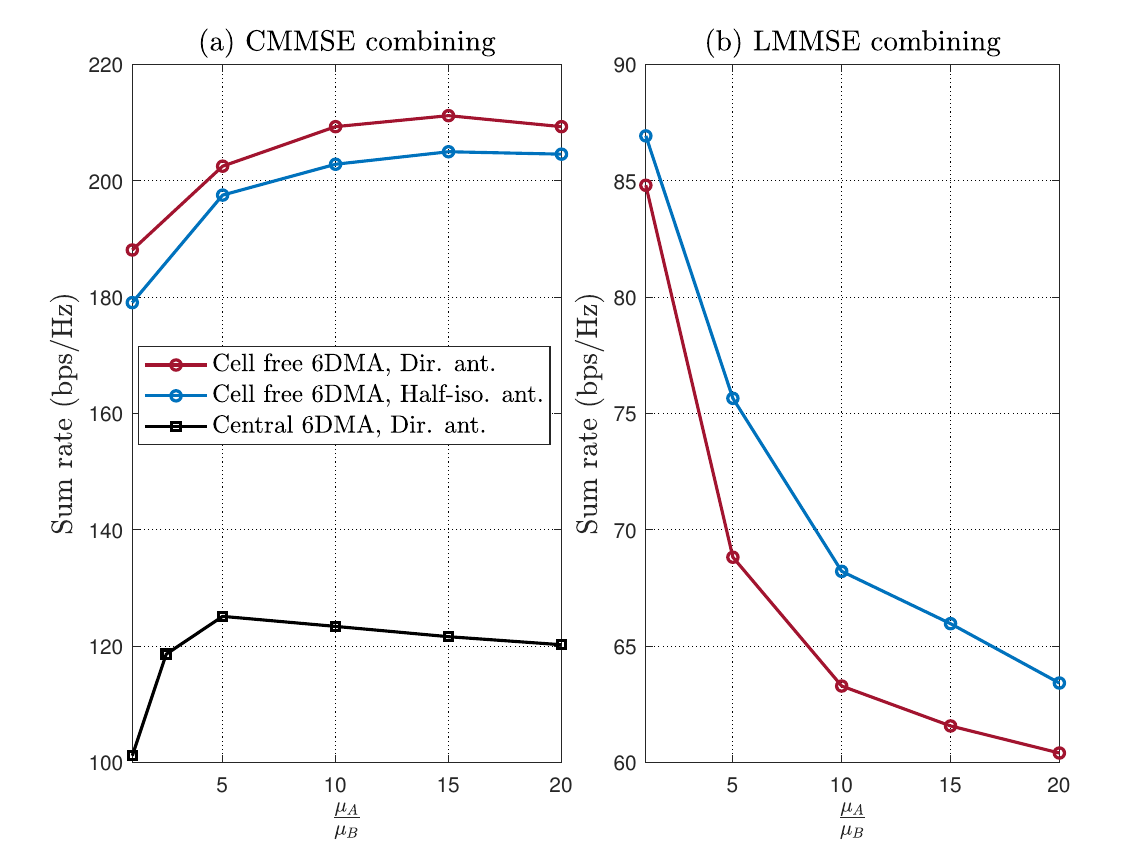}
	\vspace{-4 mm}
	\caption{Average sum-rate versus user density ratio for different combining methods \cite{free6DMA}.} 
	\label{result2}
	\vspace{-0.59cm}
\end{figure}

We assume that each AP employs the minimum mean-square error (MMSE) combining vectors using either global or local CSI, referred to as GMMSE and LMMSE combining methods, respectively. Fig. \ref{result1} shows the optimal rotations of circular 6DMA surfaces for CMMSE and LMMSE combining, with a user density ratio of coverage region $A$ to region $B$ given by $\frac{\mu_A}{\mu_B}=5$, where \( \mu_A \) and \( \mu_B \) represent the user densities in regions A and B, respectively. We see that, with the CMMSE method, all 6DMA surfaces rotate toward the high-density user region. In contrast, with the LMMSE method, the surfaces rotate toward the low-density region due to the absence of global CSI, which causes significant interference among APs. In addition, the 6DMA surfaces exhibit a non-uniform configuration based on the heterogeneous user distributions.
Fig. \ref{result2} shows the average sum-rate versus the user density ratio across different schemes. We observe that the proposed 6DMA cell-free scheme with either directional or half-space isotropic antennas outperforms centralized 6DMA with a single circular 6DMA-AP equipped with the same number of antennas as the first two schemes. This is because the adaptability of 6DMA cell-free scheme enhances the SINR for each user and achieves higher spatial multiplexing gains compared to centralized 6DMA. Furthermore, with LMMSE combining, 6DMA with half-space isotropic antennas outperforms its directional counterpart due to the broader angular coverage, which reduces channel correlation and enhances local CSI-based interference suppression (see Fig. \ref{result2} (b)).

\subsection{6DMA for Wireless Sensing}
Next-generation wireless networks require not only high-quality communication but also exceptional sensing accuracy. However, in practical scenarios with complex radio propagation, sensing performance is constrained, thereby limiting the trade-off between sensing and communication capabilities in ISAC systems.

\begin{figure}[t!]
	\centering
	\setlength{\abovecaptionskip}{0.cm}
	\includegraphics[width=2.7in]{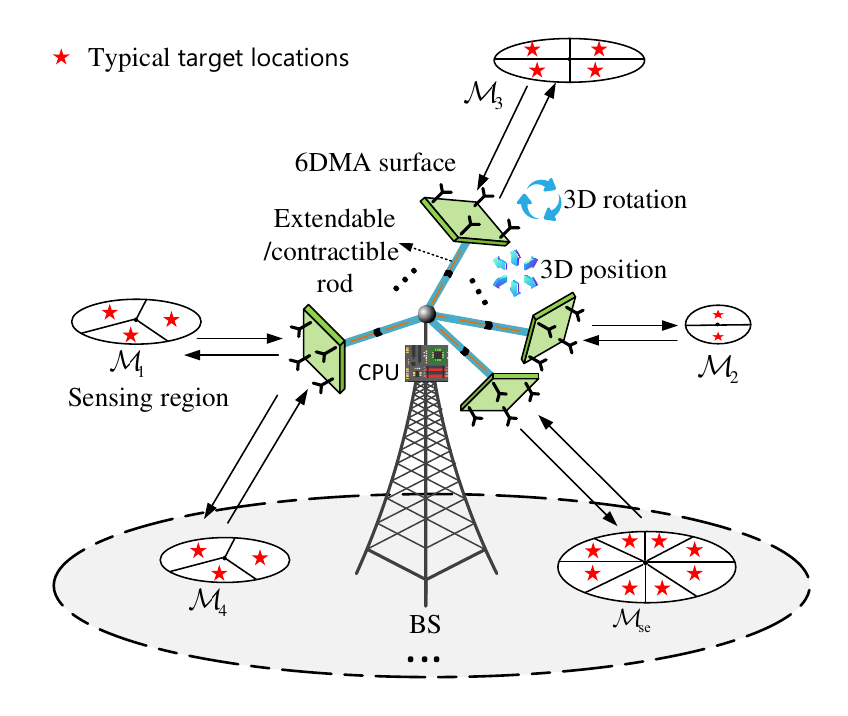}
	\caption{6DMA-enabled BS adaptable to different sensing regions.}
	\label{sensingsys}
	\vspace{-0.65cm}
\end{figure}

The 6DMA technology can potentially offer additional DoFs for enhancing the effectiveness of wireless sensing \cite{6dmaSens}. As shown in Fig. \ref{sensingsys}, a 6DMA-enhanced sensing network was studied in \cite{6dmaSens}, where a 6DMA-BS senses multiple targets distributed across \( M_{\mathrm{s}} \) 3D spatial regions, denoted as \( \mathcal{M}_m, m=1,2,\dots,M_{\mathrm{s}} \). To optimize the 6DMA’s position and orientation, each region \( \mathcal{M}_m \) is divided into \( K_m \) uniform subregions of size \( \Delta \), which is proportional to the region's area. The central position of each subregion serves as its typical target location under the assumption of LoS channels from the BS to these targets.  
Let \(\phi_k \in [-\pi,\pi]\) represent the horizontal DOA from the \( k \)-th typical target to the BS’s reference position, where \( k\in\{1,2,\dots,\tilde{K}\} \) with \( \tilde{K}=\sum_{m=1}^{M_{\mathrm{s}}} K_m \). The CRB for estimating \( \boldsymbol{\phi}=[\phi_1,\phi_2,\dots,\phi_K]^T \) is given by 
\begin{align}\label{CRB}
	&\mathrm{CRB}\left(\mathbf{q},\mathbf{u},  \boldsymbol{\phi}\right)=\nonumber\\
	&\sum_{k=1}^{\tilde{K}}\frac{\sigma^2}{2|\rho_k|^2L}
	\left[\left\|\mathbf{h}_k(\mathbf{q},
	\mathbf{u},{\phi_k})
	^H\mathbf{X}\right\|_2^2\left\|\dot{\mathbf{h}}_k(\mathbf{q},\mathbf{u},\phi_k)\right\|_2\right]^{-1}\!,
\end{align}
where $\mathbf{h}_k(\mathbf{q},\mathbf{u},\phi_k)$ represents the channel between user \( k \) and the BS as described by the basic 6DMA channel model in \eqref{uk}, $\dot{\mathbf{h}}_k(\mathbf{q},\mathbf{u},\phi_k)=\frac{\partial\mathbf{h}_k(\mathbf{q},\mathbf{u},\phi_k)}{\partial \phi_k}$, $\phi_k$ denotes the channel coefficient, $\mathbf{X}$ denotes sensing signals, and $L$ represents the duration of the sensing frame. From \eqref{CRB}, it is evident that the improvement in CRB stems from two primary factors, namely the \emph{power gain}, \( \left\|\mathbf{h}_k(\mathbf{q}, \mathbf{u},{\phi}_k)^H\mathbf{X}\right\|_2^2 \) and the \emph{geometric gain}, \( \left\| \dot{\mathbf{h}}_k(\mathbf{q},\mathbf{u},{\phi}_k)\right\|_2 \) \cite{6dmaSens}. Moreover, the CRB for estimating the DOA of each target relies heavily on the positions and rotations of the 6DMA surfaces. Thus, to minimize the CRB in \eqref{CRB} for all target locations, we jointly optimize the 3D positions \( \mathbf{q} \) and rotations \( \mathbf{u} \) of the 6DMA surfaces at the BS. The corresponding problem is formulated as follows
\begin{subequations}
	\label{MGQ3}
	\begin{align}
		\text{(P-SE)}:~~&~\min\limits_{\mathbf{q},\mathbf{u}}
		\mathrm{CRB}\left(\mathbf{q},\mathbf{u},  \boldsymbol{\phi}\right) \\
		\text {s.t.}~&\eqref{M1}, \eqref{M2}, \eqref{M3}, \eqref{jM3}.
	\end{align}
\end{subequations}

The non-convex problem \(\text{(P-SE)}\) can be solved using particle swarm optimization (PSO) \cite{6dmaSens}. It is worth noting that the optimal \( \mathbf{q} \) and \( \mathbf{u} \) in \( \text{(P1-SE)} \) depend on the typical DOAs \( \boldsymbol{\phi} \) in all subregions. The actual target locations may deviate slightly from these typical DOAs, but such deviations become negligible if the subregions are sufficiently small.

\begin{figure}[t!]
	\centering
	\setlength{\abovecaptionskip}{0.cm}
	\includegraphics[width=3.0in]{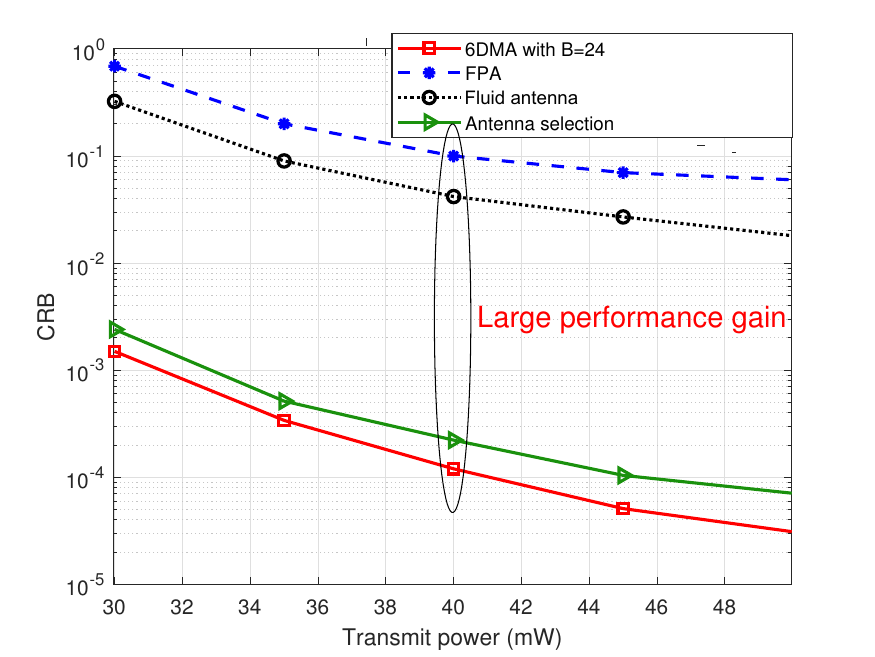}
	\caption{CRB versus transmit power in scenario with directive antennas \cite{6dmaSens}.}
	\label{direc3}
	\vspace{-0.59cm}
\end{figure}

To evaluate the effectiveness of antenna positioning and rotation for improving sensing performance, 
Fig. \ref{direc3} shows the CRB versus the transmit power for a directive antenna pattern. We set \( N = 2 \) and \( B = 24 \). We compare the 6DMA sensing scheme with FPA, antenna selection, and fluid antenna schemes. In the antenna selection approach, 128 uniformly distributed points on a spherical surface at the BS are assigned \(N=2\) antennas each. Then, 24 surfaces are selected through exhaustive search to minimize the CRB. The superior performance of 6DMA compared to benchmark schemes is evident. This is due to the greater spatial DoFs of the 6DMA system, which enables more efficient allocation of antenna resources to subregions with higher target densities. For example, when the distances between the targets and the BS are relatively uniform, the 6DMA system allocates a greater number of antennas to regions with a higher density of subregions or targets, thereby enhancing sensing performance in those areas. In contrast, the fluid antenna scheme is limited to adjusting antenna positions within each 2D surface, while the antenna selection scheme requires a larger number of physical antennas compared to the 6DMA system. 
Furthermore, the adaptability of 6DMA-aided wireless sensing proves particularly valuable for ISAC systems, where sensing accuracy and communication quality are closely interconnected. The additional DoFs provided by 6DMA make it a powerful enabler for achieving both objectives, paving the way for high-quality ISAC in future wireless networks.

\subsection{6DMA for UAV}
\begin{figure}[!t]
	\centering
	\includegraphics[width=0.43\textwidth]{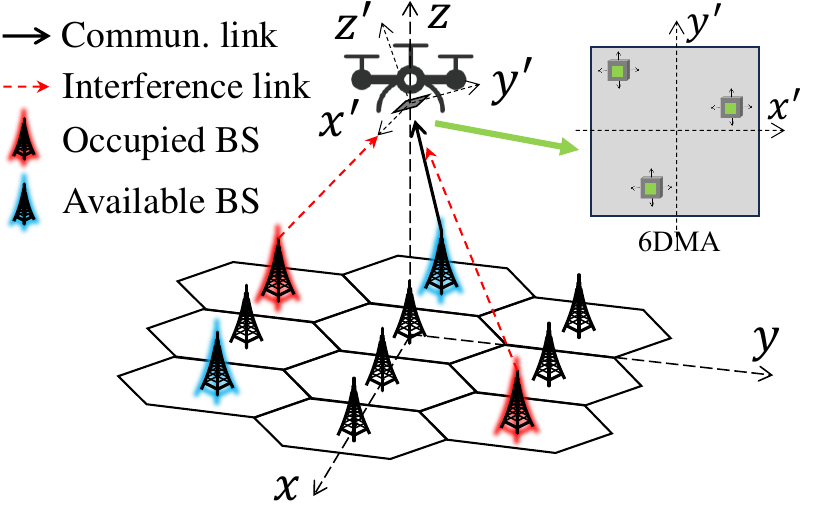}
	\caption{Aerial-ground interference mitigation via UAV-mounted 6DMAs.}
	\label{CellularUAV6DMA}
	\vspace{-0.59cm}
\end{figure}

As depicted in Fig.~\ref{CellularUAV6DMA}, 6DMA may be utilized to resolve the aerial-ground interference issue for cellular-connected UAVs \cite{cellular2019mei,mei2021aerial,ren20246D}. Specifically, due to the more favorable aerial-ground channels compared to ground channels, cellular-connected UAV users may suffer/impose more severe interference from/to co-channel terrestrial BSs in downlink/uplink transmission. By equipping each cellular-connected UAV with a 6DMA surface and carefully designing its orientation along with the positions of all antennas on the surface, the aerial-ground interference could be significantly mitigated.

In particular, the authors in \cite{ren20246D} focused on a downlink 6DMA-aided UAV system with multiple BSs operating within a given interference zone. For a specific resource block (RB) in the cellular network under consideration, each BS can serve only one user per RB. Let \( \breve{K} \) represent the number of available BSs (i.e., those not serving any users or UAVs) within this RB, with the set of all available BSs defined as \(\mathcal{\breve{K}} = \{1, 2, \dots, \breve{K}\}\). Similarly, let \( J \) denote the number of interfering BSs (i.e., those currently serving a user or UAV) in the same RB, with the set of all interfering BSs defined as \(\mathcal{J} = \{1, 2, \dots, J\}\). 
Assume that the system consists of a single 6DMA surface (\(B=1\)), which is equipped with \(N\) antennas, where \(\tilde{\mathbf{q}}_n = [x_n, y_n]^T\) denotes the position of the \(n\)-th antenna on the 6DMA surface. Consequently, the antenna position vector is expressed as \({\mathbf{q}} = [\tilde{\mathbf{q}}_1^T, \tilde{\mathbf{q}}_2^T, \cdots, \tilde{\mathbf{q}}_N^T]^T \in \mathbb{R}^{2N}\). Let \(\mathbf{u}\) represent the rotation of the entire antenna array. Suppose the \(k\)-th available BS is chosen as the BS  associated to the UAV. In this case, the received SINR at the UAV is given by
\begin{equation}\label{UAV-SINR}
	r_k = \frac{\lvert\mathbf{w}^H \mathbf{h}_k(\mathbf{q},\mathbf{u})\rvert^2 P}{\sum\limits_{j\in\mathcal{J}}\lvert\mathbf{w}^H\mathbf{h}_j(\mathbf{q},\mathbf{u})\rvert^2 P + \|\mathbf{w}\|^2\sigma^2 }, \; k\in\mathcal{\breve{K}},
\end{equation}
where $P$, $\sigma^2$, $\mathbf{w}$, and $\mathbf{h}_j(\mathbf{q},\mathbf{u})$ denote the transmit power of each BS, the noise power, the receive beamforming vector at the UAV, and the channel from the $j$-th BS to the UAV as defined by the basic 6DMA channel model in \eqref{uk}, respectively.

To mitigate the strong co-channel interference present in (\ref{UAV-SINR}), we first select a BS \(k\) from the set of \(K\) BSs for the UAV such that the SINR between the UAV and the chosen BS is maximized. Next, we optimize the position and rotation of the 6DMA to maximize \(r_k\). The resulting problem is formulated as follows
\begin{subequations}
\begin{align}
	\text{(P-UAV)}:&\max_{\mathbf{w}, \mathbf{q}, \mathbf{u}} \max_{k \in \mathcal{\breve{K}}}\;  \;\;r_k \label{cellularUAV}\\
	\mathrm{s.t.}&\; -\bar{L} \le x_n \le \bar{L}, \forall n, \\
	& \;-\bar{L} \le y_n \le \bar{L}, \forall n, \\
	& \;\| \tilde{\mathbf{q}}_m- \tilde{\mathbf{q}}_n\| \ge d_{\min}, \forall m, n, m \ne n, 
\end{align}
\end{subequations}
	where $\bar{L}$ and $d_{\min}$ denote the side length of the square 6DMA surface and the minimum distance between adjacent antennas to avoid mutual coupling. Problem $\text{(P-UAV)}$ turns out to be a highly non-convex problem that is difficult to solve. To tackle this challenge, the authors in \cite{ren20246D} first derived the optimal receive beamforming vector $\mathbf{w}$ in closed-form for given position and rotation of the 6DMA, and given the BS-UAV association. Then, the position and rotation of the 6DMA were alternately optimized based on a block coordinate decent (BCD) algorithm, with the index of the optimal associated BS, $k$, optimized lastly via enumeration. It was shown in \cite{ren20246D} that incorporating 6DMA can dramatically mitigate the co-channel interference in (\ref{UAV-SINR}) compared to schemes without 6DMA. Particularly, 6DMA can avoid the need for sophisticated symbol-level cooperation among BSs, and thus lead to an efficient implementation in practice.

\subsection{Other Related Works and Future Directions}
In addition to the aforementioned applications, 6DMA have also other promising  applications in the field of wireless communication/sensing \cite{zr1,zr2}. For example, secure sensing and secure communication are essential for future wireless systems, particularly in addressing vulnerabilities to eavesdropping and interference. Traditional FPA systems lack the flexibility to fully exploit the spatial DoFs, which limits their secure transmission and secure sensing performance \cite{shao2023enhancing}. In contrast, 6DMA enables dynamic reconfiguration of spatial channels, thereby enhancing secrecy rates for legitimate receivers while reducing interception risks from eavesdroppers. 
In the context of secure sensing, 6DMA offers significant advantages by dynamically adjusting beam patterns to enhance secure detection accuracy while reducing the ability of unauthorized radars to detect the same targets \cite{spoof}. 
In addition, to support massive communications in 6G, adopting next-generation multiple access (NGMA) methods is crucial. The interplay between 6DMA and existing access techniques, such as orthogonal multiple access (OMA), non-orthogonal multiple access (NOMA), rate-splitting multiple access (RSMA), and grant-free random access \cite{jadc2,jadc3,jadc4,jadc5}, provides valuable opportunities. For example, 6DMA-enabled NGMA can efficiently allocate resources across spatial angles, time slots, frequency bands, spreading codes, and power levels, for optimizing connectivity, spectral efficiency, energy efficiency, and latency. Furthermore, THz communications have emerged as a promising approach for 6G, complementing the role of mmWave in 5G \cite{mmw2, mmw1}. While THz offers advantages such as larger bandwidth, narrower beamwidth, and high directivity, challenges like higher propagation loss, LoS blockage, and rank-deficient channels remain prevalent\cite{ning2023beamforming}. Fortunately, incorporating 6DMA can address these challenges by dynamically reconfiguring antenna positions/rotations and radiating elements. For example, at mmWave/THz BSs, 6DMA can adapt the inter-antenna spacing to match user distributions and enhance near-field propagation for multi-stream transmission under LoS conditions.
\section{Rotatable 6DMA}

For wireless systems requiring only modest communication capacity and performance enhancement requirements, or subjected to hardware limitations that prevent antenna translation (i.e., position change), the easier-to-implement antenna rotation is a more practically viable solution than adjusting both antenna position and orientation. As such, there is a need to investigate further the unique benefits as well as design challenges for wireless systems aided by antenna rotation. In this section, we thus focus on efficient designs for 6DMA with flexible antenna rotation only, for which the radiation pattern, polarization, and array aperture can be dynamically controlled via mechanically or electronically tuned antenna rotation. In particular, by fixing the antenna position $ \mathbf{q}_b$ in the general 6DMA system model \eqref{bb1}, we obtain:
\begin{align}
	\!\!\! \cancel{\mathbf{q}_b} = 
	\begin{bmatrix}
		\cancel{x_b}, \!& \cancel{y_b}, & \!	\cancel{z_b}
	\end{bmatrix}^T,~\mathbf{u}_b = 
	\begin{bmatrix}
		{\alpha_b}, \!& {\beta_b},\! & \gamma_b
	\end{bmatrix}^T, b\in\mathcal{B}.
\end{align}

This leads to a 6DMA configuration in which the antenna can rotate flexibly while its position remains fixed. This configuration is known as the rotatable 6DMA (or rotatable/rotary antenna), which is a special case of 6DMA. Rotatable 6DMAs have the capability to reshape the
radiation pattern in the angular domain for achieving flexible beam coverage by adjusting the rotation angles of antennas.

Specifically, depending on the rotation unit and rotation mode, two types of rotatable 6DMA (R-6DMA) architectures are introduced, namely, element-wise rotatable 6DMA and array-wise rotatable 6DMA, specified as follows. 
\begin{figure}[t]
	\centering
	\includegraphics[width=0.51\textwidth]{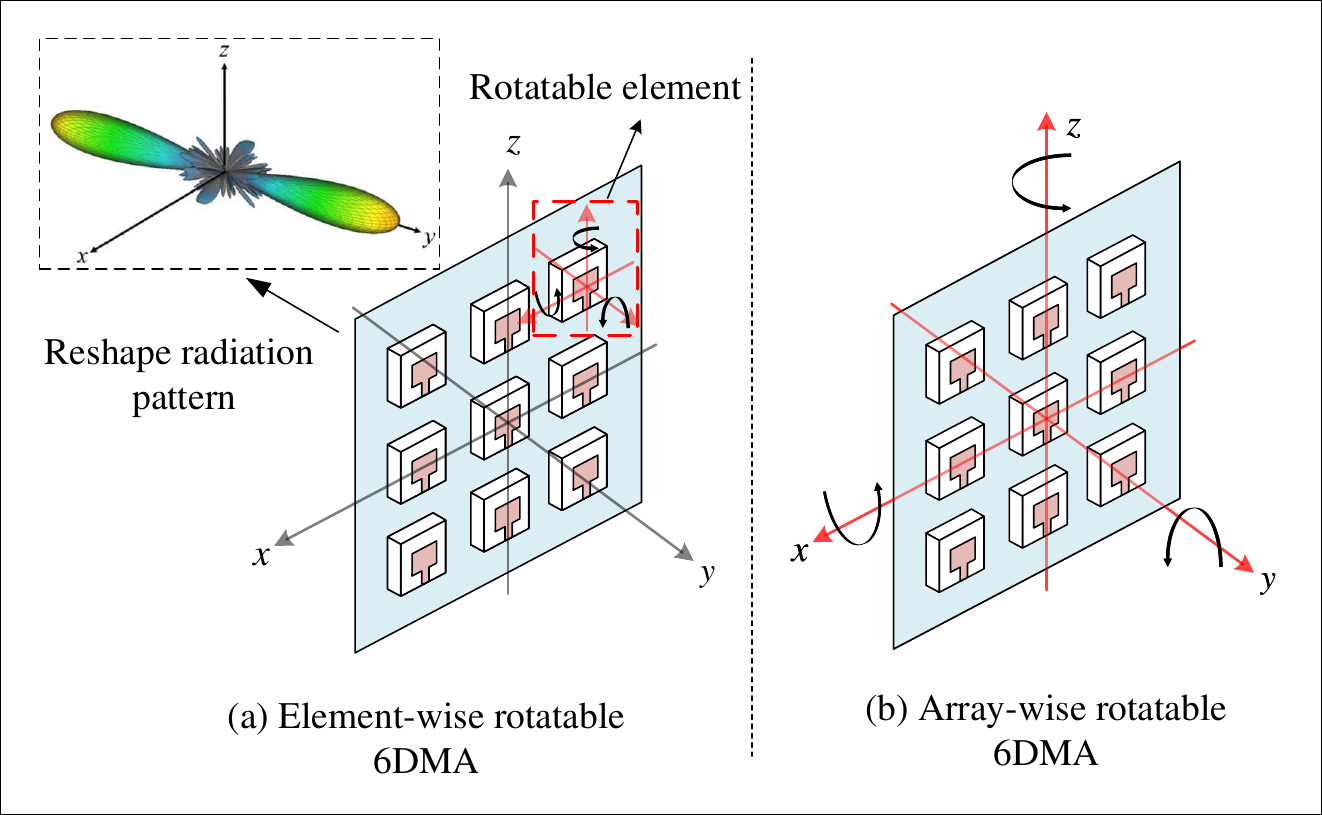}
	\caption{Two types of rotatable 6DMA architectures.} 
	\label{Fig:R-6DMA_model}
	\vspace{-0.59cm}
\end{figure}
\begin{itemize}
	\item \textbf{Element-wise rotatable 6DMA}: For element-wise R-6DMAs illustrated in Fig.~\ref{Fig:R-6DMA_model}(a), each antenna element of the array can rotate its orientation by adjusting its deflection angles~\cite{smolders2014low}, while its position does not change over time. Element-wise rotatable 6DMA offers a convenient and effective approach to adjust the radiation pattern characteristics of antenna arrays without raising array costs or decreasing aperture efficiency \cite{3dp}. Moreover, from Remark 2, we know that the polarization properties of arrays depend not only on the electrical characteristics of the antenna elements but also on the rotation of each individual element. As such, the polarization characteristics and radiation pattern can be adjusted/reshaped via individual element rotation for improving communication and sensing performance~\cite{kuhlmann2023sequential}. Additionally, wide beam-scanning in the angular domain can also be realized by employing element-wise R-6DMAs without dedicated phase control for beamforming~\cite{huang1998ka}.

	\item \textbf{Array-wise rotatable 6DMA}: For array-wise R-6DMAs shown in Fig.~\ref{Fig:R-6DMA_model}(b), the entire array can be flexibly rotated around its central reference point. For this architecture, the antenna radiation characteristics and beam coverage can also be dynamically controlled~\cite{rachang, gagnon2013using,rot}. Moreover, the array rotation can be adjusted to effectively suppress the interferences from undesired directions~\cite{sun2022anti}, hence enabling efficient spectrum sharing in cognitive radio systems~\cite{tawk2011implementation}. 
\end{itemize}
In the following, for the above two types of R-6DMAs, we introduce corresponding rotation models, design issues, and existing solutions, respectively. 

\subsection{Array-wise Rotatable 6DMA}
6DMA reduces to array-wise  R-6DMA, when the center position of this surface is fixed at $\mathbf{q}=[0,0,0]^T$, while the array itself rotates by a certain angle. Consider an antenna surface equipped with $N$ antenna elements with half-wavelength spacing. As shown in \eqref{nwqr}, the position of the \(n\)-th antenna is given by \(\mathbf{r}_{n}(\mathbf{u}^{(2)} )  = \mathbf{R}(\mathbf{u}^{(2)}) \bar{\mathbf{r}}_{n}, \forall n \in \mathcal{N}\), where \(\mathbf{u}^{(2)}=\mathbf{u}_1\), as defined in \eqref{bb1} for \(B = 1\), representing the rotation vector for the entire array. Subsequently, the channel vector between the user and the array-wise R-6DMA, denoted by \(\mathbf{h}_{k}(\mathbf{u}^{(2)})\), can be obtained from \eqref{ukr}.

Note that the array-wise R-6DMA generally has 3D spatial DoFs in rotation, which can be leveraged to enhance both communication and sensing performance. Specifically, the R-6DMA can dynamically adjust its radiation pattern through rotational motion to enable various communication functions, such as mitigating inter-user interference~\cite{sun2022anti,suzuki2020rotating}, enhancing network coverage~\cite{li2024antenna}, and increasing multiplexing gain~\cite{tawk2011implementation,wang2024power}. The generic communication optimization problem for array-wise  R-6DMA can be formulated as follows
\begin{subequations}
	\begin{align}
		\text{(P1-AW)} ~\max_{\mathbf{u}^{(2)}}&~U(\mathbf{u}^{(2)}) \nonumber \\
		\text{s.t.}~& \mathbf{u}^{(2)}  \in \mathcal{C}^{(2)},\label{R-6DMA_region1} \\
		& g_{i}(\mathbf{u}^{(2)}) \le 0, i=\{1,2,\ldots,I\},
	\end{align}
\end{subequations}
where $U(\mathbf{u}^{(2)})$ is the primary objective function  which is determined by the array rotation, $\mathcal{C}^{(2)}$ specifies the allowable rotation angle range for the entire array, and $g_{i}(\mathbf{u}^{(2)})$ represents other constraints related to communication performance.  For example,  consider the scenario, where array-wise R-6DMA is utilized to  manage interference levels for spectrum sharing in cognitive radio systems~\cite{tawk2011implementation}. The corresponding objective function can be expressed as
$U(\mathbf{u}^{(2)}) = R_{\rm s}(\mathbf{u}^{(2)})$,
where $R_{\rm s}\left(\mathbf{u}^{(2)}\right)$ represents the achievable rate of the secondary transmission. In addition, the communication performance of the primary transmission must be preserved, which can be ensured by imposing the constraint \( g_{i}(\mathbf{u}^{(2)}) \leq 0 \). This constraint guarantees that, by controlling the array rotation, the interference at the primary transmission caused by the secondary transmission, remains below a specified threshold. Moreover, the array rotation range should be confined in a finite region. For example, let $\alpha$, $\beta$, and $\gamma$ denote the 3D array rotation angles. Then, the array rotation variables should satisfy 
\begin{align}\label{P2-1_c12}
	\!\!\!\!\!\!	\alpha_{\min}  \!\le\! \alpha \! \le\! \alpha_{\max},~
	\beta_{\min} \! \le \! \beta \! \le \!\beta_{\max}, ~
	\gamma_{\min}  \!\le\! \gamma \! \le \!\gamma_{\max}. 
\end{align}

For problem \(\text{(P1-AW)}\), in the 1D array rotation case, exhaustive search  can be used to efficiently determine the optimal solution \cite{wang2024power}. However, for 3D array rotation, the computational complexity of 3D exhaustive search is demanding, and becomes even  prohibitive when the 3D rotation needs to be jointly optimized with other optimization variables such as beamforming and resource allocation~\cite{6dmaCon}. AO  algorithms and SCA method can be employed to address this issue efficiently. In \cite{6dmaCon},  an efficient joint conditional gradient and SCA techniques was proposed, where SCA was utilized to transform the non-convex parts of the problem into convex ones, and the conditional gradient technique was then employed to optimize the rotation vector.

To optimize sensing performance with array-wise R-6DMA, the authors of \cite{baig2019high} proposed an efficient algorithm for target detection and parameter estimation, in which the Doppler shift resulting from the array rotation was effectively compensated. Recently, the authors in \cite{rachang} proposed designing omnidirectional rotatable 6DMAs at the BS to enhance both communication and sensing performance by exploiting the DoF offered by array rotation. Specifically, this work considers an ISAC system where the BS employs a rotatable 6DMA array comprising \(M_t\) transmit antennas and \(M_r\) receive antennas. The BS serves \(K\) single-antenna users in the downlink communication while simultaneously detecting a target through echo signals in the monostatic mode.

Based on the rotatable 6DMA channel model \eqref{rago}, we define \(\tilde{\gamma} = \bar{\gamma} + \gamma\) as the effective spatial angle of the target by accounting for the BS array rotation, where \(\bar{\gamma}\) is the target's reference spatial angle and \(\gamma\) is the rotation angle of the antenna array (see Fig. \ref{rago}).
Then, the CRB for estimating the target angle \(\tilde{\gamma}\) is given by
\begin{align}
	\text{CRB}(\tilde{\gamma}) 
	= \frac{\chi}{\cos^2 \tilde{\gamma}} 
	\bigl(\tilde{\mathbf{a}}_{T}(\tilde{\gamma})\bigr)^H 
	\,\mathbf{W}\,\mathbf{a}_{T}(\tilde{\gamma}),
	\label{racrb}
\end{align}
where $\mathbf{a}_{T}(\tilde{\gamma})$ is transmit steering vector defined in \eqref{sra},
$\chi = \frac{3 \lambda^2 (M_r - 1)}{2 \pi^2 T \,\text{SNR}\, M_r (M_r+1) ((M_r - 1)d)^2}$, and
\(
\mathbf{W} = \sum_{k=1}^K \mathbf{w}_k \mathbf{w}_k^H
\) with $\mathbf{w}_k \in \mathbb{C}^{M_t \times 1}$ being the transmit beamforming vector for user $k$.
From \eqref{racrb}, minimizing the angle-estimation CRB in \eqref{racrb} is equivalent to maximizing
\begin{align}
	f_s\bigl(\{\mathbf{w}_k\}, \phi\bigr) 
	= \cos^2 \tilde{\gamma}\, \mathbf{a}_T^H\bigl(\tilde{\gamma}\bigr)\,\mathbf{W}\,\mathbf{a}_T\bigl(\tilde{\gamma}\bigr),
	\label{eq:fs}
\end{align}
which depends on both the transmit beamforming vectors \(\{\mathbf{w}_k\}\) and the rotation angle \(\phi\). For communication users, we use the achievable sum-rate as the performance metric $
	f_c\bigl(\{\mathbf{w}_k\}, \phi\bigr) 
	= \sum_{k=1}^K \log_2 \bigl(1 + r_k\bigr)$, where \(r_k\) is the SINR of user \(k\).

To jointly enhance communication and sensing performance, our objective is to maximize the sum-rate of the communication users while minimizing the CRB for target angle estimation. This joint optimization problem is expressed as 
\begin{subequations}
	\begin{align}
		\text{(P-ORA)}:\quad 
		&\max_{\{\mathbf{w}_k\}, \phi}\quad 
		\omega_1\, f_c\bigl(\{\mathbf{w}_k\}\bigr) 
		\;+\; \omega_2\, f_s\bigl(\{\mathbf{w}_k\}, \phi\bigr)
		\nonumber\\[1mm]
		\mathrm{s.t.}\quad 
		& \sum_{k=1}^K \|\mathbf{w}_k\|_2^2 \;\le\; P_{\max}, 
		\quad \label{P1_cons1}\\[1mm]
		& \phi \;\in\; \bigl[\gamma_{\min},\, \gamma_{\max}\bigr],
		\label{P1_cons2}
	\end{align}
\end{subequations}
where \(\omega_1\) and \(\omega_2\) are weighting coefficients for the communication performance metric \(f_c\) and the sensing performance metric \(f_s\), respectively, with \(\omega_1 + \omega_2 = 1\). Constraint \eqref{P1_cons1} limits the total transmit power at the BS, and \eqref{P1_cons2} confines the array rotation angle to the interval \([\gamma_{\min},\, \gamma_{\max}]\). Because of its non-convexity and the added rotation angle constraint, solving (P-ORA) directly is challenging. In \cite{rachang}, (P-ORA) is split into two subproblems. One subproblem optimized the beamforming vectors while the other focused on the rotation angle design.

To gain insights into the rotatable array’s performance enhancement for both communication and sensing, we examine two scenarios: the \emph{communication-only} case (i.e., \(\omega_1 = 1\) and \(\omega_2 = 0\)) and the \emph{sensing-only} case (i.e., \(\omega_1 = 0\) and \(\omega_2 = 1\)) in the following, respectively.

For the communication-only case, we focus on a single-user setup (\(K=1\)) with one LoS path and one NLoS path, employing maximal ratio transmission (MRT). The resulting maximum SNR at the user is
\begin{align} \label{kp}
	r_1 
	&= \bigl(\lvert \xi_{1,0}\rvert^2 + \lvert \xi_{1,1}\rvert^2\bigr) M_t + 2\,\lvert \xi_{1,0}\,\xi_{1,1}\rvert\,\Re\bigl\{\mathbf{a}^H(\tilde{\gamma}_{1,0})\,\mathbf{a}(\tilde{\gamma}_{1,1})\bigr\}.
\end{align}

Based on \eqref{kp}, we state the following proposition.

\textbf{Proposition 1:} In the communication-only case, define
\begin{align}
	G_R(\gamma) 
	= \frac{\bigl|\mathbf{a}^H\bigl(\tilde{\gamma}_{1,0}\bigr)\,\mathbf{a}\bigl(\tilde{\gamma}_{1,1}\bigr)\bigr|}
	{\bigl|\mathbf{a}^H\bigl({\gamma}_{1,0}\bigr)\,\mathbf{a}\bigl({\gamma}_{1,1}\bigr)\bigr|},
	\label{eq:rotation_gain}
\end{align} 
as the \emph{rotation gain}, which is the ratio of the rotatable array’s gain to that of a fixed antenna. Then, the maximum rotation gain for communication is
\begin{align}\label{kp0}
	G_R\bigl(\gamma^*\bigr)
	= M_t \,\frac{\bigl|\sin\bigl(\pi\,\sin\zeta_1\,\cos\zeta_2\bigr)\bigr|}
	{\sin\bigl(M_t \,\pi\,\sin\zeta_1\,\cos\zeta_2\bigr)}.
\end{align}
where the optimal array rotation is set as  \(\gamma^* = \frac{n\pi}{2}-\zeta_2 \) with \(n\) being an integer. In \eqref{kp0}, \(\zeta_1 = \frac{\gamma_{1,0} - \gamma_{1,1}}{2}\) and \(\zeta_2 = \frac{\gamma_{1,0} + \gamma_{1,1}}{2}\).

Note that \(G_R(\gamma) = 1\) always holds if \(\zeta_1 \neq 0\). Intuitively, tuning \(\gamma\) can enhance the correlation between \(\mathbf{a}^H({\gamma}_{1,0})\) and \(\mathbf{a}^H({\gamma}_{1,1})\), thereby increasing the SNR.

For the sensing-only case, we have the following proposition.

\textbf{Proposition 2:} The optimal beamforming solution for the sensing-only optimization is 
$
\mathbf{W}^* 
= \frac{P_{\max}}{M_t} \,\mathbf{a}_T(	\tilde{\gamma}) \,\mathbf{a}_T^H(	\tilde{\gamma}),
$
based on which, we have 
$
\mathbf{a}_T^H(	\tilde{\gamma}) \,\mathbf{W}^* \,\mathbf{a}_T(	\tilde{\gamma})
= P_{\max}\,M_t$.
Hence, to maximize \(f_s\bigl(\{\mathbf{w}_k\}, \gamma\bigr)\), the rotation angle must be chosen to maximize \(\cos^2(\bar{\gamma} + \gamma)\). It can be shown that the corresponding optimal angle satisfies \(\gamma^* = 0\) or \(\pi\).

\begin{figure}[t!]
	\centering
	\setlength{\abovecaptionskip}{0.cm}
	\includegraphics[width=2.8in]{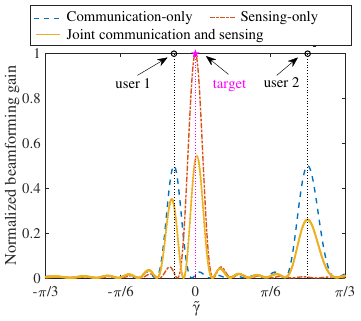}
	\caption{Normalized beamforming gain for the array-wise rotatable 6DMA system \cite{rachang}.}
	\label{racs}
	\vspace{-0.59cm}
\end{figure}

To illustrate how the rotatable array balances communication and sensing performance, we plot the normalized beamforming gain over the effective spatial angle in Fig.~\ref{racs} with \(K=2\), $SNR=-10$ dB, $M_t=M_r=16$, and $P_{\max}=1$ W. We can see that for both the sensing-only and communication-only scenarios, the transmit beam is oriented toward the target or the users to either minimize the CRB or maximize the achievable sum-rate. However, the normalized beamforming gain in the joint communication and sensing case is slightly reduced, which reflects the inherent performance trade-off between communication and sensing.

In addition, Fig.~\ref{racrb} presents the communication and sensing performance of the array-wise rotatable 6DMA system under different strategies. Two benchmarks are evaluated: (1) Beamforming optimization only, where the rotation angle remains fixed while the beamforming vectors are optimized using the method in \cite{rachang}. (2) Rotation optimization only, where the beamforming vectors follow the zero-forcing (ZF) technique and the rotation angle is optimized.
 We observe that, compared to these two benchmarks, the proposed rotatable array approach consistently achieves a lower CRB and/or a higher sum-rate. This result demonstrates the effectiveness of rotatable arrays in enhancing both communication and sensing performance.

\subsection{Element-wise Rotatable 6DMA}
First, to characterize the impact of antenna element rotation on the polarization characteristics and radiation pattern, we consider a planar array consisting of $N$ antennas, positioned at $\mathbf{r}_{n}=[x_n,y_n,z_n]^T$, $ n \in \{1,2,,\ldots,N\}$. Let $\mathbf{a}(\theta, \phi) = [e^{j\frac{2\pi}{\lambda} \mathbf{f}^T \mathbf{r}_{1}},\ldots,e^{j\frac{2\pi}{\lambda} \mathbf{f}^T \mathbf{r}_{N}} ]^T$  denote the 
spatial steering vector corresponding to the signal direction $\left( \theta, \phi\right)$, where  
$\mathbf{f}^T$ is the DOA vector towards direction ($\theta, \phi$), given by
$
	\mathbf{f} = [\sin\theta\cos\phi,\sin\theta\sin\phi,\cos\theta]^T$.
According to~\cite{dai2023linear}, the LoS spatial polarization manifold vector can be expressed as
\begin{align}\label{SP_vec}
	\mathbf{v}(\theta, \phi,\mathbf{R}^{(1)}) = \text{diag}(\mathbf{a}(\theta, \phi) ) \mathbf{R}^{(1)} \mathbf{G}(\theta, \phi) \mathbf{p}(\vartheta, \eta),
\end{align}
where $\mathbf{R}^{(1)} = [\mathbf{u}_{1}^{(1)},\ldots,\mathbf{u}_{N}^{(1)} ]^T \in \mathbb{R}^{N\times3} $ is the rotation matrix for the element-wise R-6DMA with $\mathbf{u}_{n}^{(1)} = [s_{\alpha_{n}} c_{\beta_{n}}, s_{\alpha_{n}} s_{\beta_{n}},c_{\alpha_{n}}]^T$ representing the rotation vector of the $n$-th antenna element. $\mathbf{G}(\theta, \phi)$ is the polarization transformation  matrix, which is given by
	\begin{align}
		\mathbf{G}(\theta, \phi) =
		\begin{bmatrix}
			\cos \theta \cos \phi & -\sin \phi \\
			\cos \theta \sin \phi & \cos \phi \\
			-\sin \theta & 0
		\end{bmatrix},
	\end{align}
	and	$\mathbf{p}(\vartheta, \eta) = [\cos \vartheta,  \sin \vartheta e^{j \eta}]^T$ is polarization impinge on the array \cite{dai2023linear}, where $\vartheta$ and $\eta$ denote the polarization auxiliary angle and polarization phase difference, respectively. 

In the considered rotatable 6DMA-aided communication system, $K$ far-field polarized signals, denoted by $s_k$ for $k = 1, 2, \ldots, K$, arrive at the antenna array from directions $(\theta_k, \phi_k)$. The corresponding received signal at the antenna array, denoted by $\mathbf{y} \in \mathbb{C}^{N \times 1}$, is given by
\begin{align}\label{yv0}
	\mathbf{y} = \sum_{k=1}^{K}\mathbf{v}(\theta_k, \phi_k,\mathbf{R}^{(1)}) s_k + \mathbf{n},
\end{align}
where $\mathbf{n} \in \mathbb{C}^{N \times 1}$ is the AWGN noise. 
Based on~\eqref{yv0}, it can be shown that rotating antenna elements allows to flexibly 
control the power distribution in the observation plane, hence providing new opportunities to enhance  communication  performance by manipulating the spatial radiation pattern and polarization characteristics~\cite{kuhlmann2023sequential}. In addition, element rotation also helps to enhance the information sensing dimensions, thus enabling ambiguity-free 2D DOA estimation by exploiting the sensitivity of polarization ~\cite{dai2023linear}.
\begin{figure}[t!]
	\centering
	\setlength{\abovecaptionskip}{0.cm}
	\includegraphics[width=3.2in]{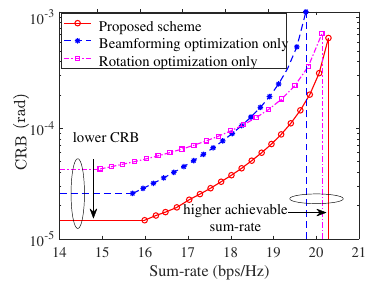}
	\caption{Sensing performance for the array-wise rotatable 6DMA system \cite{rachang}.}
	\label{racrb}
	\vspace{-0.59cm}
\end{figure}

Next, we present a generic framework for the rotation optimization of element-wise R-6DMA, where the main performance metrics capture the impacts of the antenna’s polarization characteristics and radiation patterns~\cite{kuhlmann2023sequential}. This optimization problem can be formulated as follows
\begin{subequations}
\begin{align}
	\text{(P1-EW)}~\max_{\mathbf{R}^{(1)} }~&U(\mathbf{R}^{(1)} )   \nonumber \\
	\text{s.t.}~~& \mathbf{u}_{n}^{(1)}  \in \mathcal{C}^{(1)},~n\in\mathcal{N}, \label{R-6DMA_region2} \\
	& f_{\rm CO}(\mathbf{R}^{(1)}) \le \Gamma_{\rm XPL},\label{C_xlp}  \\
	& f_{\rm SL}(\mathbf{R}^{(1)}) \le \Gamma_{\rm SL}.\label{C_SL},
\end{align}
\end{subequations}
where the objective function aims to maximize the communication performance $U(\mathbf{R}^{(1)})$ (e.g., achievable rates) given a specified angular region $\mathcal{C}^{(1)}$, where each pair of allowable angles $\{\theta,\phi\}$ belongs to $\mathcal{C}^{(1)}$ as imposed by \eqref{R-6DMA_region2} to satisfy the maneuverability restriction and avoid antenna coupling. In addition, it is necessary to consider other relevant constraints to optimize the polarization characteristics and radiation pattern. For example, constraint~\eqref{C_xlp} ensures that the cross-polarization level (XPL), denoted by $f_{\rm CO}(\mathbf{R}^{(1)})$, remains below a predefined threshold $\Gamma_{\rm XPL}$ within the specified angle region. Similarly, constraint~\eqref{C_SL} limits the side-lobe (SL) value, denoted by $f_{\rm SL}(\mathbf{R}^{(1)})$, to be below a predefined threshold $\Gamma_{\rm SL}$. Optimization problem (P1-EW) is generally challenging to solve, since it involves a larger number of rotation variables for all elements. Moreover, the expressions of $U(\mathbf{R}^{(1)} )$, $f_{\rm CO}(\mathbf{R}^{(1)})$, and $f_{\rm SL}(\mathbf{R}^{(1)})$ are functions of the spatial polarization vector in complicated forms.

To solve this problem, convex relaxation methods can be employed to obtain suboptimal solutions. Suboptimal rotation matrix $\mathbf{R}^{(1)}$ for problem (P1-EW) can be obtained by optimizing the pointing vector of each antenna element through the successive convex optimization (SCA) technique and the semidefinite relaxation (SDR) method. Moreover, a variety of heuristic methods, such as PSO, differential evolution, and genetic algorithms, can also be used for obtaining high-quality solutions with low complexity.
For example, the authors in~\cite{li2020design} employed the PSO method to generate sum and difference patterns with reduced SL and XPL. In addition, the dynamic differential evolution methods was utilized in~\cite{li2018shaped} to optimize the element rotation angles to design a targeted beam pattern. Besides stochastic optimization algorithms, random sequential rotation random sequential rotation was proposed in \cite{smolders2014low} to efficiently design the radiation pattern. Specifically, random sequential rotation combines the advantages of random array and sequential rotation technique to simultaneously maximize the antenna gain and control the side-lobe. Furthermore, the authors in \cite{eler, eler1} reduce sidelobe and cross-polarization levels while enhancing the co-polarization level of the main beam in the desired direction. They solve the resulting optimization problem using both genetic algorithms and gradient-based methods.
\begin{figure*}[t]
	\centering
	\includegraphics[width=0.90\textwidth]{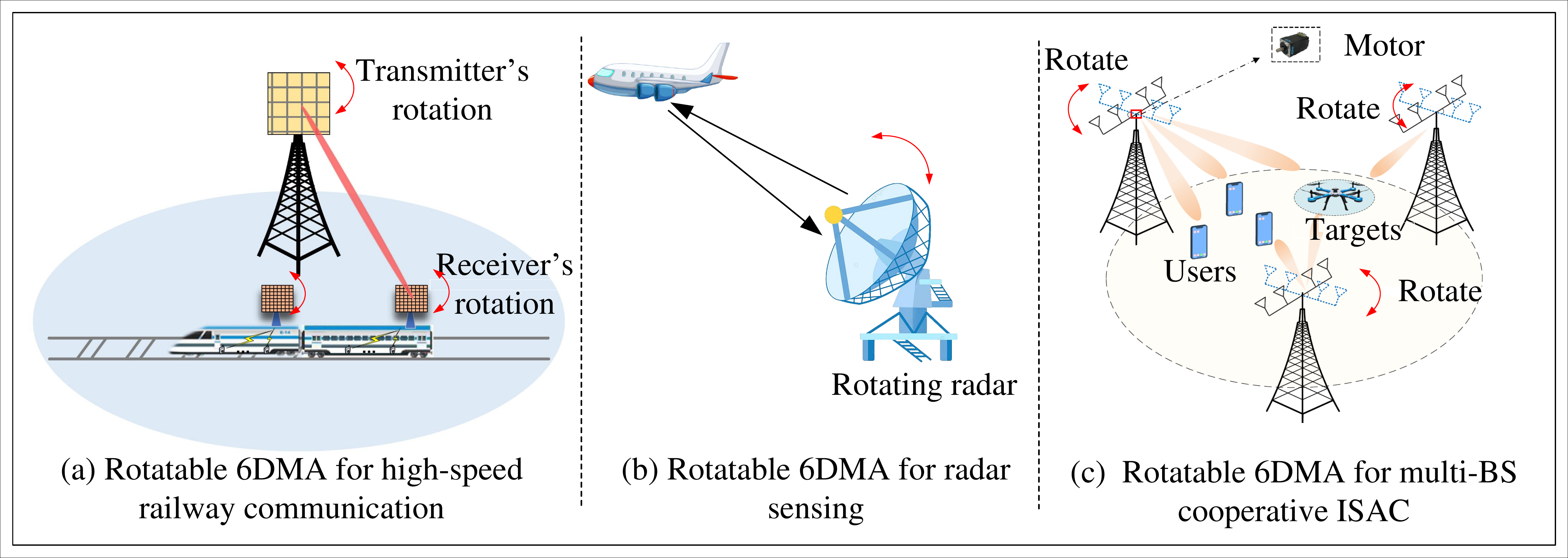}
	\caption{Applications of rotatable 6DMA.} 
	\label{R6DMA_application}
	\vspace{-0.59cm}
\end{figure*} 

On the other hand, for element-wise R-6DMA assisted target sensing systems, the received signal vector at the array over $T$ snapshots can be expressed as~\cite{dai2023linear}
\begin{align}
	\mathbf{Y} =\mathbf{v}(\theta, \phi,\mathbf{R}^{(1)})\mathbf{v}(\theta, \phi,\mathbf{R}^{(1)})^H \mathbf{X}^T+ \mathbf{N},
\end{align}
where $\mathbf{X}\in \mathbb{C}^{T \times N} $ denotes the sensing signal matrix and $\mathbf{N}$ represents the AWGN matrix.
For parameter sensing, such as $\theta$ and $\phi$ estimation, primary performance metrics include the deterministic CRB and stochastic CRB~\cite{zhao2019manifold}. Accordingly, a generic objective function can be written as
\begin{align}
	\text{(P2-EW)}~ \min_{\mathbf{R}^{(1)}}~ \mathbb{E}_{\mathbf{q}_{\rm target}} \Big[ {\rm CRB}_{\rm \iota}(\mathbf{R}^{(1)} ) \Big],
\end{align}
where ${\rm CRB}_{\rm \iota}(\mathbf{R}^{(1)})$ represents the deterministic or stochastic CRB for target sensing, with $\iota \in \{{\rm deterministic}, {\rm stochastic} \}$, and  $\mathbf{q}_{\rm target} \in \mathcal{C}_{\rm target}$ with  $\mathcal{C}_{\rm target}$ denoting the spatial region occupied by the target.  The aforementioned optimization/heuristic methods can be employed to solve this problem. For example, to minimize the asymptotic NMSE in DOA estimation, the authors in~\cite{dai2023linear} optimized the orientation of each antenna element by using a genetic algorithm. Moreover, the receive beamforming and the 3D rotations of all antenna elements can be jointly optimized to maximize the minimum SNR/SINR in both single-user and multi-user uplink systems. An optimal solution was obtained in closed form for the single-user case, while an AO algorithm was proposed for the multi-user scenario in \cite{wu2024modeling}. They theoretically and numerically demonstrated that element-wise R-6DMA with optimized orientations significantly outperforms conventional FPA-based arrays that use either directional or isotropic antennas. 

\subsection{Other Related Works and Future Directions}
In  Fig. \ref{R6DMA_application}, we illustrate several typical applications of rotatable 6DMA in wireless networks. A rotatable 6DMA can be utilized in high-speed rail wireless communication systems to enhance performance in terms of array gain and spatial multiplexing when installed on both the BS and the train roof (see Fig. \ref{R6DMA_application}(a)).
The rapidly changing channel conditions in high-speed rail scenarios pose significant challenges for configuring rotatable 6DMA. In this context, the authors in \cite{rot1} proposed an analytical framework for configuring rotatable 6DMA-based extremely large-scale MIMO systems in high-speed rail environments by leveraging spatial correlation. Additionally, they developed an optimization algorithm based on the differential evolution method to determine the optimal rotation angles of the transceiver panels and validated the proposed framework and algorithm through experimental results.

It is also practically appealing to apply rotatable radar in wireless sensing networks to enhance performance. By leveraging the adaptability of rotatable radar, the system can dynamically adjust its scanning angles and beam directions to optimize sensing coverage and accuracy (see Fig. \ref{R6DMA_application}(b)). For example, the authors in \cite{ro33} studied spectral coexistence between
rotating radar and power-controlled cellular networks in the radar bands. For ISAC systems, rotatable 6DMA can also be exploited to improve both the communication and sensing performance. For example, the authors in \cite{rot6} considered a rotatable 6DMA system, where multiple BSs equipped with rotatable 6DMAs collaboratively design their rotations to improve system performance (see Fig. \ref{R6DMA_application}(c)). An optimization problem was formulated to jointly optimize the multi-BS array rotation angles and the beamforming matrix to achieve optimal sensing beam pattern matching while meeting communication performance requirements and power constraints. Their results demonstrate the effectiveness of the proposed approach in achieving low beampattern mean square error (MSE) for given communication requirements.

Tables \ref{6DMApr} and \ref{tab:6dma_summary} summarize representative studies on rotatable 6DMA rotation optimization and channel estimation, respectively. For example, in \cite{15}, the authors introduced a DOA estimation scheme based on a rotatable 6DMA, which demonstrates robust performance for under-determined scenarios in which the number of sources exceeds the number of receive antenna elements. In \cite{7}, the achievable rate in point-to-point LoS channels was studied, where both the transmitter and receiver were equipped with a rotatable 6DMA. The reported findings indicated that rotatable 6DMA-enabled systems can approach the LoS capacity at any desired SNR. Furthermore, the authors of \cite{rot} and \cite{10} explored the use of rotatable 6DMAs for wireless energy transfer by examining a setting in which a power beacon equipped with a rotatable 6DMA rotates continuously and sends energy signals to multiple low-power devices. These devices harvest the transmitted energy to recharge their batteries. In \cite{11}, a prototype was developed and tested for hybrid mechanical-electrical beamforming in mmWave WiFi. The experiments for a point-to-point configuration highlighted that optimal array rotation can significantly boost throughput under both LoS and NLoS conditions.

\section{Positionable 6DMA}
In contrast to rotatable 6DMA with fixed antenna position, positionable 6DMA refers to the other simplified implementation of 6DMA with fixed antenna rotation (e.g., due to hardware limitation), but allowing 
antenna translation/movement along a line or a plane. Similar to rotatable 6DMA, positionable 6DMA generally performs inferior to fully adaptive 6DMA, while it also has its unique design challenges as well as suitable applications. Thus, in this section, we study positionable 6DMA dedicatedly. In particular, by fixing the antenna rotation $ \mathbf{u}_b$ in the general 6DMA system model \eqref{bb1}, we obtain:
\begin{align}
	\!\!\!\! \mathbf{q}_b = 
	\begin{bmatrix}
		{x_b}, & {y_b}, & z_b
	\end{bmatrix}^T, ~\cancel{\mathbf{u}_b} = 
	\begin{bmatrix}
		\cancel{\alpha_b}, & \cancel{\beta_b}, & 	\cancel{\gamma_b}
	\end{bmatrix}^T, b\in\mathcal{B}.
\end{align}

This setup is referred to as positionable 6DMA (P-6DMA), which is a special case of 6DMA. We present the main design issues, including joint antenna position optimization and resource allocation, channel estimation, antenna movement path planning, ISAC, etc., and propose potential solutions for them. 

\subsection{Position Optimization and Resource Allocation}
In the subsection, we present three typical examples for position optimization and resource allocation for P-6DMA.

\subsubsection{Positionable 6DMA-Aided Multi-Group Multicast System}
\begin{figure}[!t]
	\centering
	\includegraphics[scale=0.52]{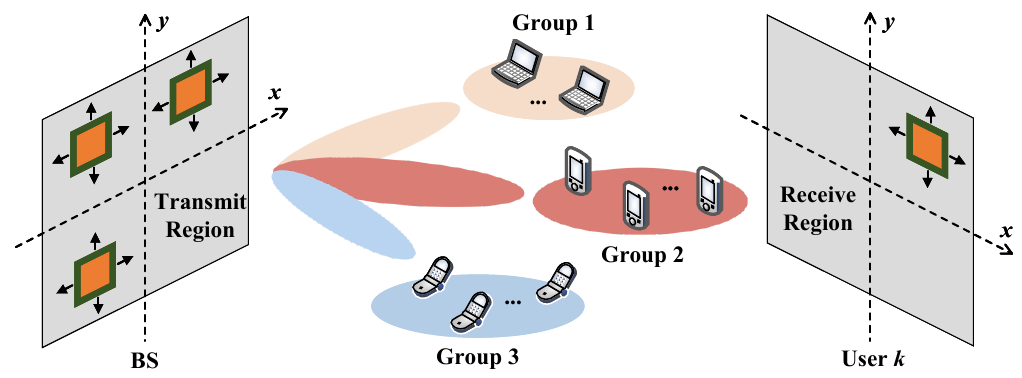}
	\caption{Positionable 6DMA-aided multi-group multicast system.} \label{multicast_model}
	\vspace{-0.59cm}
\end{figure}
%For single-group multicast systems, antenna positions can be adjusted to pursue better channel conditions and enhance the correlation among user channel vectors within the group, thereby improving signal quality for all users. In multi-group scenarios, the antenna position optimization builds on these benefits by additionally mitigating inter-group interference. This combined effect improves both intra-group signal quality and inter-group isolation, leading to enhanced overall system performance in complex multicast settings. To illustrate these concepts, 

First, we consider a positionable 6DMA-aided multi-group multicast MISO communication system comprising a BS equipped with \(Q\) transmit antennas, with a total of \(K\) single-antenna users divided into \(Z\) groups. As depicted in Fig. \ref{multicast_model}, both the transmit and receive antennas can move within a 2D planar region. The transmit antennas are represented by set $\mathcal{Q}\triangleq \{1, \dots, Q\}$ with cardinality $|\mathcal{Q}| = Q$. Denote by $\mathcal{G}_i$ the set of users in group $i$, where $i \in \mathcal{Z} \triangleq \{1, \dots, Z\}$. Each user is assigned to exactly one group, so for any distinct groups $i$ and $q$, we have $\mathcal{G}_i \cap \mathcal{G}_q = \emptyset$, and the union of all groups covers all users, i.e., $\bigcup_{i=1}^Z \mathcal{G}_i = \mathcal{K}$.
Let $\mathcal C^t$ denote the given 2D movement region for the $Q$ transmit antennas. The 2D movement region for the single antenna at user $k$ is denoted by $\mathcal C_k^r$. Without loss of generality, we assume that both $\mathcal C^t$ and $\mathcal C_k^r$ are square regions of size $A \times A$. The positions of the $m$-th transmit antenna at the BS and the single receive antenna at the $k$-th user are represented by 
$\mathbf{q}^{\mathrm{t}}_m = [x_m^t, y_m^t]^T \in \mathcal C^t $ and $\mathbf{q}_k^{\mathrm{r}} = [x_k^r, y_k^r]^T \in \mathcal C_k^r$, respectively. Based on the positionable 6DMA channel model described in \eqref{p6}, the channel vector from the BS to the \( k \)-th user is given by \cite{2024_Ying_multicast}  
\begin{align}\label{p66}
	\bm{h}_k(\mathbf{q}^{\mathrm{t}},\mathbf{q}_k^{\mathrm{r}}) = \bm{G}_k(\mathbf{q}^{\mathrm{t}})^H \mathbf{\Sigma}_k \bm{c}_k(\mathbf{q}_k^{\mathrm{r}}) \in \mathbb{C}^{Q \times 1},
\end{align}
where $\mathbf{q}^{\mathrm{t}}\triangleq \{\mathbf{q}^{\mathrm{t}}_m\}$, \(\bm{G}_k(\mathbf{q}^{\mathrm{t}}) = \left[\bm{t}_k(\mathbf{q}_1^{\mathrm{t}}),  \cdots, \bm{t}_k(\mathbf{q}_Q^{\mathrm{t}})\right] \in \mathbb{C}^{L_k \times Q}\), \(\bm{t}_k(\mathbf{q}_b^{\mathrm{t}}) = \left[e^{-j\frac{2\pi}{\lambda} (\mathbf{f}_{k,1}^{\mathrm{t}})^T \mathbf{q}_b^{\mathrm{t}}}, \cdots, e^{-j\frac{2\pi}{\lambda} (\mathbf{f}_{k,L_k}^{\mathrm{t}})^T \mathbf{q}_b^{\mathrm{t}}}\right]^T\) and \(\!\bm{c}_k(\mathbf{q}_k^{\mathrm{r}})\! =\! \left[e^{-j\frac{2\pi}{\lambda} (\mathbf{f}_{k,1}^{\mathrm{r}})^T \mathbf{q}_k^{\mathrm{r}}}, \! \cdots,\! e^{-j\frac{2\pi}{\lambda} (\mathbf{f}_{k,L_k}^{\mathrm{r}})^T \mathbf{q}_k^{\mathrm{r}}}\right]^T\!\) denote the steering vector for all channel paths corresponding to the antennas at the BS and the position of the single antenna at user \( k \), respectively. Here, $L_k$ denotes the number of paths between user $k$ and the BS.
Vectors \(\mathbf{f}_{k,l}^{\mathrm{t}} = [\cos(\theta_{k,l}^{\mathrm{t}})\cos(\phi_{k,l}^{\mathrm{t}}), \cos(\theta_{k,l}^{\mathrm{t}})\sin(\phi_{k,l}^{\mathrm{t}}), \sin(\theta_{k,l}^{\mathrm{t}})]^T\) and \(\mathbf{f}_{k,l}^{\mathrm{r}} = [\cos(\theta_{k,l}^{\mathrm{r}})\cos(\phi_{k,l}^{\mathrm{r}}), \cos(\theta_{k,l}^{\mathrm{r}})\sin(\phi_{k,l}^{\mathrm{r}}), \sin(\theta_{k,l}^{\mathrm{r}})]^T\) represent the transmit and receive pointing vectors, respectively, where \(\phi_{k,l}^{\mathrm{t}}\) and \(\theta_{k,l}^{\mathrm{t}}\) denote the azimuth and elevation AoD, respectively, for the \( l \)-th channel path between the BS and the scatterers. Similarly, \(\phi_{k,l}^{\mathrm{r}} \) and \(\theta_{k,l}^{\mathrm{r}} \) denote the azimuth and elevation angles of arrival
(AoA), respectively, for the \( l \)-th channel path between user \( k \) and the scatterers. Finally, \(\mathbf{\Sigma}_k = \mathrm{diag}(\boldsymbol{\tau}_k) \in \mathbb{C}^{L_k \times L_k}\) represents the diagonal channel coefficient matrix, as shown in \eqref{p6}.

At the BS, we assume linear transmit precoding with $\bm w_i$ denoting the beamforming vector for group $i$. Then, the SINR of user $k\in\mathcal G_i, i \in \mathcal{Z}$, can be written as
\begin{align}
	r_k\left(\mathbf{q}^{\mathrm{t}}, \mathbf{q}^{\mathrm{r}}_k, \bm W\right)  = \frac{\left|\bm h_k(\mathbf{q}^{\mathrm{t}}, \mathbf{q}^{\mathrm{r}}_k)^H\bm w_i\right|^2}{\sum_{q=1,q\neq n}^N\left|\bm h_k(\mathbf{q}^{\mathrm{t}}, \mathbf{q}^{\mathrm{r}}_k)^H\bm w_q\right|^2 + \sigma_k^2},
\end{align}
where $\bm W \triangleq \left\lbrace\bm w_i\right\rbrace$ and $\sigma_k^2$ is the noise power at user $k$. 

For multicast communication systems, the max-min fairness (MMF) design problem and the quality of service (QoS) design problem are a dual pair. Let $f\left(\mathbf{q}^{\mathrm{t}}, \mathbf{q}^{\mathrm{r}}_k, \bm W\right)$ denote the MMF utility, which can take the form of: (i) minimum SINR, $f_1\left(\mathbf{q}^{\mathrm{t}}, \mathbf{q}^{\mathrm{r}}_k, \bm W\right) = \min_{k\in\mathcal G_i, i \in \mathcal{Z}}r_k\left(\mathbf{q}^{\mathrm{t}}, \mathbf{q}^{\mathrm{r}}_k, \bm W\right)$, or (ii) minimum rate, $f_2\left(\mathbf{q}^{\mathrm{t}}, \mathbf{q}^{\mathrm{r}}_k, \bm W\right) = \min_{k\in\mathcal G_i, i \in \mathcal{Z}}\log_2\left(1+r_k\left(\mathbf{q}^{\mathrm{t}}, \mathbf{q}^{\mathrm{r}}_k, \bm W\right)\right)$. The MMF utility $f\left(\mathbf{q}^{\mathrm{t}}, \mathbf{q}^{\mathrm{r}}_k, \bm W\right)$ can be improved by jointly optimizing the position of each transmit/receive antenna and the transmit beamforming, which is formulated as the following problem \cite{2024_Ying_multicast}:
\begin{subequations}\label{P-MMF}
	\begin{align}
		\text{(P-MMF)}: \ &\underset{\mathbf{q}^{\mathrm{t}}, \{\mathbf{q}^{\mathrm{r}}_k\},\boldsymbol W}{\max} \ f\left(\mathbf{q}^{\mathrm{t}}, \mathbf{q}^{\mathrm{r}}_k, \bm W\right) \\[1mm]
		&\text{s.t.} \ \sum_{i=1}^Z\left\|\bm w_i\right\|^2 \leq P_{\max}, \label{P-MMF_cons:b}\\[1mm]
		&\mathbf{q}^{\mathrm{t}}_m \in\mathcal C^{\rm t}, \ \forall m\in\mathcal Q, \label{P-MMF_cons:c}\\[1mm]
		&\left\|\mathbf{q}^{\mathrm{t}}_m - \mathbf{q}^{\mathrm{t}}_p\right\| \geq d_{\min}, \ \forall m,p\in\mathcal Q, m\neq p, \label{P-MMF_cons:d}\\[1mm]
		&\mathbf{q}^{\mathrm{r}}_k \in\mathcal C^{\rm r}_k, \ \forall k\in\mathcal G_i, i \in \mathcal{Z}, \label{P-MMF_cons:e}
	\end{align}
\end{subequations}
where $P_{\max} > 0$ in \eqref{P-MMF_cons:b} denotes the maximum instantaneous transmit power of the BS, and \eqref{P-MMF_cons:d} guarantees that the distance between every pair of transmit antennas is at least $d_{\min} > 0$, preventing coupling effects between them. The dual QoS problem of minimizing the transmit power, subject to a given QoS threshold on $f\left(\mathbf{q}^{\mathrm{t}}, \mathbf{q}^{\mathrm{r}}_k, \bm W\right)$, can be formulated as 
\begin{subequations}\label{P-QoS}
	\begin{eqnarray}
		\text{(P-QoS)}: \hspace{-3mm}&\underset{\mathbf{q}^{\mathrm{t}}, \{\mathbf{q}^{\mathrm{r}}_k\},\boldsymbol W}{\max}& \hspace{-2mm} \sum_{n=1}^N\left\|\bm w_i\right\|^2 \\
		&\text{s.t.}& \hspace{-5mm} f\left(\mathbf{q}^{\mathrm{t}}, \mathbf{q}^{\mathrm{r}}_k, \bm W\right) \geq \eta, \label{P-QoS_cons:b}\\
		&& \hspace{-5mm} \eqref{P-MMF_cons:c}-\eqref{P-MMF_cons:e}.
	\end{eqnarray}
\end{subequations}
Note that if each group consists of only a single user, these two problems simplify to those examined for positionable 6DMA-aided multiuser MISO downlink communication systems. 
\begin{figure}[!t]
	\centering
	\includegraphics[scale=0.65]{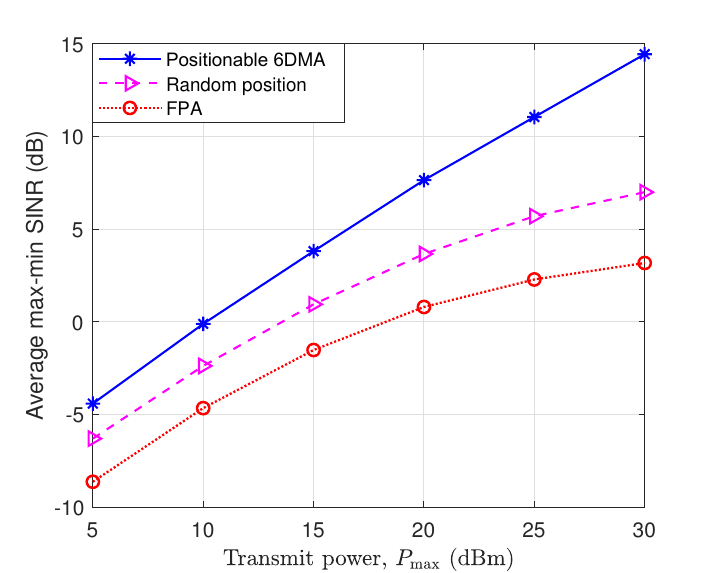}
	\caption{Average max-min SINR versus the maximum transmit power at the BS.} \label{fig:multi_vs_Pmax}
	\vspace{-0.59cm}
\end{figure}

Problems (P-MMF) and (P-QoS) are challenging to solve due to the coupling of optimization variables. To tackle these problems, the AO method can be applied by dividing the optimization variables into three blocks: $\mathbf{q}^{\mathrm{t}}$, $\{\mathbf{q}^{\mathrm{r}}_k\}$, and $\boldsymbol W$. By iteratively updating each block in succession until convergence, suboptimal solutions for (P-MMF) and (P-QoS) can be obtained. In each iteration, $\mathbf{q}^{\mathrm{t}}$ and $\{\mathbf{q}^{\mathrm{r}}_k\}$ can be updated using techniques such as SCA \cite{2024_Ying_multicast}, gradient descent \cite{mawide}, or PSO \cite{6dmaSens,zr14}. In addition, $\boldsymbol W$ can be updated through SCA, SDR \cite{2008_Eleftherios_multicast,zr15}, or the alternating direction method of multipliers (ADMM) \cite{2017_Erkai_multicast}, as commonly applied in conventional multicast systems. Besides, it is worth noting that due to the duality between (P-MMF) and (P-QoS), a solution to (P-MMF) can be found by iteratively solving (P-QoS) \cite{2008_Eleftherios_multicast}.  

To assess the benefits of antenna position optimization in improving multicast performance, we plot in Fig. \ref{fig:multi_vs_Pmax} the max-min SINR achieved by different schemes versus the maximum transmit power at the BS. The BS is positioned at $[0,0]^T$, and users are randomly distributed within a disk centered at $[60,0]^T$ (in meters (m)) with a radius of $20$ m. The distance between user $k$ and the BS is denoted as $d_k$. We set $L_k = 10$, $\forall k \in \mathcal K$ and  $\tau_{k,\ell} \sim \mathcal{CN}(0, \frac{c_k^2}{10})$, $\ell = 1,\ldots,10$, where $c_k^2 = C_0d_k^{-\epsilon}$, with $C_0 = -40$ dB (reference power gain at 1 m) and $\epsilon = 2.8$ (path-loss exponent). The elevation and azimuth AoDs/AoAs are assumed to be i.i.d., uniformly distributed over $\left[-\frac{\pi}{2}, \frac{\pi}{2}\right]$. The transmit/receive antenna regions are defined as $\mathcal C^{\rm t} = \mathcal C_k^{\rm r} = \left[ - \frac{A}{2}, \frac{A}{2}\right]\times \left[ - \frac{A}{2}, \frac{A}{2}\right]$, where $A = 4\lambda$, $\forall k\in\mathcal K$. Other parameters are set as follows: $M = 4$, $d_{\min} = \frac{\lambda}{2}$, $\sigma_k^2 = -80$ dBm, $K = 9$, $Z = 3$, and $\left|\mathcal G_1\right| = \left|\mathcal G_2\right| = \left|\mathcal G_3\right| = 3$. We also evaluate two benchmark schemes for comparison: (i) Random position: For each channel realization, 100 independent samples of ${\mathbf{q}^{\mathrm{t}}_m}$ and ${\mathbf{q}^{\mathrm{r}}_k}$ are randomly generated, satisfying constraints \eqref{P-MMF_cons:c}-\eqref{P-MMF_cons:e}. Transmit beamforming is optimized for each sample, and the best-performing solution is selected. (ii) FPA: The BS employs a ULA with $M$ FPAs spaced at $\frac{\lambda}{2}$, while each user's antenna is fixed at the reference point in its receive region. 

As shown in Fig. \ref{fig:multi_vs_Pmax}, the proposed antenna position optimization scheme consistently achieves a higher max-min SINR than the two benchmark schemes, as it maximizes the use of the spatial DoFs to leverage both channel diversity and interference mitigation gains in the spatial domain. The performance loss caused by random antenna positioning underscores the importance of optimizing antenna placement to enhance system performance. Moreover, it is observed that the max-min SINR achieved by the FPA scheme tends to saturate as $P_{\max}$ increases, primarily due to the increased severity of inter-group interference. This is consistent with the theoretical and numerical results in \cite{2017_Hamdi_multicast} showing that classical beamforming cannot provide interference-free streams for all multicast groups when the condition $Q \geq 1 + K - \min_{i\in\mathcal Z}\left|\mathcal G_i\right|$ is violated. In contrast, antenna position optimization facilitates an almost linear growth in max-min SINR with increasing  $P_{\max}$, showcasing its ability to effectively eliminate or, at the very least, significantly mitigate inter-group interference even when the above condition is not satisfied.  

\subsubsection{Positionable 6DMA-Aided Interference Networks}
\begin{figure}[!t]
	\centering
	\includegraphics[width=0.39\textwidth]{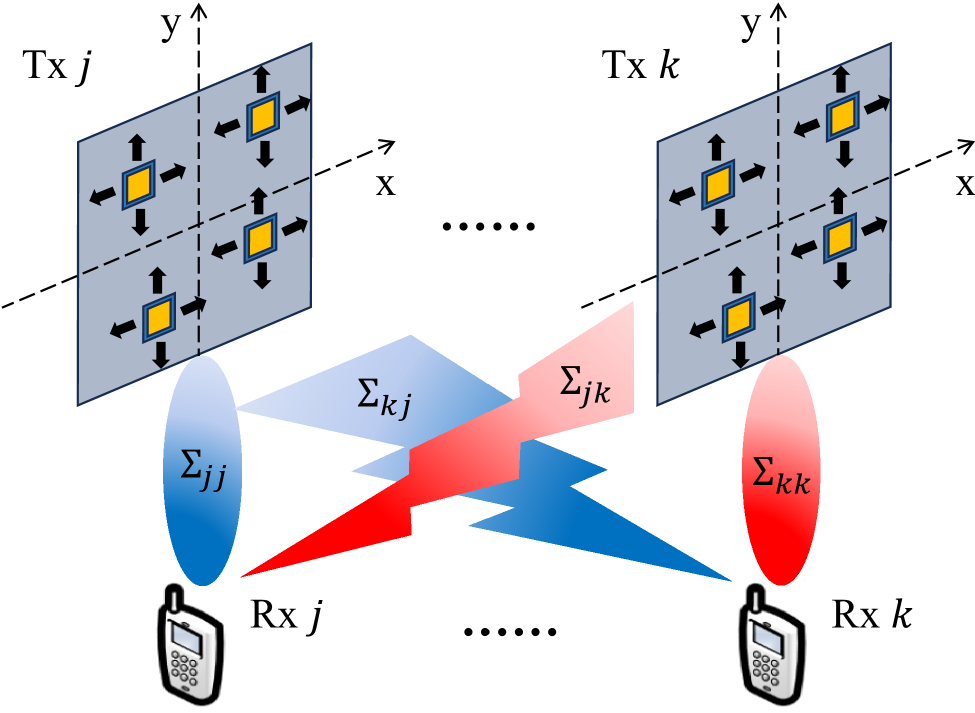}
	\caption{MISO interference network aided by positionable 6DMAs.}
	\label{fig:IF_model}
	\vspace{-0.59cm}
\end{figure}

%The growing demand for spectrum sharing renders the overall network performance limited by co-channel interference (IF), whereas optimizing antenna positions to reconfigure either instantaneous or statistical channels between the transmitter and receiver can overcome this inherent constraint for two reasons, i.e., increasing the desired signal power and mitigating the interference. In typical IF networks, each BS serves a corresponding group of users but interferes with other users. To demonstrate the performance advantages antenna translation provide, 
Next, we consider another typical communication scenario, namely a P-6DMA-aided MISO interference (IF) network with $K$ users. In particular, we consider a scenario with multiple transmitters (i.e., BSs), where each transmitter's antennas can move, while the receiver's antenna remains fixed. As shown in Fig. \ref{fig:IF_model}, each BS is equipped with $Q$ antennas, which can translate their positions in the 2D plane and simultaneously transmit signals to their corresponding single-FPA users. 
Let $\mathbf{w}_j\in\mathbb{C}^{N\times1}$ denote the transmit beamforming vector of BS $j$, with $\sum_{j=1}^{K}\left|\left|\mathbf{w}_j\right|\right|^2\le{P_{\max}}$, where $P_{\max}$ denotes the total transmit power. Let $\mathbf{T}_j=\left[\mathbf{q}^{\mathrm{t}}_{j,1},\ldots,\mathbf{q}^{\mathrm{t}}_{j,Q}\right]\in\mathbb{R}^{2{\times}Q}$ denote the collection of the coordinates of the $Q$ antennas at BS $j$, with $\mathbf{q}^{\mathrm{t}}_{j,i}=\left[x_{j,i},y_{j,i}\right]^T\in\mathcal{C}_j$ for $i\in\left\{1,\ldots,Q\right\}$ indicating the position of the $i$-th antenna at BS $j$, and $\mathcal{C}_j$ being the antenna movement region of BS $j$ \cite{WangHH_interference_MA}.
Then, the resulting SINR at user $k$ can be expressed as $r_k=\frac{\left|\mathbf{h}_{kk}^H\mathbf{w}_k\right|^2}{\sum_{j\ne{k}}\left|\mathbf{h}_{kj}^H\mathbf{w}_j\right|^2+\sigma_k^2},~\forall{k}$, where $\mathbf{h}_{kj}$ denotes channel vector from BS $j$ to
user $k$ as described in \eqref{p6}. To ensure QoS fairness, the minimum achievable rate across all users is enhanced by jointly optimizing antenna positions $\left\{\mathbf{T}_j\right\}_{j=1}^K$ and beamforming vectors $\left\{\mathbf{w}_j\right\}_{j=1}^K$, which leads to the following optimization problem \cite{WangHH_interference_MA}:
\begin{subequations}\label{eq_PIF1}
	\begin{align}
		\text{(P-IF1)}:~~&~\underset{\left\{\mathbf{T}_j,\mathbf{w}_j\right\}_{j=1}^K}{\max}~\underset{k\in\mathcal{K}}{\min} \log_2\left(1+r_k\right)\label{eq_PIF1_a}\\
		\text{s.t.}~&~ \sum_{j=1}^{K}\left|\left|\mathbf{w}_j\right|\right|^2\le{P_{\max}},\label{eq_PIF1_b}\\
		~&~ \hspace{1pt}\mathbf{q}^{\mathrm{t}}_{j,i}\in\mathcal{C}_j,~{\forall}j,i,\label{eq_PIF1_c}\\
		~&~ \left|\left|\mathbf{q}^{\mathrm{t}}_{j,i}-\mathbf{q}^{\mathrm{t}}_{j,\tilde{n}}\right|\right|\ge{d_{\min}},~{\forall}j,\tilde{n}{\ne}i.\label{eq_PIF1_d}
	\end{align}
\end{subequations}
The solution to (P-IF1) can be obtained by iteratively solving the following total transmit power minimization problem \cite{WangHH_interference_MA}: 
\begin{subequations}\label{eq_PIF2}
	\begin{eqnarray}
		\text{(P-IF2)}:&\underset{\left\{\mathbf{T}_j,\mathbf{w}_j\right\}_{j=1}^K}{\min}& \sum_{j=1}^{K}\left|\left|\mathbf{w}_j\right|\right|^2\label{eq_PIF2_a}\\
		&\text{s.t.}& \hspace{-4mm} r_k\ge\gamma_\text{min},~\forall{k},\label{eq_PIF2_b}\\
		&& \hspace{-4mm} \eqref{eq_PIF1_c},\eqref{eq_PIF1_d}.
	\end{eqnarray}
\end{subequations}
In particular, the minimum SINR (or rate) requirement, $\gamma_{\text{min}}$ in \eqref{eq_PIF2_b}, is iteratively adjusted based on the solution obtained for (P-IF2). The iteration terminates when the power consumption for (P-IF2) satisfies the power constraint $P_{\max}$ specified in (P-IF1). At this point, the final value of $\gamma_{\text{min}}$ serves as the objective value of (P-IF1). Nevertheless, (P-IF2) is still intractable due to the non-convexity of the minimum SINR constraint \eqref{eq_PIF2_b} and the minimum distance constraint \eqref{eq_PIF1_d}. To address this challenging problem, local optimization techniques (e.g., gradient ascent and SCA) or global optimization techniques (e.g., PSO) can be employed to find suboptimal antenna positions.

\begin{figure}[t]
	\centering
	\includegraphics[scale=0.63]{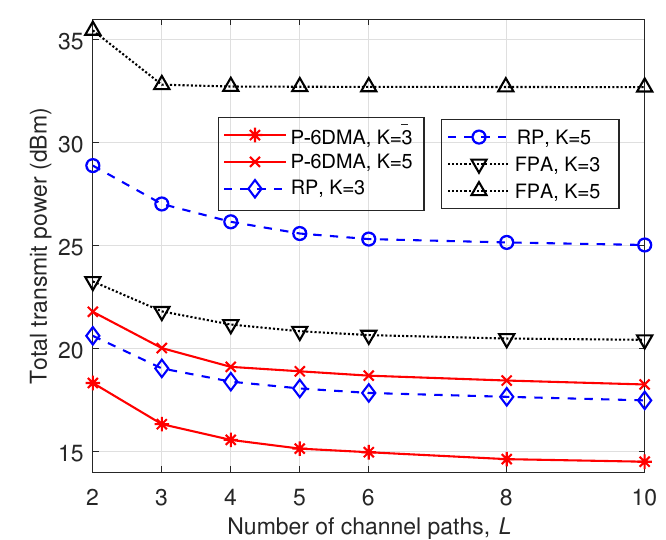}
	\caption{Total transmit power of MISO interference network aided by positionable 6DMAs.}
	\label{fig:power_vs_L}
		\vspace{-0.59cm}
\end{figure}

To evaluate the effectiveness of antenna position optimization for power-saving and reducing inter-cell interference, we show in Fig. \ref{fig:power_vs_L} the total transmit power of P-6DMA-IF, random position (RP)-IF, and FPA-IF networks versus the number of channel paths. Specifically, the considered networks with $K=3$ or $K=5$ BS-user pairs employ $Q=4$ transmit antennas at each BS. $\mathcal{C}_j$ is set as a 2D square region of size $4\lambda\times4\lambda$. 
We set \(L_{kj}\) to be the same as \({L}\) for all \(k\) and \(j\), with \( \gamma_{\text{min}} = 10~\text{dB} \), \(d_{\min} = \frac{\lambda}{2} \), and \( \sigma_k^2 = -80~\text{dBm} \). The elevation angle \( \theta_{kj,l} \in [0, \pi] \) and azimuth angle  \( \phi_{kj,l} \in [0, \pi] \) of all channel paths follow the joint probability density function (PDF) \( f_A(\theta_{kj,l}, \phi_{kj,l}) = \frac{\sin{\theta_{kj,l}}}{2\pi} \). Assuming a limited number of scatterers, the ${L} \times K$ channel path angles for each BS are randomly drawn from a shared set of 10 angle pairs generated using this PDF.

Fig. \ref{fig:power_vs_L} illustrates that the powers of all considered schemes decrease with increasing $L$, while the positionable 6DMA-aided system surpasses the benchmarks for the considered values of $K$ owing to the interference mitigation benefits derived from antenna position optimization. Note that the reduction in transmit power for the FPA system is not due to a higher average channel gain, which remains constant (normalized by $L$), but rather due to the interference reduction. As $L$ increases, the spatial diversity provided by antenna position adjustments enhances the channel variation, thereby reducing correlation among channel vectors. However, when $L$ exceeds 5, the power reduction becomes slower as the channel correlation becomes limited by the finite set of elevation and azimuth angles. Consequently, further antenna position adjustments offer diminishing returns in reducing channel correlation. These results demonstrate that antenna position optimization significantly enhances the ability to suppress interference and strengthen desired signals. For instance, with $K=5$ BS-user pairs, the total transmit power of the system with antenna position movement is even lower than that of the FPA system with $K=3$ pairs, indicating that position optimization enables support for more cells without increasing transmit power.  

\subsubsection{Two-Timescale Transmission Scheme}
\begin{figure}[!t]
	\centering
	\includegraphics[width=0.49\textwidth]{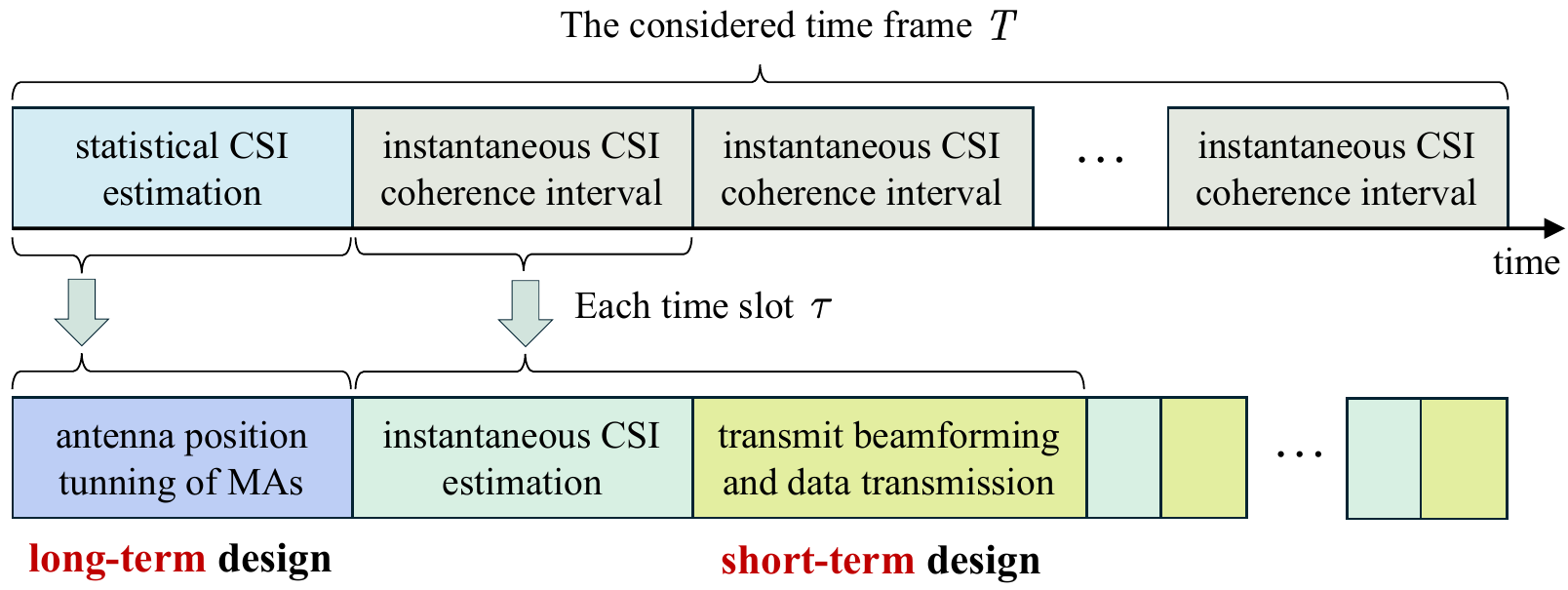}
	\caption{Illustration of the two-timescale framework.}
	\label{fig:twotime_framework}
	\vspace{-0.59cm}
\end{figure}

In 6DMA wireless systems, acquiring accurate instantaneous CSI across the entire movement region is challenging, especially for fast-fading channels. Instead of dynamically repositioning antennas to track instantaneous channel variations, a hierarchical two-timescale transmission scheme can be employed, as illustrated in Fig.~\ref{fig:twotime_framework}. In the large timescale, antenna positions at the BS are optimized based on relatively stable statistical CSI, enhancing long-term system performance without frequent repositioning. Once the optimal antenna positions are fixed, the beamforming matrix is designed in the short-term timescale to address dynamic channel fluctuations, where instantaneous CSI is acquired using conventional estimation techniques. This two-timescale design reduces the updating frequency of the antennas' positions, lowering both channel estimation overhead and movement energy consumption while retaining beamforming gains from fast-fading components.

We consider a downlink MIMO system capable of antenna position movement, where a BS equipped with $Q$ transmit antennas serves $K$ single-FPA users. Under the two-timescale transmission framework, the ergodic sum rate for all users is maximized by jointly optimizing the antenna positions $\{\mathbf{q}_i=\left[x_{i},y_{i}\right]^T\}_{i=1}^Q$ in the large timescale and transmit beamforming vectors $\{\boldsymbol{w}_k\}_{k=1}^K$ in the short timescale. The general optimization problem is formulated as follows \cite{2024_Ziyuan_twotime}
\begin{subequations}
	\begin{align}
		\text{(P-TT1)}:\,\underset{\{\mathbf{q}_i\}_{i=1}^Q}{\max}&\quad \mathbb{E}\bigg[ \underset{\{\boldsymbol{w}_k\}}{\max}\sum_{k=1}^K \log_2\left(1+r_k(\{\mathbf{q}_i\},\{\boldsymbol{w}_k\})\right) \bigg] \nonumber\\
		\text{s.t.}\quad & \sum_{m=1}^M \lVert \boldsymbol{w}_k \rVert^2 \le P_{\max}, \label{P1b}\\
		& \lVert \mathbf{q}_n - \mathbf{q}_i \rVert \ge d_{\min},\quad \forall n \ne i, \label{P1c}\\
		& \mathbf{q}_i \in \mathcal{C},\quad \forall i\in \mathcal{Q}, \label{P1d}
	\end{align}
\end{subequations}
where $r_k(\{\mathbf{q}_i\},\{\boldsymbol{w}_k\})$ denotes the receive SINR at user $k$, $\mathcal{C}$ is the 2D movement region, the expectation is taken over all possible channel realizations, \eqref{P1c} ensures a minimum separation distance between any two antennas, and \eqref{P1d} specifies the feasible region for antenna positions.

Solving $\text{(P-TT1)}$ is challenging due to the intertwined variables and the lack of a closed-form expression for the ergodic sum rate. Stochastic optimization or Monte Carlo methods can approximate the solution using randomly generated channel samples but incur high computational complexity. Alternatively, adopting classical beamforming schemes such as MRT and ZF reduces complexity \cite{2024_Ziyuan_twotime, schober,zr16}. 

\begin{figure}[!t]
	\centering
	\includegraphics[width=3.0in,height=2.2in]{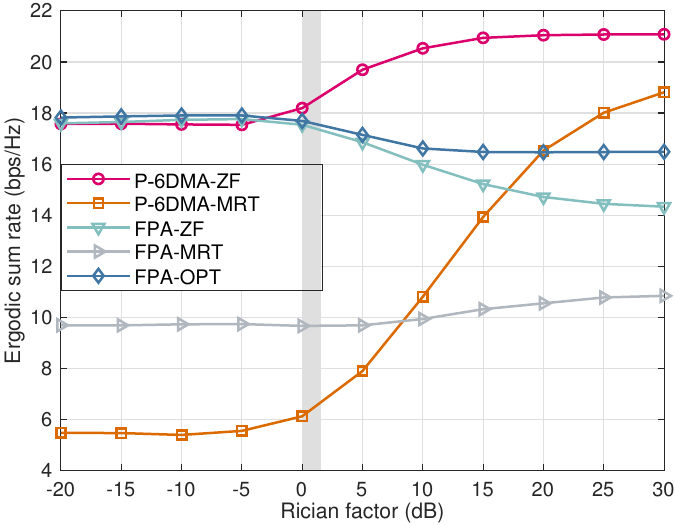}
	\caption{Ergodic sum rate versus Rician factor $\kappa$ with $N=4$, $M=3$.}
	\label{fig:sum_rate_vs_kappa}
	\vspace{-0.59cm}
\end{figure}

To evaluate the performance of two-timescale schemes, Fig.~\ref{fig:sum_rate_vs_kappa} shows the ergodic sum rate as a function of the channel Rician factor $\kappa$. Users are randomly distributed around the BS with the BS-user distances $d_k$ uniformly sampled in [50\,m, 70\,m]. The Rician fading channel model with a common Rician factor $\kappa_k=\kappa$ is employed for all BS-user channels. The elevation and azimuth AoD and AoA for each path are uniformly distributed within $\left[ -\frac{\pi}{2},\frac{\pi}{2} \right]$. The movable regions for the transmit antennas are set as $\mathcal{C}=\left[ -\frac{N_rA\lambda}{2},\frac{N_rA\lambda}{2} \right] \times \left[ -\frac{N_cA\lambda}{2},\frac{N_cA\lambda}{2} \right]$ with $N_rN_c=Q$, expanding adaptively with $Q$, and $A$ is set to 2. We set $Q=4$, $K=3$, and $P_{\max} = 30$\,dBm.
The following schemes are compared: positionable 6DMA with ZF (P-6DMA-ZF) and positionable 6DMA with MRT (P-6DMA-MRT), which employ ZF and MRT beamforming, respectively, with fixed power allocation and optimized antenna positions based two-timescale CSI; FPA-ZF and FPA-MRT, which employ ZF and MRT beamforming, respectively, with optimized power allocation based on instantaneous CSI; and FPA-OPT, which applies optimal beamforming using instantaneous CSI.

As shown in Fig.~\ref{fig:sum_rate_vs_kappa}, the positionable 6DMA schemes significantly benefit from increasing $\kappa$, as the channel becomes more deterministic and there is more potential for antenna position optimization. Despite using equal power allocation and a suboptimal beamforming scheme, P-6DMA-ZF demonstrates remarkable performance under moderate to strong LoS conditions, consistently achieving the highest ergodic sum rate and even surpassing optimal beamforming. P-6DMA-MRT also improves with increasing $\kappa$, benefiting from stronger deterministic LoS components. When $\kappa$ approaches $30$\,dB, P-6DMA-MRT reaches performance levels comparable to that of FPA-ZF and even FPA-OPT, offering a simpler alternative to more complex beamforming designs. However, P-6DMA-MRT remains inferior to P-6DMA-ZF across all values of $\kappa$, highlighting the limitations of MRT beamforming in scenarios where interference suppression is crucial. These results underscore the importance of wireless channel reshaping and exploiting the spatial DoF enabled by positionable 6DMAs and indicate the synergy of combining beamforming and positionable 6DMA techniques for effective interference management. 

\subsection{Channel Estimation for Positionable 6DMA}
In this subsection, we discuss the existing channel estimation techniques for P-6DMA, which can be categorized into two branches, i.e., those for sparse channels and those for rich-scattering channels.
\subsubsection{Rich-Scattering Channels}
Due to the complexity of the environment, positionable 6DMA systems often operate in rich-scattering channels. Traditional channel estimation techniques designed for sparse MIMO channels may not perform effectively in such environments. To solve this problem, some prior works have proposed to exploit some learning-based tools such as successive Bayesian reconstructor and deep neural networks.
For example, the authors in \cite{zhang2024successive} studied an uplink channel estimation problem in a positionable 6DMA (i.e., fluid antenna) system comprising a BS equipped with $Q$ P-6DMAs and a single-FPA user. The receive region at the BS is equally discretized into $M$ sampling points/ports, to which the $Q$ antennas can be repositioned. Let $\mv{h}\in\mathbb{C}^{M\times 1}$ and $P$ denote the channels from the user to the $M$ points/ports as described in \eqref{p6} and the number of transmit pilots within a coherence time frame, respectively. Define $\mv{S}\in\{0,1\}^{Q\times M}$ as the antenna-port association matrix in time slot $p, p=1,2,\cdots,P$, where the $(i,m)$-th entry of $\mv{S}$ is equal to one if and only if the $i$-th antenna is located at the $m$-th port. As such, the received signal at the BS in time slot $p$ is given by
\begin{align}
	\mv{y}_p=\mv{S}_p\mv{h}s_p+\mv{z}_p,
\end{align}
where $s_p$ denotes the pilot transmitted by the user and $\mv{z}_p\sim\mathcal{CN}(\mv{0}_Q,\sigma^2\mv{I}_Q)$ denotes the AWGN at the $Q$ selected ports in time slot $p$. For simplicity, we assume that $s_p=1, p\in\{1,2,\cdots,P\}$. Thus, over the $P$ time slots for pilot transmission, we have
\begin{equation}\label{rxpilot}
	\mv{y}=\mv{S}\mv{h}+\mv{z},
\end{equation}
where $\mv{y}=[\mv{y}_1^T,\cdots,\mv{y}_p^T]^T$, $\mv{S}\in[\mv{S}_1^T,\cdots,\mv{S}_P^T]^T$ and 
$\mv{z}=[\mv{z}_1^T,\cdots,\mv{z}_p^T]^T$. The aim in \cite{zhang2024successive} was to reconstruct the $M$-dimensional channel $\mv{h}$ based on the noisy received signal $\mv{y}$.

To this end, we can move the $Q$ antennas to different positions over the $P$ time slots and measure their channels with the user accordingly. It is noted that this is similar to a successive sampling process. Moreover, as the spacing between nearby sampling points/ports is small, their channels with the user show high correlation. Inspired by the above, the authors in \cite{zhang2024successive} proposed to reconstruct $\mv{h}$ by using successive Bayesian regression, where an experiential kernel was built to facilitate channel reconstruction. Furthermore, the channel $\mv{h}$ is modeled as a sample of a Gaussian process with zero mean and prior covariance $\mv{\Omega} \in {\mathbb{C}}^{M \times M}$, and the uncertainty of this stochastic process is successively eliminated by kernel-based sampling and regression. Specifically, the successive Bayesian regression is divided into two stages, i.e., an offline stage and an online regression. In the offline stage, the authors determined the association matrix $\mv{S}$ and combining weights $\mv{w}\in\mathbb{C}^{PQ\times 1}$ by following the principle of maximum posterior variance. In the online regression stage, the designed switch matrix $\mv{S}$ is employed at the BS to obtain the received pilots as in (\ref{rxpilot}). Then, the channel $\mv{h}$ can be reconstructed as $\hat{\mv{h}}=\mv{w}^H\mv{y}$.
\begin{figure}[!t]
	\centering
	\includegraphics[scale=0.48]{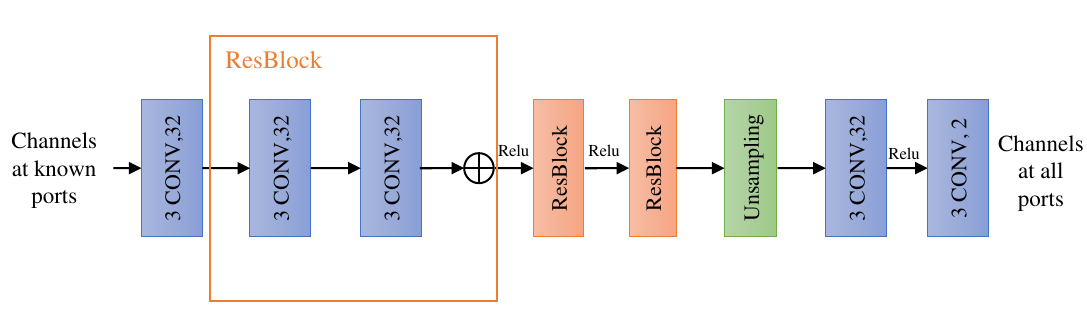}
	\caption{Machine learning network architecture for positionable 6DMA channel estimation.}\label{NN_est}
	\vspace{-0.59cm}
\end{figure}

Alternatively, the authors in \cite{ji2024correlation} considered a single-input-single-output (SISO) communication system with a single P-6DMA receiver and a single-FPA transmitter. The antennas could be switched to one of the $M$ sampling points/ports within the receive region. Since the total number of points/ports $M$ in a positionable 6DMA system is huge, the channel estimation overhead is costly. Estimating a subset of the points/ports first and then recovering others via interpolation is a more practical solution. To this end, the authors in \cite{ji2024correlation} introduced a new machine learning technique, as depicted in Fig. \ref{NN_est}. The network is based on ResNet, a lightweight network with 9 conventional layers and 22000 tunable parameters, which is implementable on mobile devices. In particular, the estimated channels at a set of selected points/ports are used as input, and are reshaped into their real and imaginary parts for subsequent processing. The first layer is a convolutional layer comprised of 32 filters with a kernel size of 3, followed by three ResBlocks which consist of two convolutional layers, a Relu activation function and an identity connection. The convolutional layers in the ResBlocks use 32 filters with a kernel size of 3. Following these there is an unsampling layer and a convolutional layer comprised of 2 filters with kernel size 3. The output of the network provides the estimated channels at all points/ports\cite{ji2024correlation}.

\subsubsection{Sparse Channels}
With the advancement of millimeter-wave and terahertz technologies, high-frequency positionable 6DMA channels often exhibit sparsity. For such sparse channels, the existing works have mostly focused on leveraging their structural properties to recover the channel map within the transmit/receive movement region with various algorithms, e.g., via compressed sensing. For example, consider a P-6DMA-assisted SISO communication system where the transmitter/receiver is equipped with a single antenna. The antenna can flexibly adjust its position in a specified 2D transmit/receive region denoted by $\ca{C}^t$/$\ca{C}^r$. Based on the positionable 6DMA channel model described in \eqref{p6}, the end-to-end transmitter-receiver channel is given by \cite{zhumo, wu, Zhu2024MovableAntennas, mawide}
\begin{equation}\label{eqn_SISO_Ch}
	h(\mathbf{q}^{\mathrm{t}},\mathbf{q}^{\mathrm{r}})=\bm c^H(\mathbf{q}^{\mathrm{r}})\bs{\Sigma}\bs{t}(\mathbf{q}^{\mathrm{t}}),
\end{equation}
where receive steering vector $\mathbf{c}(\mathbf{q}^{\mathrm{r}})$,  channel coefficient matrix $\bs{\Sigma}$, and transmit steering vector $\bs{t}(\mathbf{q}^{\mathrm{t}})$ are defined in \eqref{p66}.

It is unaffordable to estimate $h(\mathbf{q}^{\mathrm{t}},\mathbf{q}^{\mathrm{r}})$ directly for each pair of $(\mathbf{q}^{\mathrm{t}},\mathbf{q}^{\mathrm{r}})$ in the  transmit/receive antenna movement regions $\ca{C}^t$/$\ca{C}^r$, as the pilot resources for channel estimation are limited. However, it can be observed from \eqref{eqn_SISO_Ch} that the end-to-end channel depends on the channel paths' AoDs $\left\{\theta_l^t,\phi_l^t\right\}_{l=1}^{L_t}$, channel paths' AoAs $\left\{\theta_l^r,\phi_l^r\right\}_{l=1}^{L_r}$, and channel coefficient matrix $\bs{\Sigma}$, where \(L_t\) and \(L_r\) denote the number of transmitter-side and receiver-side multipath components, respectively. Motivated by this, a compressed sensing-based method is proposed in \cite{ma2023compressed}, where the AoAs/AoDs/channel coefficient matrix are successively estimated. Specifically, let $M_t$ denote the total number of sampled positions for the transmit antenna, which are denoted by $\bs{\tilde{t}}=\left[\mathbf{q}^{\mathrm{t}}_1,\cdots,\mathbf{q}^{\mathrm{t}}_{M_t}\right]\ib{R}^{2\times M_t}$ with $\mathbf{q}^{\mathrm{t}}_m=[x_m^t,y_m^t]^T\ic{C}^t$. We assume that one pilot is transmitted for each $\mathbf{q}^{\mathrm{t}}_m$, and the position of receive antenna is fixed as $\mathbf{q}^{\mathrm{r}}_1\ic{C}_r$. Then, the received signals are given by
\begin{equation}
	\begin{aligned}
		\left(\bs{y}^t\right)^T=(\bs{x}^t)^H\bs{G}(\bs{\tilde{t}})+\left(\bs{z}^t\right)^T,
	\end{aligned}
\end{equation}
where $\bs{G}(\bs{\tilde{t}})\triangleq\left[\bs{t}(\mathbf{q}^{\mathrm{t}}_1),\bs{t}(\mathbf{q}^{\mathrm{t}}_2),\cdots,\bs{t}(\mathbf{q}^{\mathrm{t}}_{M_t})\right]\ib{C}^{L_t\times M_t}$, $\bs{x}^t\triangleq\sqrt{\bar{P}}\bs{\Sigma}^H\bm c(\mathbf{q}^{\mathrm{r}}_1)$ with $\bar{P}$ being the transmit power, and $\bs{z}^t\sim\ca{CN}(0,\sigma^2\bs{I}_{M_t})$ is the AWGN vector with $\sigma^2$ denoting the average noise power. To efficiently estimate the AoDs $\left\{\theta_p^t,\phi_p^t\right\}_{p=1}^{L_t}$ by compressed sensing, we discretize $\theta\in[-1,1]$ and $\phi\in[-1,1]$ into $G$ grid points with $G\gg L_t$, and can show  $(\bs{x}^t)^H\bs{G}(\bs{\tilde{t}})\approx(\bs{\bar{x}}^t)^H\bs{\bar{G}}(\bs{\tilde{t}})$, where $\bs{\bar{G}}(\bs{\tilde{t}})\ib{C}^{G^2\times M}$ is an over-complete matrix and its $\left(g_1+\left(g_2-1\right)\right)$-th row is in the form of $\left[e^{j\frac{2\pi}{\lambda}\bar{\bs{u}}^T_{g_1,g_2}\mathbf{q}^{\mathrm{t}}_1},e^{j\frac{2\pi}{\lambda}\bar{\bs{u}}^T_{g_1,g_2}\mathbf{q}^{\mathrm{t}}_2},\cdots,e^{j\frac{2\pi}{\lambda}\bar{\bs{u}}^T_{g_1,g_2}\mathbf{q}^{\mathrm{t}}_{M_t}}\right]\ib{C}^{1\times M_t}$ with $\bs{\bar{u}}_{g_1,g_2}\triangleq[-1+2g_1/G,-1+2g_2/G]^T$, $1\le g_1,g_2\le G$, and $\bs{\bar{x}}^t \in \mathbb{C}^{G^2 \times 1}$ is a sparse vector containing $L_t$ non-zero elements. When $G \gg L_t$, the problem of estimating AoDs can be reformulated as a sparse signal recovery task. Specifically, it involves finding a sparse vector $\bs{\bar{x}}^t$ that minimizes the norm $||\left(\bs{y}^t\right)^T - (\bs{\bar{x}}^t)^H \bs{\bar{G}}(\bs{\tilde{t}})||_2$. This optimization can be effectively addressed using classical compressed sensing algorithms such as OMP. Similarly, to estimate the AoAs \(\{\theta_q^r, \phi_q^r\}_{q=1}^{L_r}\), the transmit antenna is fixed at \(\mathbf{q}_1^{\mathrm{t}}\) while the receive antenna moves to \(M_r\) different positions. The same method can then be applied.
Finally, based on the channel measurements in the previous two steps and the estimated AoDs/AoAs, the channel coefficient matrix is estimated using the LS algorithm.

However, the above sequential parameter estimation may lead to a cumulative error and require high channel measurement overhead. To overcome this limitation, a new channel estimation framework based on compressed sensing was proposed in \cite{xiao2024channel}, where the AoDs, AoAs, and complex channel coefficients are jointly estimated via a smaller number of channel measurements. In addition, the measurement matrix is fundamentally determined by the antenna measurement positions, which has a significant influence on the channel estimation performance. To address this issue, the authors investigated the mutual coherence of the measurement matrix via Fourier transform, and proposed two criteria for the antenna measurement positions to successively resolve the multi-path components in the angular domain.

It is worth noting that the above two compressed-sensing based channel estimation algorithms for SISO systems are also applicable to MIMO systems. Nonetheless, both approaches depend on discrete angle quantization for sparse channel representations, which inherently restricts the accuracy of channel estimation. To enhance channel estimation precision while minimizing the pilot training overhead, the authors in \cite{zhang2024channel} introduced a method based on tensor decomposition to estimate multi-path component parameters. Specifically, they implemented a two-stage transmit-receive antenna movement strategy during pilot training, ensuring that the received pilot signals in each stage form a third-order tensor. The factor matrices of this tensor are then derived using canonical polyadic (CP) decomposition, which facilitates the estimation of AoDs, AoAs, and the complex channel coefficients. This process enables the reconstruction of the wireless channel for any position pair within $\ca{C}^t$/$\ca{C}^r$.
\begin{figure}[!t]
	\centering
	\includegraphics[width=3.1in]{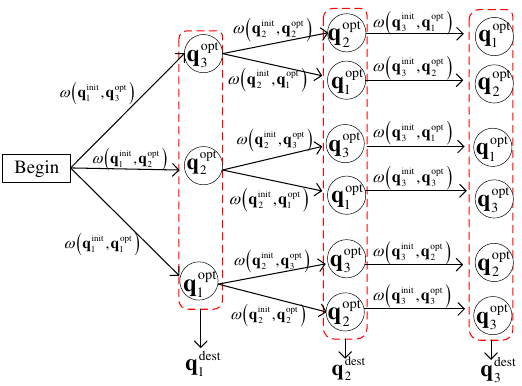}
	\caption{An example of antenna position matching with $\widetilde{N}=3$.}
	\label{pathplan}
	\vspace{-0.59cm}
\end{figure}

\subsection{Path Planning for Positionable 6DMA}
\begin{figure*}[!t]
	\centering
	\includegraphics[width=7.2in]{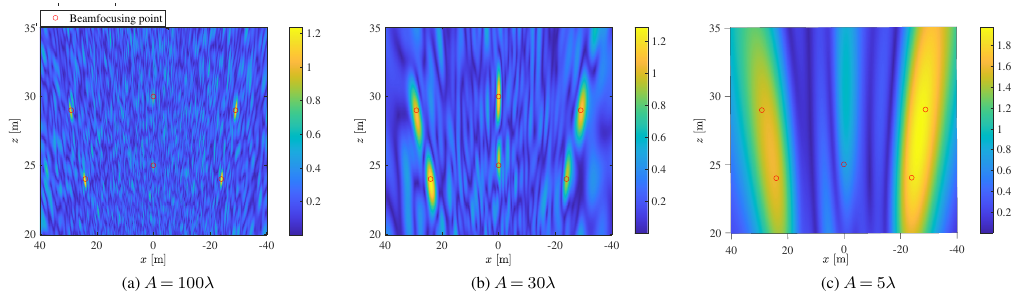}
	\caption{Beampattern for different sizes of the movement region.}
	\label{MAISAC}
	\vspace{-0.59cm}
\end{figure*}
After determining the optimized positions of P-6DMAs, a practical challenge arises in scheduling their movement. This involves transitioning the 6DMAs to their target positions over time while minimizing disruptions to communication channels, which may degrade system performance. Efficient scheduling across all transmitters and receivers is critical to minimizing performance loss, as well as reducing energy consumption and time costs.

To address this challenge, an antenna position matching method was proposed in \cite{jing}, which employs a greedy algorithm to minimize the total movement distance of antennas. Specifically, the initial and optimized positions of the $N$ antennas, before and after position optimization, are denoted as $\mathbf{q}_n^{\text{init}} \in \mathbb{R}^{3\times1}$ and $\mathbf{q}_n^{\text{opt}} \in \mathbb{R}^{3\times1}, 1 \leq n \leq N$, respectively. Specifically, the problem of minimizing the total movement distance can be modeled as a shortest-path search in a tree graph. For instance, as illustrated in Fig.~\ref{pathplan}, when $N = 3$, the movement distance between different positions is given by
\begin{align}
	\omega\left(\mathbf{q}_n^{\text{init}}, \mathbf{q}_n^{\text{opt}}\right) = \|\mathbf{q}_n^{\text{init}} - \mathbf{q}_n^{\text{opt}}\|_2, \quad 1 \leq n \leq N.
\end{align}
Each antenna selects a destination position $\mathbf{q}_n^{\text{dest}} \in \mathbb{R}^{3\times1}$ based on the movement distance. Once a destination is assigned to an antenna, it becomes unavailable for the others.
Exhaustive evaluation of all possibilities is computationally infeasible for large $N$. Instead, a greedy strategy is adopted to find a suboptimal solution. The algorithm initializes the antenna index set as $\mathcal{N}^0 = \{1, \dots, N\}$. For each antenna, the destination $\mathbf{q}_n^{\text{dest}}$ is chosen from the updated index set to minimize the movement distance
\begin{align}
	\mathbf{q}_n^{\text{dest}} = \arg \min_{\mathbf{q}_n^{\text{opt}}} \left\{ \omega\left(\mathbf{q}_n^{\text{init}}, \mathbf{q}_n^{\text{opt}}\right) \mid n \in \mathcal{N}^{n-1} \right\}.
\end{align}
Once a destination is selected, the corresponding antenna index is removed from set $\mathcal{N}^{n-1} \setminus \{\hat{n}\}$.
This process continues until all $N$ antennas are assigned destinations. By focusing on minimizing the movement distance, the algorithm helps reduce energy and time costs in positionable 6DMA systems, particularly when antennas operate in large movement regions.

In addition, an extended approach was proposed in \cite{pathplan2} to address movement delay minimization by considering not only position assignments but also trajectory planning for antennas moving from their initial positions to their destinations within a 2D region. The goal is to minimize the movement delay while maintaining a minimum distance between antennas to avoid collisions. This problem is formulated as a continuous-time mixed-integer linear programming  problem, which is computationally demanding. To simplify it, a two-stage framework is used, where position associations are first determined to map initial and destination positions optimally, followed by trajectory optimization to design efficient movement paths while avoiding collisions. The proposed algorithm was shown to achieve near-optimal performance in delay minimization \cite{pathplan2}. Furthermore, the study emphasizes that moving antennas along straight-line paths may result in excessive delays due to the extended waiting times required for collision avoidance. This finding indicates that alternative trajectory designs, such as curved paths, could further enhance scheduling efficiency. Other strategies such as parallel movement, sequential movement, or collaborative movement among 6DMAs are promising but require further investigation to improve scheduling efficiency.

\subsection{Positionable 6DMA-aided ISAC}
%ISAC is a key enabler of next-generation wireless networks, integrating communication and radar sensing into a unified system that utilizes shared frequency bands and hardware. In ISAC systems, sensing functions aim to acquire critical information about targets and their environment. A significant advancement in this area is the adoption of MIMO technology, which improves spatial beamforming and waveform shaping through advanced precoding techniques. However, traditional ISAC systems often depend on FPA arrays, which limits their ability to fully utilize the continuous spatial domain's DoFs. One of the primary advantages of positionable 6DMA is its ability to unify communication and sensing by establishing ISAC channels where the two functions complement each other. 

The introduction of P-6DMA also provides a transformative improvement for ISAC systems. At the transmitter side, positionable 6DMA allows joint optimization of antenna positions and precoders, facilitating the creation of favorable communication and sensing channels at the same time. At the receiver side, positionable 6DMA mitigates interference caused by radar sensing by leveraging channel envelope variations\cite{zr17}. This technology also enables flexible spatial resource allocation, which ensures a balance between radar sensing and communication functions. However, P-6DMA-aided ISAC systems also face significant challenges, particularly in optimizing antenna positions and beamforming, which is an NP-hard problem. To tackle this issue, advanced techniques, including deep reinforcement learning with pointer networks and the gradient projection method, have been proposed in \cite{p61} and \cite{p101}.

In addition, the traditional far-field assumption may no longer hold in scenarios involving large antenna movement regions \cite{zr6,zr7}. To enhance the performance of such systems, a full-duplex BS equipped with multiple P-6DMA arrays for near-field ISAC was proposed in \cite{jing}. The P-6DMA arrays at the BS could simultaneously sense multiple targets and serve multiple users in the uplink and downlink, respectively. The authors also proposed a two-layer random positioning algorithm to maximize the weighted sum of sensing and communication rates by designing transmit beamformers, sensing signal covariance matrices, receive beamformers, and antenna positions. In the inner layer, these variables are updated iteratively for given antenna positions using an AO method, while the outer layer assigns random transmit and receive antenna positions, choosing the configuration that maximizes system performance rate. 

To demonstrate the advantages of large movement regions in near-field ISAC systems, Fig.~\ref{MAISAC} presents beampatterns for a positionable 6DMA array with 128 transmit antennas. When the side length of the transmit movement region, denoted by \(A\), is \(100\lambda\) with \(\lambda\) as the wavelength (see Fig.~\ref{MAISAC}(a)), the beam is precisely focused at multiple desired locations, with extremely narrow main lobes. This shows that the beams focused on the desired locations cause minimal interference to other areas, enabling the near-field ISAC system to achieve superior performance in multi-target sensing and multi-user communication through joint resolution in both the distance and angle domains. However, as \(A\) decreases (see Fig.~\ref{MAISAC}(b) and Fig.~\ref{MAISAC}(c)), the main lobes of the beams become wider due to the reduced maximum aperture provided by the positionable 6DMA array, which results in significant interference to targets and users at undesired locations and degrades the overall system performance. In other words, when multiple targets and users share the same direction relative to the BS, a far-field ISAC system fails to provide effective sensing and communication services.
\begin{figure}[!t]
	\centering
	\includegraphics[width=2.7in]{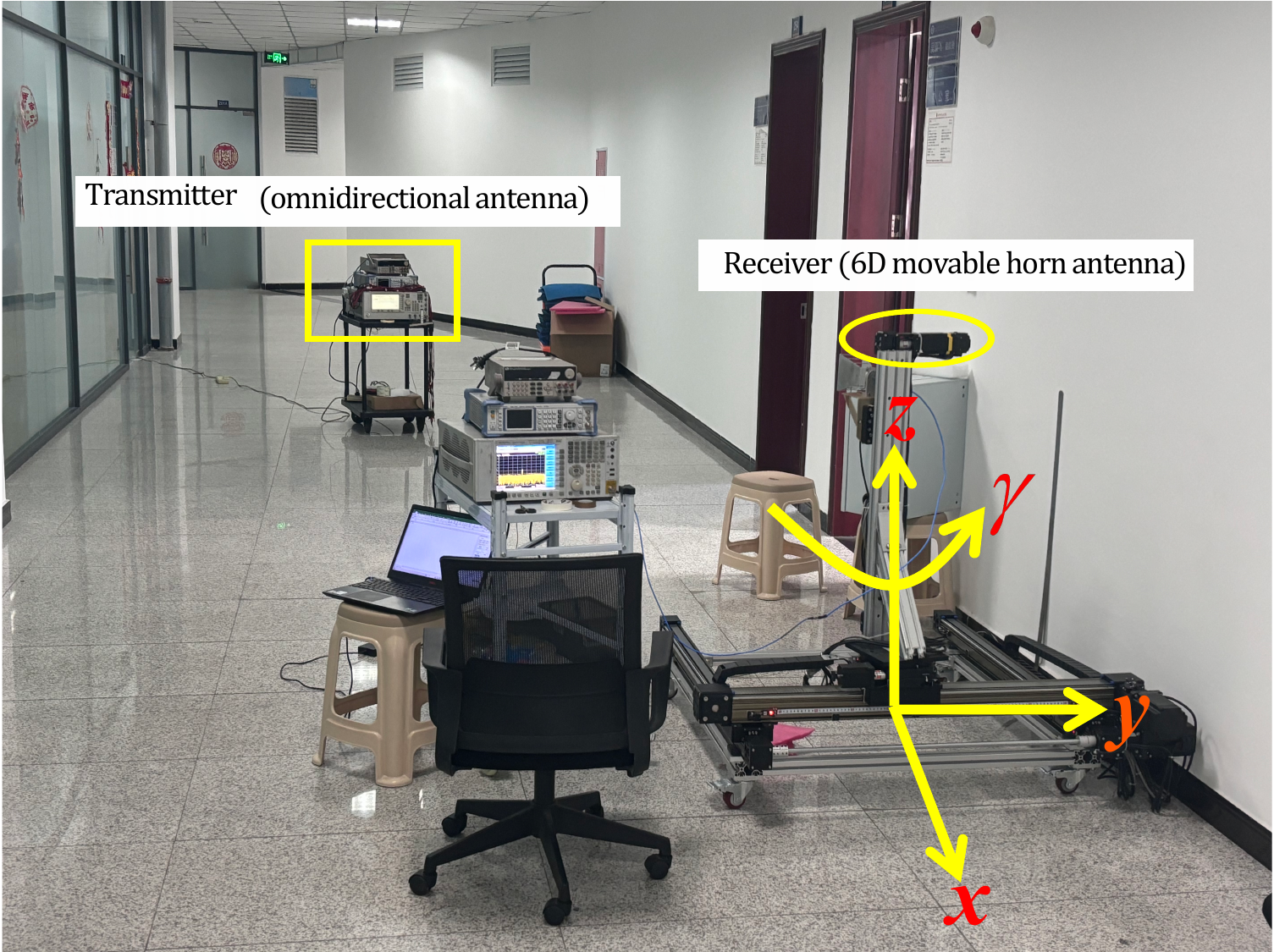}
	\caption{Mechanically-driven 6DMA communication prototype.}
	\label{weidongp}
	\vspace{-0.3cm}
\end{figure}
\subsection{Other Related Works and Future Directions}
Tables \ref{6DMApr} and \ref{tab:6dma_summary} summarize representative works on position optimization and channel estimation for P-6DMA, respectively. While these studies highlight critical areas, there are numerous other potential applications for P-6DMA that remain unexplored in this tutorial \cite{zr9}. 
For instance, P-6DMA also offers unique advantages in managing co-channel interference without relying on precoding, which makes it an excellent fit for cognitive radio networks\cite{wei2024joint}. Moreover, it can function as a hybrid beamforming solution in the analog domain without requiring phase shifters. This suggests that hybrid signal processing techniques could be employed to further optimize the performance of P-6DMA architectures\cite{zhang2024hybrid}.
Another potential application lies in the emerging field of semantic communications, which is often deployed in challenging environments characterized by narrowband and interference-rich conditions. In such scenarios, P-6DMA could improve communication reliability by mitigating interference and reducing reliance on heavy channel coding. In addition, P-6DMA's ability to enhance wireless relaying is particularly noteworthy, as it can improve both reception and transmission processes at source, relay, and destination. Unlike antenna position optimization for other scenarios, P-6DMA-assisted relaying introduces a two-stage optimization challenge, i.e., adjusting positions for both source-to-relay and relay-to-destination transmissions. In \cite{li2024movable}, the authors examined a P-6DMA-assisted amplify-and-forward relay system and proposed an AO algorithm to find a locally optimal solution for antenna positioning across both stages. 
These examples underscore the untapped potential of P-6DMA in wireless communication systems, which offers opportunities for further innovation and exploration.
\begin{figure}[!t]
	\centering
	\includegraphics[width=3.3in]{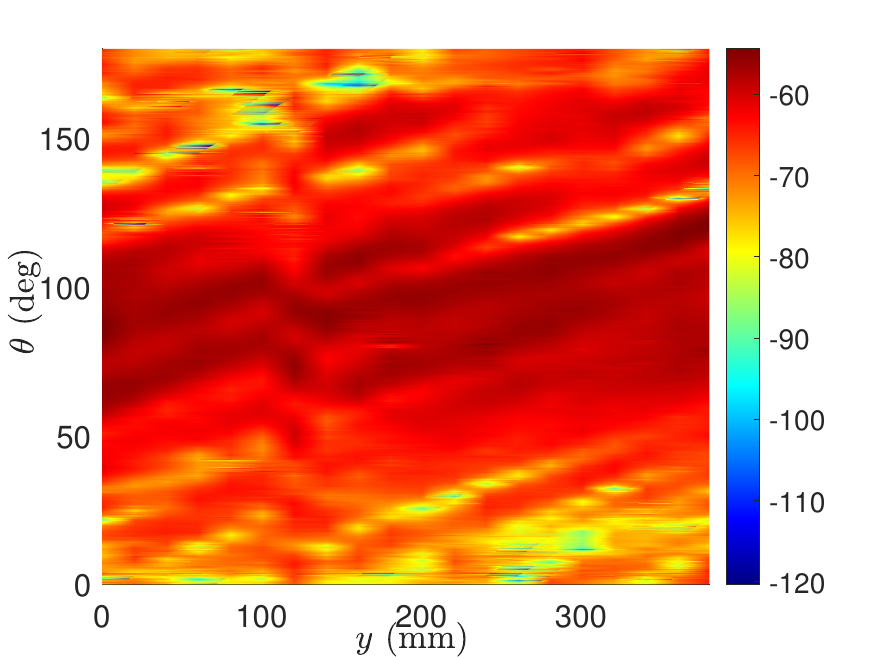}
	\caption{The measured signal power (in dBm) with the 6DMA prototype.}
	\label{weidongp2}
	\vspace{-0.59cm}
\end{figure}
\section{Prototypes and Experiments}
This section provides an overview of the prototypes developed for 6DMA-enabled wireless systems, including a 6DMA prototype with both flexible antenna positions and rotations, as well as rotatable 6DMA and positionable 6DMA prototypes.
\subsection{6DMA Prototype}
Recently, \emph{Mei et al.} from the University of Electronic Science and Technology of China (UESTC) introduced a mechanically driven 6DMA prototype capable of repositioning a horn receive antenna along three spatial axes (\(x\), \(y\), \(z\)) and rotating it by an angle \(\gamma\), as depicted in Fig.~\ref{weidongp}. The transmitter antenna is an omnidirectional antenna. By mechanically adjusting both the position and orientation, the horn’s main lobe can be steered to align with various spatial angles. Fig.~\ref{weidongp} illustrates the testing scenario for this prototype in an indoor environment. In this setup, the horn antenna, serving as the receiver, achieves a main-lobe gain of 12~dBi, while the transmitter operates with 0~dBm of power at a carrier frequency of 3.5~GHz.

The color map in Fig.~\ref{weidongp2} depicts how the received signal power (in dBm) in the 6DMA prototype varies with changes in the antenna’s \(y\)-position and rotation angle \(\gamma\). Notably, even when the antenna is held at the same location along \(y\), sweeping \(\gamma\) can alter the received power by more than 60~dB, with the peak power typically observed between \(60^\circ\) and \(100^\circ\). These measurements underscore that adjusting the antenna rotation, on top of simply changing its position, yields extra DoFs for enhancing channel quality.

To enable 6DMA with flexible antenna positioning and rotation, the hardware design and implementation remain at an early stage and are deliberately largely left open for engineers to determine the optimal configurations. For example, future experiments could extend the current single-user 6DMA prototype to a multi-user setup, with multiple freely rotating and positioning 6DMA surfaces at the transmitter controlled by robotic arms. This approach will further validate how 6DMAs adapt to the user distribution to achieve array, spatial multiplexing, and geometric gains while effectively suppressing interference.

\subsection{Positionable 6DMA Prototype}
The authors of \cite{hanchong} developed a wideband channel measurement platform with a 0.2~mm-precision positionable 6DMA system, which enables adjusting the antenna's position (i.e., translation) along a 2D plane while keeping its rotation angle fixed. This setup consists of a computer for control, a displacement system, and a measurement system comprising transmitter/receiver modules and a Ceyear 3672C vector network analyzer for RF and local oscillator (LO) signals. The transmitter, mounted on the displacement system, moves in the $x–z$ plane at a precision of 0.02~mm. The computer alternates between positioning the transmitter and triggering the vector network analyzer. Measurements start from the lower-left corner and proceed horizontally then vertically, taking about 2.3~s per position.
\begin{figure}[!t]
	\centering
	\includegraphics[width=2.8in]{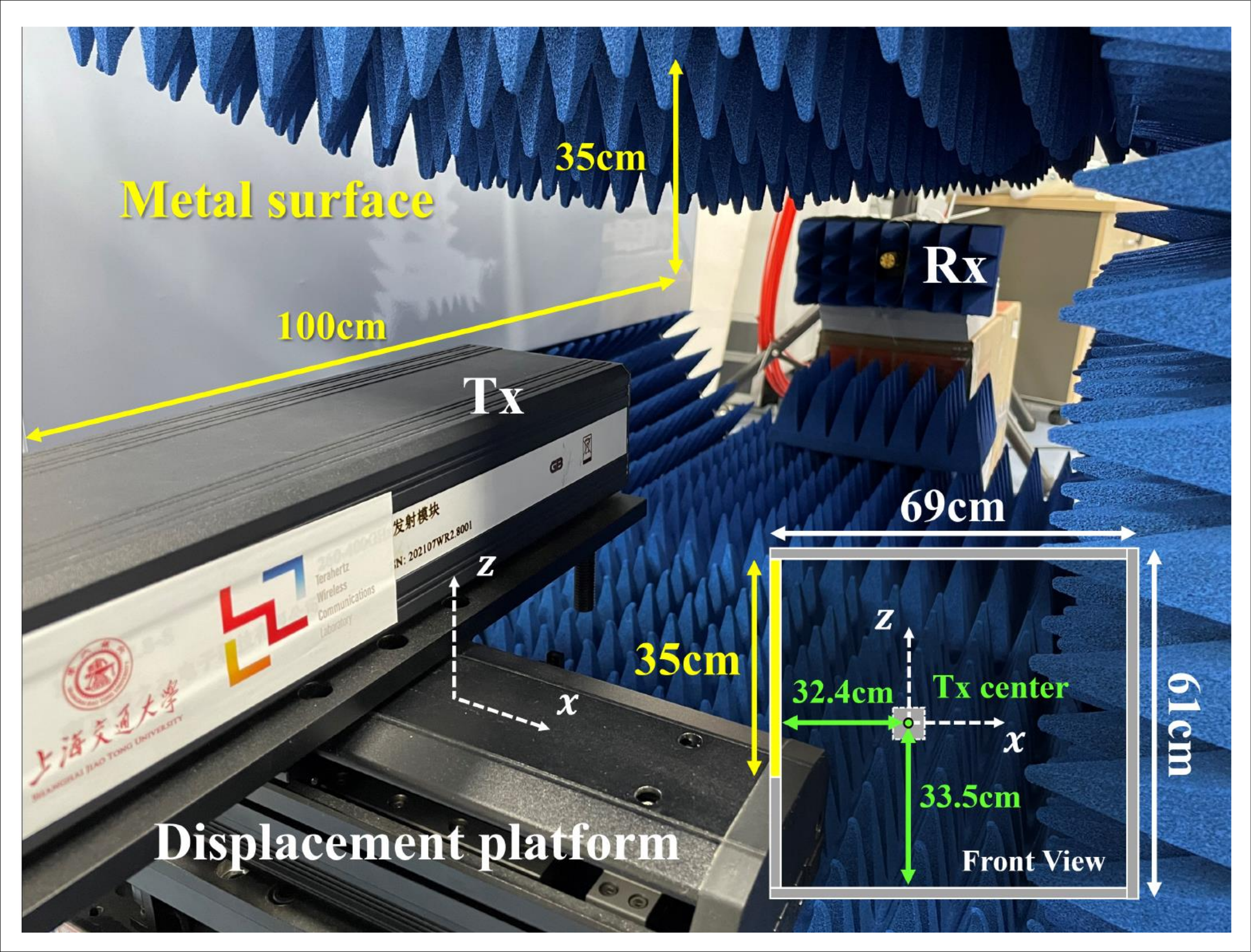}
	\caption{Port selection-driven positionable 6DMA communication prototype \cite{hanchong}.}
	\label{hanchong}
	\vspace{-0.59cm}
\end{figure}

As shown in Fig.~\ref{hanchong}, a 0.69~m\,\(\times\)\,0.61~m\,\(\times\)\,1.00~m anechoic chamber confines the wireless channel, with one side covered by a metal plate and the others lined with absorbers to reduce multipath. The transmitter and receiver stand at opposite ends, with the receiver fixed at the center of the transmitter. The transmitter has 32\,\(\times\)\,32 ports (1~mm spacing). The transmitter and receiver are separated by 0.86~m. The system operates at 300~GHz. Experiments show that the use of P-6DMAs mitigates multipath fading more effectively than FPAs. In addition, the P-6DMA system combined with a uniform-region SINR-optimized position selection algorithm improves spectral efficiency by 11.48\% within the \(32 \times 32~\text{mm}^2\) area at THz frequencies, compared to a FPA.

Other positionable 6DMA prototypes have been investigated. For instance, the authors in \cite{Dong2024Movable} developed a positionable 6DMA communication system prototype, where the antenna is capable of moving along a one-dimensional horizontal line with a step size of $0.01\lambda$ or within a two-dimensional square region with a step size of $0.05\lambda$. The prototype employs feedback control to guarantee that each power measurement is conducted only after the antenna has reached the specified position.

\subsection{Rotatable 6DMA Prototype}
\begin{figure}[!t]
	\centering
	\includegraphics[width=3.3in]{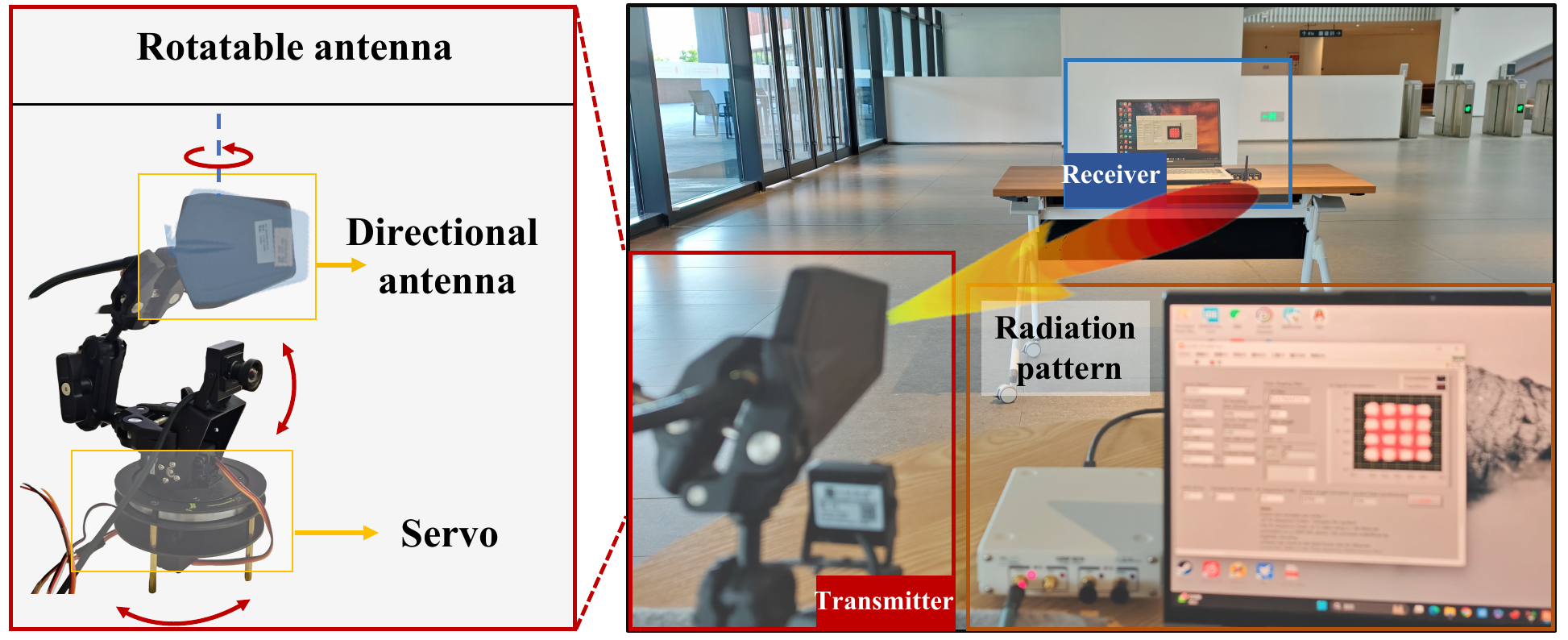}
	\caption{Mechanically-driven rotatable 6DMA communication prototype.}
	\label{Fig_RAP}
	\vspace{-0.3cm}
\end{figure}

\begin{figure}[!t]
	\centering
	\includegraphics[width=3.2in]{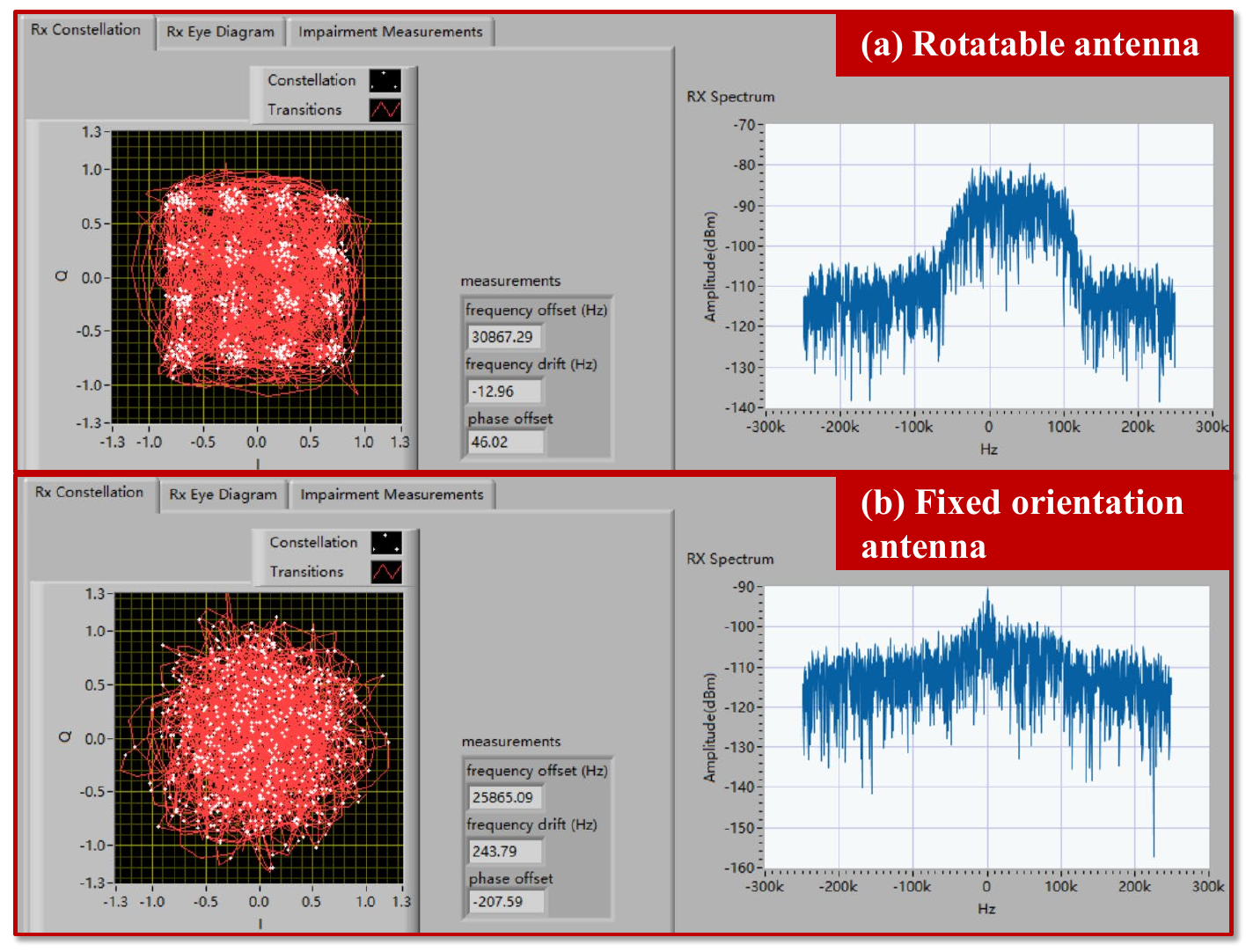}
	\caption{Measured receive constellation and power spectrum for the rotatable 6DMA prototype and a fixed orientation antenna, respectively.}
	\label{Fig_RAM}
	\vspace{-0.59cm}
\end{figure}

More recently, a mechanically-driven rotatable 6DMA prototype was demonstrated, which enables adjusting the antenna’s rotation angle while keeping its position fixed. As shown in Fig.~\ref{Fig_RAP}, the prototype allows the 3D rotation of a directional antenna to be mechanically adjusted by a servo, enabling dynamic alignment of its main lobe to different spatial directions. Experimental results for the rotatable 6DMA prototype in an indoor environment are presented in Fig.~\ref{Fig_RAM}. Specifically, the rotatable antenna serves as a transmitter with a main lobe beam gain of 10 dBi and operates at a carrier frequency of 5.2 GHz (within Wi-Fi bands). The modulation scheme adopted at the transmitter is 16-quadrature amplitude modulation (16QAM), with a transmit power of 10 dBm and a transmission rate of 0.5 Mbps.

During the experiment, as the receiver moved to another position, the rotatable antenna adjusted the orientation of its main lobe in real time to track the receiver's current position. This adjustment was controlled by visual equipment that utilized a deep learning model for the identification and tracking of the receiver (communication object). When the rotatable antenna was properly configured, the receiver was able to reliably receive the signal and accurately decode the constellation diagram, as shown in Fig.~\ref{Fig_RAM}(a). In contrast, for the fixed orientation antenna system, the receiver's movement resulted in a discrepancy between the antenna orientation of the transmitter and the receiver, thereby leading to a distorted decoding of the constellation diagram, as shown in Fig.~\ref{Fig_RAM}(b). It was observed that the orientation adjustment capability of the rotatable antenna enhanced the received signal power compared to the system with a fixed orientation, offering a gain of over 10 dBm. Consequently, the experimental results further verified that by dynamically adjusting the orientation of the rotatable 6DMA, additional spatial DoF can be achieved to significantly enhance the received signal strength \cite{demora, wu2024modeling,zheng2025rotatable}.

Other methods for implementing rotatable 6DMA prototypes have also been explored. For example, the authors of \cite{rotimp} proposed using parasitic elements to implement a rotating antenna for MIMO receivers. Furthermore, a rotatable 6DMA prototype with \(15 \times 15 \) manually rotated elements has been fabricated and presented in \cite{mpro}, where full-wave simulation results demonstrated that the gain of the beam focused in the user direction was 25.8~dB, with a 1-dB gain bandwidth exceeding $28.6\%$.  

\subsection{Other Related Works and Future Directions}
Various strategies have been explored for implementing 6DMA systems. One promising approach is the use of pixel-reconfigurable antennas for positionable 6DMA, as proposed in \cite{225ant}, where an antenna with 49 switches theoretically supports up to \(2^{49}\) possible states. Furthermore, a pixel-based positionable 6DMA antenna prototype described in \cite{zhang2024pixel} achieved a signal power enhancement of approximately 30 dB at 2.5 GHz. In addition, liquid-based positionable 6DMA prototypes for wireless communication systems, as demonstrated in \cite{Shen2024Design}, showed that in a 4-user system, double-channel fluid antennas could reduce the outage probability by 57\% compared to static omnidirectional antennas. 
Each 6DMA control method comes with unique advantages and limitations. Current prototypes predominantly rely on a single control mechanism, such as mechanical control, liquid-based control, or electronic tuning. However, hybrid control prototypes for 6DMA systems remain largely unexplored, representing a significant opportunity for further innovation. Moreover, using prototypes for extensive channel measurements could be instrumental in developing empirical channel models and supporting performance analysis. Effectively utilizing these prototypes for such purposes remains a critical area for future research. These efforts would substantially enhance our understanding and optimization of 6DMA systems, paving the way for improved performance and practical deployment.

\section{Conclusion}
In this paper, we have provided a comprehensive tutorial on the emerging 6DMA technique as a promising enabler for exploiting the wireless channel spatial variations. 6DMA-aided wireless communication and sensing have led to a fundamental paradigm shift in the design of wireless system/network from the traditional one with fixed position antennas to a new
architecture comprising agile and adaptable antennas that operate in an intelligent way. Although research on 6DMA-assisted wireless networks is still in an early stage, we have
detailed their new challenges and potential research directions for future developments, covering topics ranging from theoretical 6DMA signal and channel modeling to practical 6DMA position and rotation optimization, channel estimation, and path planning. In addition, we have discussed two important special cases of 6DMA, namely, rotatable 6DMA with fixed antenna position and positionable 6DMA with fixed antenna rotation. Furthermore, we have demonstrated the effectiveness of 6DMA-enhanced communication systems through prototype development and experimental results. We hope this tutorial will inspire further 6DMA research to unlock its full potential in future generation (6G/Beyond-6G) wireless communication and sensing networks.

\bibliographystyle{IEEEtran}
\bibliography{fabs}
\end{document}